\pdfoutput=1 
\documentclass[a4paper,11pt]{article}
\usepackage[T1]{fontenc}
\usepackage{amsmath,amsfonts,amssymb,graphics}
\usepackage[normalem]{ulem}
\usepackage{graphicx}
\usepackage{mathrsfs}
\usepackage{bbm}
\usepackage{pifont}
\usepackage{braket}
\usepackage{footmisc}
\usepackage[dvipsnames]{xcolor}
\usepackage{mathtools}
\usepackage{xspace}
\usepackage{booktabs}
\usepackage{siunitx}
\usepackage{tikz}
\usetikzlibrary{positioning,shapes,arrows,calc}
\usetikzlibrary{decorations.pathmorphing}
\usetikzlibrary{decorations.markings}
\usepackage{standalone}
\usepackage{csquotes}
\usepackage{slashed}
\usepackage{braket}
\usepackage[title]{appendix}
\usepackage{jcappub}
\usepackage{hyperref}

\title{\boldmath \Large \bf{S.M.A.S.H.E.D.\;:}\\ 
\it{\textbf{S}tandard \textbf{M}odel \textbf{A}xion \textbf{S}eesaw  \textbf{H}iggs Inflation\\ \textbf{E}xtended for \textbf{D}irac Neutrinos }\unboldmath}

\author[a]{Maximilian Berbig}
\affiliation[a]{Bethe Center for Theoretical Physics und Physikalisches Institut der Universit\"at Bonn,\\
Nussallee 12, Bonn, Germany}
\emailAdd{berbig@physik.uni-bonn.de}

\abstract{
\noindent 
Inspired by the S.M.A.S.H. framework we construct a model that addresses the strong CP problem, axion dark matter, inflation and Dirac neutrino masses as well as leptogenesis. The model   possesses only two dynamical scales, namely the SM breaking scale $v_H$ and the Peccei Quinn (PQ) breaking scale $v_\sigma$.
We introduce heavy vector-like quarks in the usual KSVZ fashion to implement the PQ mechanism for the strong CP problem. To generate neutrino masses via a dimension six operator scaling as $m_\nu \sim v_H^3 / v_\sigma^2$ we add heavy triplet and doublet leptons, which are vector-like under the SM but chiral under PQ symmetry. The model is free from the cosmological domain wall problem and predicts an axion to photon coupling which is about an order of magnitude larger than in conventional DFSZ and KSVZ models. Thus our scenario can be probed and potentially excluded by current and next generation axion experiments such as ORGAN or MADMAX.
In addition we numerically demonstrate that our construction can generate the observed baryon asymmetry by realizing a   version of the Dirac-Leptogenesis scenario. As a consequence of our neutrino mass mechanism we find that the asymmetry in triplet fermion decays can also be significantly enhanced by up to six orders of magnitude when compared to typical Seesaw       scenarios without needing to invoke a resonant enhancement.
In passing we note that a decaying Dirac fermion with multiple decay modes contains all the necessary ingredients required for the \enquote{quasi optimal efficiency}-scenario previously encountered in the context decaying scalar triplets. The impact of the  right handed neutrinos and the axion on $\Delta N_\text{eff}$ is estimated and lies within current bounds.}

\begin{document}
\maketitle
\flushbottom

\section{Introduction}

Reductionism has been one of the most widely used approaches to building particle physics models that are supposed to address the theoretical, aesthetical and phenomenological gaps in the Standard Model (SM). For many decades the most common top-down approach consisted in unifying the SM gauge symmetries into single larger non-abelian Lie groups. This strategy led to the discovery of economical scenarios such as the Type I Seesaw-mechanism \cite{Minkowski:1977sc,Yanagida:1979as,Gell-Mann:1979vob,Glashow:1979nm,10.1143/PTP.64.1103, PhysRevLett.44.912} addressing both laboratory observations like neutrino masses and mixing as well as important cosmological issues such as the baryon asymmetry of the universe via leptogenesis \cite{FUKUGITA198645}. In more recent years the focus has shifted to bottom-up approaches  realizing the wanted phenomenology often via amending the SM with only a single additional  global or gauged $\text{U}(1)$ factor. The most prominent examples of this latter category are the $\nu$MSM \cite{Asaka:2005an,Asaka:2005pn,Shaposhnikov:2006xi}, which consists of a Type I Seesaw  with the lightest right handed neutrino being a good dark matter (DM) candidate, as well as the S.M.A.S.H. proposal \cite{Ballesteros:2016euj,Ballesteros:2016xej,Ballesteros:2019tvf}. Here solutions to the strong CP problem, neutrino masses, electroweak vacuum stability, dark matter, inflation and baryogenesis via leptogenesis are possible by combining a Type I Seesaw with a global anomalous  $\text{U}(1)_\text{PQ}$ Peccei-Quinn  symmetry playing the role of spontaneously broken lepton number. Building on an earlier construction \cite{Shin:1987xc} inspired by the KSVZ  \cite{PhysRevLett.43.103,SHIFMAN1980493} axion model  this framework identifies the mass scale of the heavy right handed neutrinos with the PQ breaking scale that in turn corresponds to the decay constant of the QCD axion (see also \cite{Salvio:2015cja} for a similar setup where the right handed neutrino mass does not arise from PQ breaking). There also exists a class of related models based on the DFSZ \cite{Zhitnitsky:1980tq,DINE1981199} approach see \cite{Sopov:2022bog} for a recent example.
Only two mass scales are present in the S.M.A.S.H. scenario: the electroweak breaking scale from the vacuum expectation value (vev) of the Higgs doublet scalar and the much larger vev of the PQ breaking singlet whose imaginary part is the axion. No new physics other than the PQ charged sector is needed  up to the Planck scale. Most theories beyond the SM address the neutrino mass issue via mechanisms inducing parametrically light Majorana masses since this usually involves the smallest amount of new unknown coupling constants and Weyl spinors. However a priori in the absence of any experimental signal there is no reason to focus only on Majorana neutrinos, which is why there has been renewed interest in building Dirac neutrino mass model (see \cite{Ma:2014qra,Ma:2015mjd,Ma:2015raa,Ma:2018bow,Gu:2007ug,Farzan:2012sa,CentellesChulia:2016rms} for some explicit models and \cite{CentellesChulia:2018gwr,CentellesChulia:2018bkz,CentellesChulia:2019xky} for systematic studies just to name a few). In this work we set out to extend the S.M.A.S.H. class of models for light Dirac neutrinos.
We outline the particle content and the most important interactions for the low energy phenomenology in section \ref{sec:model}. Section \ref{sec:smaschcosmo} serves a brief summary of the cosmological history and most important parameters for the original S.M.A.S.H. scenario.
The main focus of this work is Dirac-Leptogenesis in section \ref{eq:DiracLep}. A novel way to enhance the leptonic asymmetry parameter from heavy fermion decays is presented in subsection \ref{sec:boost}. Analytical estimates in \ref{sec:analy} help us narrow down the relevant parameter space and we show the validity of our scenario by numerically solving the Boltzmann equations  from section \ref{sec:Boltz} in subsection \ref{sec:num}.
In the aforementioned section we also demonstrate that the efficacy for asymmetry production from Dirac fermions can be larger than for Majorana fermions similar to \cite{Hambye:2005tk} for decaying scalar triplets.
After estimating the amount of dark radiation in section \ref{sec:darkrad} we summarize our findings in \ref{sec:sum}.
Additional relevant information was collected in appendices \ref{sec:lim}-\ref{sec:rates}.

\section{The model}\label{sec:model}
\begin{table}[t]
\centering
 \begin{tabular}{|c||c|c|c|c||c|} 
 \hline
  field&   $\text{SU(3)}_\text{C}$ & $\text{SU}(2)_\text{L}$ & $\text{U}(1)_\text{Y}$ & $\text{U}(1)_\text{PQ}$ & \text{generations}\\
 \hline
 \hline 
    $q_L$&  3 & 2& $1/6$ & 0 &3\\
    $u_R$& 3 & 1& $2/3$ & 0 &3\\
    $d_R$& 3 & 1& $-1/3$ & 0 &3\\
 \hline 
    $L$&  1 & 2& $-1/2$ & 1 &3\\
    $e_R$ & 1 & 1 & $-1$& 1 &3\\
    \hline
    $H$& 1& 2& $1/2$& 0 & 1 \\
 \hline
 \hline
    \addlinespace[0.3ex]
    $Q_L^{(1,2)}$ & 3 & 1 & $2/3$ or $-1/3$ & 1&2\\
    $Q_R^{(1,2)}$ & 3 & 1& $2/3$ or $-1/3$ & 0&2 \\
    $Q_L^{(3)}$ & 3 & 1 & $2/3$ or $-1/3$ & -1&1\\
    $Q_R^{(3)}$ & 3 & 1& $2/3$ or $-1/3$ & 0&1 \\
\hline    
    $T_L$ & 1 & 3 & 0 & 2&3\\
    $T_R$ & 1 & 3 & 0 & 1&3\\
    $D_L$& 1 & 2 & $1/2$ & 3&3\\
    $D_R$ & 1 & 2 & $1/2$ & 2&3\\
    $\nu_R$ & 1 & 1 & 0 & 3&3\\
     \hline
    $\sigma$ & 1 & 1 & 0 & 1 & 1  \\
\hline    
\end{tabular}
\caption{Charges and Representations under the SM gauge group and $\text{U}(1)_\text{PQ}$.}
\label{tab:charges-reps}
\end{table} 

One way to generate tiny Dirac masses is the Type I Dirac-Seesaw       scheme pioneered in \cite{RONCADELLI1983325}. In general one starts out by imposing a symmetry to forbid the tree-level Dirac neutrino mass term 
\begin{align}
    \overline{L}\epsilon H^\dagger \nu_R
\end{align}
 with $\epsilon = i \tau_2$ being the second Pauli matrix, as well as all possible Majorana masses. Then heavy vector-like SM singlet fermions coupling to both the SM leptons and $\nu_R$ are integrated out at energies below their mass scale leading to light neutrino masses from the threshold correction. Most models realize the neutrino mass via a dimension five operator similar to the Weinberg operator \cite{PhysRevLett.43.1566} of the schematic form $(LH)^2$ for Majorana neutrinos. Since the $\nu_R$ are SM singlets this necessitates the inclusion of a scalar singlet $\phi$ to form the required operator $(\overline{L}\epsilon H^\dagger) \phi \nu_R$. The vev of $\phi$  introduces a third scale $v_\phi$ apart from the SM Higgs vev $v_H$ and the heavy mediator scale $M\gg v_\phi, v_H$. Dirac masses then scale as $m_\nu \sim v_H v_\phi /M$ and the additional parameters are the reason why these scenarios are considered to be less minimal than Majorana models. If we wish to generate this operator via PQ charged particles and a singlet scalar $\sigma$ for spontaneous symmetry breaking of $\text{U}(1)_\text{PQ}$ there are essentially two options: One can either identify the heavy mass scale with the PQ breaking scale $M\sim v_\sigma$  as was the case for combining the Type I Seesaw       with PQ symmetry in \cite{Shin:1987xc}. This then requires that $\phi$ is a third scalar field and $v_H\lesssim v_\phi \ll v_\sigma$. The other option is to identify $\phi$ with the PQ breaking field $\sigma$ \cite{Peinado:2019mrn} and assume a separate source for the heavy vector-like fermion masses.
However since cosmological and astrophysical arguments require $v_\sigma > \SI{e8}{\giga\electronvolt}$ or even $v_\sigma \simeq \SI{e11}{\giga\electronvolt}$ (see subsection \ref{sec:axionDM}) the mediator mass scale must be potentially close to the Planck scale even for small Yukawa couplings \cite{Peinado:2019mrn}. Our model will be able to avoid these complications altogether. The key idea is that we can connect $L$ and $\nu_R$ by integrating out two different species of vector-like fermions transforming non-trivially under the electroweak gauge symmetry. To avoid a third scale besides $v_H$ and $v_\sigma$ we will generate the required threshold correction via a dimension six operator of the schematic form $(\overline{L} \epsilon H^\dagger)(H H^\dagger) \nu_R$. The heavy fermion masses scale with $v_\sigma$ and the presence of three Higgs doublets follows from the required $\text{SU}(2)_\text{L}$ contractions. In addition to that the active neutrino masses scale as $m_\nu \sim v_H^3 / v_\sigma^2$. Before we face the neutrino sector we briefly review the KSVZ-axion model, which solves the strong CP problem via heavy vector-like fermions with PQ charge.

\subsection{KSVZ-axion}\label{sec:KSVZ}
 In order to implement the Peccei-Quinn solution  \cite{PhysRevLett.38.1440,Peccei:1977ur} to the strong CP problem in the KSVZ model \cite{PhysRevLett.43.103,SHIFMAN1980493} we introduce a pair of color triplet quarks $\left(Q_L,Q_R\right)$, which are vector-like under the SM but chiral under PQ as well as a singlet scalar $\sigma$ coupling via
 \begin{equation}
      \mathcal{L}_\text{KSVZ} = - Y_Q \; \sigma \;\overline{Q_L}Q_R + \text{h.c.}\;\;.
\end{equation}
The charges and representations under all symmetries can be found in table \ref{tab:charges-reps}.
The scalar potential reads 
\begin{equation}
    V(H,\sigma) = V(H)- \mu_\sigma^2 \left|\sigma\right|^2 + \lambda_\sigma \left|\sigma\right|^4+ \lambda_{\sigma H} \left|\sigma\right|^2 \left|H\right|^2,
\end{equation}
with $\mu_\sigma^2>0$ for the SSB of PQ symmetry and $ V(H)$ the SM scalar potential.
We expand the singlet scalar as
\begin{equation}
    \sigma = \frac{1}{\sqrt{2}}\left(v_\sigma + \rho_\sigma\right)e^{i\frac{a}{v_\sigma}} \quad \text{with}\quad v_\sigma \gg v_H = \SI{246}{\giga\electronvolt},
\end{equation}
where $a$ denotes the axion field and we see that the exotic quarks have a  mass term consisting of
\begin{equation}
    M_Q\equiv Y_Q \frac{v_\sigma}{\sqrt{2}}.
\end{equation}
After rotating the axion field away by an anomalous Peccei-Quinn transformation of the quarks \cite{Fujikawa:1983bg} we can integrate them out and obtain the axion coupling to the QCD anomaly term. For $\text{SU}(2)_\text{L}$ singlets the QCD anomaly coefficient reads \cite{DiLuzio:2020wdo}
\begin{equation}\label{eq:color-anomaly-coeff}
    N = \sum_{\psi} \chi_\psi \;  T_d(\psi),
\end{equation}
where $T_d$ is the Dynkin color index for a $d$-dimensional representation with $T_3(Q_L) = T_3(Q_R)=1/2$ and $\chi_\psi$ denotes the PQ charge of the particle $\psi$. If we assume only one generation of exotic quarks then
\begin{equation}
     N = \frac{1}{2} \left(\chi_{Q_L}-\chi_{Q_R}\right)= \frac{1}{2}\chi_\sigma.
\end{equation}
The non-linearly realized $\text{U}(1)_\text{PQ}$ symmetry is explicitly broken by the non-perturbative QCD effects down to a $\mathcal{Z}_{2N}$ once the temperature of the universe cools below the QCD phase transition at $T=\Lambda_{QCD}=\mathcal{O}(200\;\text{MeV})$.
The aforementioned QCD effects manifest themselves in an effective cosine potential and thus a mass for the axion $a$, which dynamically relaxes to its minimum to cancel the strong CP violation \cite{PhysRevLett.38.1440,Peccei:1977ur} encoded in $\theta_\text{QCD}+\theta_\text{weak}$.
In this context $\theta_\text{QCD}$ is the topological angle of QCD and $\theta_\text{weak}$ is the contribution from the chiral transformations needed to diagonalize the SM quark masses.
After the angular mode $a$   relaxes in one of the $2N$ equivalent vacua, topological defects in the form of domain walls are formed from the spontaneous breaking of this discrete symmetry \cite{PhysRevLett.48.1867,VILENKIN1985263,Vilenkin:2000jqa}.
The cosmological domain wall number is given by $N_\text{DW} = 2 N$ and stable domain walls could overclose the universe \cite{Press_1980,PhysRevLett.48.1156}. 
The domain walls form a network with axionic strings produced during the SSB of PQ symmetry via the Kibble mechanism \cite{Kibble_1976,PhysRevD.26.435,KIBBLE1980183}, and the network will in general be stable for $N_\text{DW}>1$. For $N_\text{DW}=1$ the network eventually decays to low momentum axions \cite{PhysRevD.24.2082,DAVIS1986225} and contributes to their relic density \cite{Hiramatsu:2012gg,Gorghetto:2018myk,Vaquero_2019}. Pre-inflationary PQ breaking can dilute the domain walls and  explicitly PQ breaking bias-terms in the scalar potential \cite{PhysRevLett.48.1156,PhysRevLett.113.241301,Reig:2019vqh,Caputo:2019wsd} could make the domain walls decay. However we will see in  section \ref{sec:axionDM} that S.M.A.S.H. is only compatible with post-inflationary PQ breaking. Bias terms have to be large enough to make the domain walls decay before they dominate  the energy density of the universe \cite{Sikivie:2006ni}. On the other hand, they have the drawback of contributing to the axion mass, so that one needs to ensure, that they do not spoil the PQ solution to the strong CP problem, leading to an upper limit on the corresponding coupling \cite{Sikivie:2006ni,Kawasaki:2013ae}. There exists a parameter space that satisfies both conditions.
The last class of solutions to the domain wall problem embeds the  $\mathcal{Z}_{2N}$ into the center of a larger continuous global or local group \cite{LAZARIDES198221,BARR1982227}.
However we prefer to avoid  these complications altogether by simply normalizing the PQ charges of the quarks properly. We demand $N_\text{DW}=1$, from which we deduce that $\chi_\sigma=1$. In this scenario we have an axion decay constant of 
\begin{equation}\label{eq:ax-dec}
    f_a\equiv \frac{v_\sigma}{N_\text{DW}}.
\end{equation}
If one wishes to incorporate more generations of exotic quarks without generating additional domain walls then one has to make sure that the QCD anomaly coefficients for the  additional generations cancel each other, for example by choosing equal and opposite PQ charges for those two generations. This explains the charge assignments for the third generation of exotic quarks in \ref{tab:charges-reps}.
At the present stage the exotic quarks would be absolutely stable owing to their separately conserved baryon number \cite{PhysRevLett.43.103}. This would lead to exotic hadrons which could also overclose the universe and are tightly constrained relative to ordinary baryons by dedicated searches  \cite{DiLuzio:2016sbl,DiLuzio:2017pfr}. In order to make the exotic quarks decay  we introduce a  renormalizable coupling to the SM doublet quarks $q_L$  and consider the following operators for $\chi_{Q_R}=0$ \cite{DiLuzio:2016sbl,DiLuzio:2017pfr}
\begin{equation}\label{eq:mass-mix}
    \mathcal{L}_\text{decay} = - Y_{qQ} \begin{cases} \overline{q_L }\epsilon H^\dagger Q_R \quad \text{for} \quad  Y_{Q_R} = \frac{2}{3} \\ \overline{q_L} H Q_R\quad \text{for} \quad Y_{Q_R} =-\frac{1}{3} \end{cases} + \text{h.c.},
\end{equation}
where $Y_{qQ}$ is a dimensionless Yukawa coupling to the SM Higgs. There will be a lower limit on the Yukawa coupling in \eqref{eq:mass-mix} from demanding that decay rate (assuming $m_Q\gg m_H$)
\begin{equation}
    \Gamma\left(Q\rightarrow q_L H\right) \simeq \frac{Y_{qQ}^2 m_Q}{16\pi}
\end{equation}
is faster than the Hubble rate at the temperature $T=m_Q$ implying
\begin{equation}
    Y_{qQ} \gtrsim  10^{-5}\; \sqrt{\frac{m_Q}{10^8\;\text{GeV}}},
\end{equation}
so that the abundance of vector-like quarks is actually depleted and an epoch of intermediate era of matter domination  \cite{Scherrer:1984fd} from   very long-lived vector-like quarks is avoided. In the above we used a value for $m_Q$ that will be motivated in section \ref{sec.vac}. Vector-like quarks could be produced at colliders, either in pairs from a gluon or together with an SM quark via the coupling in \eqref{eq:mass-mix}. Searches for new colored fermions exclude vector-like quark masses below about $1\;\text{TeV}$ \cite{CMS:2018zkf,ATLAS:2018ziw}. The  large couplings together with heavy large masses are the reason, why we expect the life-time of the  vector-like quarks, if kinematically accessible at colliders, to be very short.

\subsection{Neutrino masses}

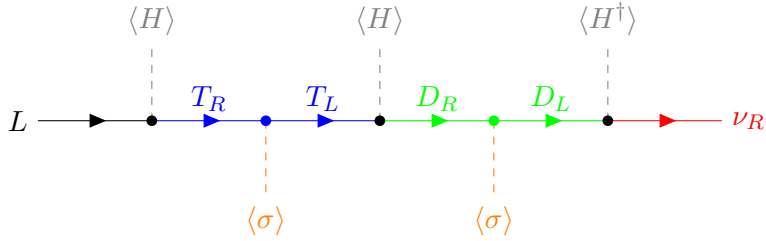
\begin{figure}[t]
 \centering
  \tikzset{
  blackline/.style={thin, draw=black, postaction={decorate},
    decoration={markings, mark=at position 0.6 with {\arrow[black]{triangle 45}}}},
    blueline/.style={thin, draw=blue, postaction={decorate},
    decoration={markings, mark=at position 0.6 with {\arrow[blue]{triangle 45}}}},
    redline/.style={thin, draw=red, postaction={decorate},
    decoration={markings, mark=at position 0.6 with {\arrow[red]{triangle 45}}}},
    greenline/.style={thin, draw=green, postaction={decorate},
    decoration={markings, mark=at position 0.6 with {\arrow[green]{triangle 45}}}},    
    graydashed/.style={dashed, draw=gray, postaction={decorate},
    decoration={markings}},
   yellowdashed/.style={dashed, draw=orange, postaction={decorate},
    decoration={markings}},
    photon/.style={decorate, draw=red,
    decoration={coil,amplitude=12pt, aspect=0}},
  gluon/.style={dashed, decorate, draw=black,
    decoration={coil, segment length=5pt, amplitude=8pt}}
  line/.style={thick, draw=black, postaction={decorate},
    decoration={markings}}
}

\begin{tikzpicture}[node distance=1cm and 1cm]

%particles 
\coordinate[label = left: $L$] (start1);
\coordinate[right=9cm of start1,label=right: $\color{red}\nu_R$] (end);

%vertices
\coordinate[right=1.5cm of start1] (H1);
\coordinate[right=1.5cm of H1] (sigma1);
\coordinate[right=1.5cm of sigma1] (H2);
\coordinate[right=1.5cm of H2] (sigma2);
\coordinate[right=1.5cm of sigma2 ] (H3);

%vev insertions 
\coordinate[above=1cm of H1,label=above: $\color{gray} \braket{H}$] (vevH1);
\coordinate[above=1cm of H2,label=above: $\color{gray} \braket{H}$] (vevH2);
\coordinate[above=1cm of H3,label=above: $\color{gray} \braket{H^\dagger}$] (vevH3);
\coordinate[below=1cm of sigma1,label=below: $\color{orange} \braket{\sigma}$] (vevS1);
\coordinate[below=1cm of sigma2,label=below: $\color{orange} \braket{\sigma}$] (vevS2);

%axis
\draw[blackline] (start1)   -- (H1);
\draw[blueline] (H1)   -- (sigma1);
\draw[blueline] (sigma1)   -- (H2);
\draw[greenline] (H2)   -- (sigma2);
\draw[greenline] (sigma2)   -- (H3);
\draw[redline] (H3)   -- (end);

%vev insertions
\draw[graydashed] (H1)   -- (vevH1);
\draw[graydashed] (H2)   -- (vevH2);
\draw[graydashed] (H3)   -- (vevH3);
\draw[yellowdashed] (sigma1)   -- (vevS1);
\draw[yellowdashed] (sigma2)   -- (vevS2);

%particle labels
\coordinate[right=0.75cm of H1, label=above: $\color{blue} T_R$];
\coordinate[left=0.75cm of H2, label=above: $\color{blue} T_L$];  
\coordinate[right=0.75cm of H2, label=above: $\color{green} D_R$];
\coordinate[left=0.75cm of H3, label=above: $\color{green} D_L$];    
  
%dots
\fill (H1) circle (2pt);
\fill (H2) circle (2pt);
\fill (H3) circle (2pt);
\fill[blue] (sigma1) circle (2pt);
\fill[green] (sigma2) circle (2pt);
  
\end{tikzpicture}
  \caption{Diagrammatic representation of the dimension 6 operator giving rise to Dirac masses for the active neutrinos.}
  \label{fig:mass-gen}
\end{figure}

\noindent 
In a similar spirit we now introduce vector-like leptons as well to generate the Dirac neutrino masses in the Seesaw fashion. We give vector-like PQ charges to the SM leptons and charge $\nu_R$ in such a way that the tree level mass term $\overline{L}\epsilon H^\dagger \nu_R$ with $\epsilon = i \tau_2$ being the second Pauli matrix is absent.
Since the cosmologically preferred PQ breaking scale $f_a \simeq10^{11}\;\text{GeV}$ is lower than the typical Seesaw-scale (for order one Yukawas) of $M_N\simeq 10^{14}\;\text{GeV}$ we choose to integrate out two distinct fermions instead of a single messenger. However these two fermion species will have comparable masses so this \textit{sequential Seesaw} depicted in   figure \ref{fig:mass-gen} does not lead to a double Seesaw-mechanism \cite{PhysRevLett.56.561,PhysRevD.34.1642}.
The resulting operator for neutrino masses will have mass dimension six (see \cite{CentellesChulia:2018bkz} for a compendium of possible Dirac dimension six operators) compared to the usual Weinberg operator at dimension five \cite{PhysRevLett.43.1566}.
We start with introducing three generations of vector-like pairs of triplets $(T_L,T_R)$ and doublets $(D_L,D_R)$. The multiplets can be expanded into their components as
\begin{equation}
    T_L \equiv  \frac{T_L^a \tau^a}{2} =   \begin{pmatrix} \frac{T_L^0}{\sqrt{2}} & T_L^+ \\ T_L^- & - \frac{T_L^0}{\sqrt{2}} \end{pmatrix}, \quad 
    T_R \equiv  \frac{T_R^a \tau^a}{2}= \begin{pmatrix} \frac{T_R^0}{\sqrt{2}} & T_R^+ \\ T_R^- & - \frac{T_R^0}{\sqrt{2}} \end{pmatrix}  
\end{equation}
and
\begin{equation}
    \quad   D_L \equiv  \begin{pmatrix} E_L^+ \\N_L  \end{pmatrix},
    \quad   D_R \equiv  \begin{pmatrix} E_R^+ \\-N_R  \end{pmatrix}.
\end{equation}
We also introduce the following notation of $ \tilde{H} \equiv \epsilon H^\dagger$. 
A combination of chiral PQ charges, Hypercharge and non-trivial $\text{SU(2)}_\text{L}$ representations allows only the following mass 
\begin{equation}
    \mathcal{L}_\text{mass} = -Y_T \sigma \overline{T_L}^a T_R^a - Y_{D} \sigma \overline{D_L}  D_R + \text{h.c.}
\end{equation}
and mixing terms 
\begin{equation}\label{eq:main-model}
    \mathcal{L}_\text{int} = -Y_{LT}\; \overline{L}\; T_R^a \tau^a   \tilde{H} -  Y_{TD}\;   D_R  \epsilon\; \overline{T_L}^a\tau^a  \tilde{H} - Y_{DR}\; \overline{D_L} H \nu_R + \text{h.c.}\;.
\end{equation}
All charges and representations for the four component spinors have been summarized in table \ref{tab:charges-reps}.
If we had given $D_{L,R}$ the opposite hypercharge $-1/2$ an operator of the schematic form $\overline{L} D_R\sigma^*$ would be allowed by all imposed symmetries and this operator would ruin the sequential nature of our mass generation mechanism by coupling $\nu_L$ directly to the exotic $N_R$ neutrino with a large vev $v_\sigma$. We use triplets $T_{L,R}$ instead of  singlet fermions $S_{L,R}$  because the PQ charge assignment would allow for a term $\overline{S_L}  \nu_R \sigma^*$, which would also spoil the intended mass generation mechanism.
Note that if one identifies our unconventional  chiral choice of PQ charges with lepton number or $\text{B-L}$, which are usually taken to have vector-like charges normalized to $\pm 1$, one can understood the \enquote{Diracness} of the neutrinos as follows:
Since $\sigma$ breaks PQ symmetry by only a single unit all the renormalizable  Majorana mass terms which would require breaking by two, four or six units (see table \ref{tab:charges-reps}) are forbidden. This is in a similar spirit to the argument that breaking conventionally assigned lepton number or $\text{B-L}$ by any number other than two allows only for Dirac neutrinos \cite{Ma:2014qra}.
Of course PQ  symmetry does not forbid the following non-renormalizable operators
\begin{align}
     \frac{c_{T_R}}{\Lambda_\text{UV}}\left(\sigma^*\right)^2 \overline{T_R^c}^a T_R^a,\quad  
    \frac{c_{T_L}}{\Lambda_\text{UV}^3}\left(\sigma^*\right)^4 \overline{T_L^c}^a T_L^a,\quad 
    \frac{c_{\nu_R}}{\Lambda_\text{UV}^5}\left(\sigma^*\right)^6 \overline{\nu_R^c}\nu_R,
\end{align}
as well as 
\begin{align}
    \frac{c_L}{\Lambda_\text{UV}^3}\left(\sigma^*\right)^2 \left(\overline{L^c}\epsilon H\right) \left(L \epsilon H\right),\quad
    \frac{c_{D_R}}{\Lambda_\text{UV}^5}\left(\sigma^*\right)^4 \left(\overline{D_R^c} H^\dagger\right) \left(D_R H^\dagger\right),\quad
    \frac{c_{D_L}}{\Lambda_\text{UV}^7}\left(\sigma^*\right)^6 \left(\overline{D_L^c} H^\dagger\right) \left(D_L H^\dagger\right),\quad 
\end{align}
where the $c_i$ are dimensionless Wilson-coefficients and $\Lambda_\text{UV}$ is some mass scale above the PQ scale. Evidently the dimension five operator for $T_R$ is the least suppressed and the mass term for the $D_{L}$ at dimension eleven has the largest suppression factor due to SM gauge invariance and PQ breaking by six units.
We have checked that these operators are not generated at loop level for the given particle content in field theory, but if one includes quantum gravity they might arise.  Non-perturbative quantum gravitational effects could lead to a low energy effective field theory which will contain all the terms  allowed by only the local gauge symmetries  \cite{DiLuzio:2020wdo} such as the above ones.
On top of that quantum gravity is  expected to violate global symmetries \cite{COLEMAN1988643,GIDDINGS1988854,GILBERT1989159} like PQ symmetry.
These quantum gravity effects are heuristically\footnote{there might be an additional suppression factor $e^{-S_\text{wh}}$, where the large number $S_\text{wh}$ is the wormhole action \cite{Kallosh:1995hi,Alonso:2017avz} } encoded in Planck-mass suppressed explicitly PQ violating operators leading  to the well known \enquote{axion quality problem} \cite{GEORGI1981409,DINE1986109,PhysRevD.46.539,KAMIONKOWSKI1992137,Holman:1992us,GHIGNA1992278} that could spoil the solution to the strong CP problem. PQ violating Majorana masses could arise in the same way too \cite{deGouvea:2000jp}. Since we have nothing to add to the solution of these \enquote{quality problems} we will assume that the Wilson coefficients of both sets of  hypothetical effective operators (PQ conserving or  violating) are either negligibly small or that some other mechanism prevents their existence.\footnote{After the completion of this work reference \cite{Penedo:2022gej} was released, in which the authors manage to avoid the PQ conserving higher dimensional operators by choosing the PQ charge of $L$ to be a non-integer $\chi_L=1/3$ and shifting all other fermionic charges accordingly so that $\chi_{\nu_R}=2+ 1/3 = 7/3$. In this case the estimate for the axion to photon coupling in \eqref{eq:ENtot} remains unchanged, because the difference in PQ charge between the different chiralities  $(T_L,T_R)$ and $(D_L,D_R)$ is still one.}\\
\\
Before EWSB   the triplets and doublets are decoupled and each component of an $\text{SU(2)}_\text{L}$ multiplet has a common mass set by the PQ breaking scale, which we call
\begin{equation}
    M_T \equiv Y_T \frac{v_\sigma}{\sqrt{2}} \quad \text{and} \quad     M_D \equiv Y_D \frac{v_\sigma}{\sqrt{2}}.
\end{equation}
After integrating them out and applying a Fierz-transformation we find
\begin{equation}\label{eq:dim-six}
  \mathcal{L}_\text{eff.} =  \frac{1}{2}\; Y_{LT}M_T^{-1} Y_{TD}M_D^{-1} Y_{DR}\;\;\left(\overline{L}\epsilon H^\dagger \right)\left( H H^\dagger \right) \nu_R + \text{h.c.}\;.
\end{equation}
In the one flavor approximation we find  the following relation for the active neutrino mass scale after electroweak symmetry breaking (EWSB) 
\begin{equation}\label{eq:nu-mass}
    m_\nu \simeq \SI{0.05}{\electronvolt}\cdot Y_{LT} Y_{TD} Y_{DR}\cdot \left(\frac{\SI{e9}{\giga\electronvolt}}{M_T}\right) \cdot \left(\frac{\SI{e8}{\giga\electronvolt}}{M_D}\right).
\end{equation}
If we choose $M_T, M_D$ lighter than about $\mathcal{O}\left(\SI{e8}{\giga\electronvolt}\right)$ we can maintain the light neutrino mass scale by  decreasing the Yukawa coupling $Y_{LT} Y_{TD} Y_{DR}$. On the other hand the overall neutrino mass scale could be lowered too far if we chose 
$M_T, M_D\gg\mathcal{O}\left(\SI{e8}{\giga\electronvolt}\right)$, which is why we work in the previously mentioned regime. Since we expect $f_a\simeq \mathcal{O}\left(\SI{e11}{\giga\electronvolt}\right)$ (see \eqref{sec:axionDM}) this means that Yuakwa couplings of the $T,D$ fields to the PQ breaking field must satisfy $Y_{T,D}\lesssim 10^{-3}$.
Here there are only two dynamical scales $v_H$ and $v_\sigma$ involved in the neutrino mass generation, which comes at the price of introducing five Yukawa matrices $Y_{LT},Y_{TD},Y_{DR},Y_T,Y_D$. In order to generate the two mass splitting needed to explain the neutrino oscillation data we need to introduce at least two generations of $T_{L,R}, D_{L,R}$ and in the following we will assume the existence of three such generations. 
We can estimate the axion decay constant 
\begin{equation}
    f_a \simeq \SI{4e8}{\giga\electronvolt}  \cdot \sqrt{\frac{\SI{0.1}{\electronvolt}}{m_\nu}}\cdot \sqrt{\frac{Y_{LT} Y_{TD} Y_{DR}}{ Y_T Y_D}}
\end{equation}
as a function of the active neutrino mass.
If   we drop the previous assumption about the Yukawa couplings and allow all five of them to vary between $\mathcal{O}(1)$ and $\mathcal{O}\left(10^{-6}\right)$ (which are the largest and smallest Yukawa couplings in the SM of the top quark and electron respectively), we find 
\begin{equation}\label{eq:fa-bounds}
    \SI{0.4}{\giga\electronvolt}\cdot\sqrt{\frac{\SI{0.1}{\electronvolt}}{m_\nu}}\; \lesssim\; f_a\; \lesssim\;  \SI{4e14}{\giga\electronvolt}\cdot \sqrt{\frac{\SI{0.1}{\electronvolt}}{m_\nu}}.
\end{equation}
The lower range of $f_a \sim \SI{0.4}{\giga\electronvolt} $ obtained for the extreme choice $Y_{LT}\sim Y_{TD}\sim Y_{DR} \sim 10^{-6}$ and $Y_T \sim Y_D \sim 1$ would correspond to the Weinberg-Wilczek \cite{PhysRevLett.40.223,PhysRevLett.40.279} axion which has been ruled out experimentally via meson decays a long time ago \cite{Bardeen:1986yb}. Furthermore astrophysical arguments based on stellar cooling demand $f_a > \SI{e8}{\giga\electronvolt}$, so that the region of small $Y_{LT}\sim Y_{TD}\sim Y_{DR}$ is already excluded. Note that having such small couplings would defeat the purpose of building a rather involved Seesaw  model to begin with.
We depict the  decay constants that would lead to a too small neutrino mass as the grey region in the  figures \ref{fig:Axion-param-space-limits} and \ref{fig:Axion-param-space-projections}, which shall be the focus of the next subsection.

\subsection{Axion to photon coupling}
\begin{figure}
    \centering
    \includegraphics[width=0.81\textwidth]{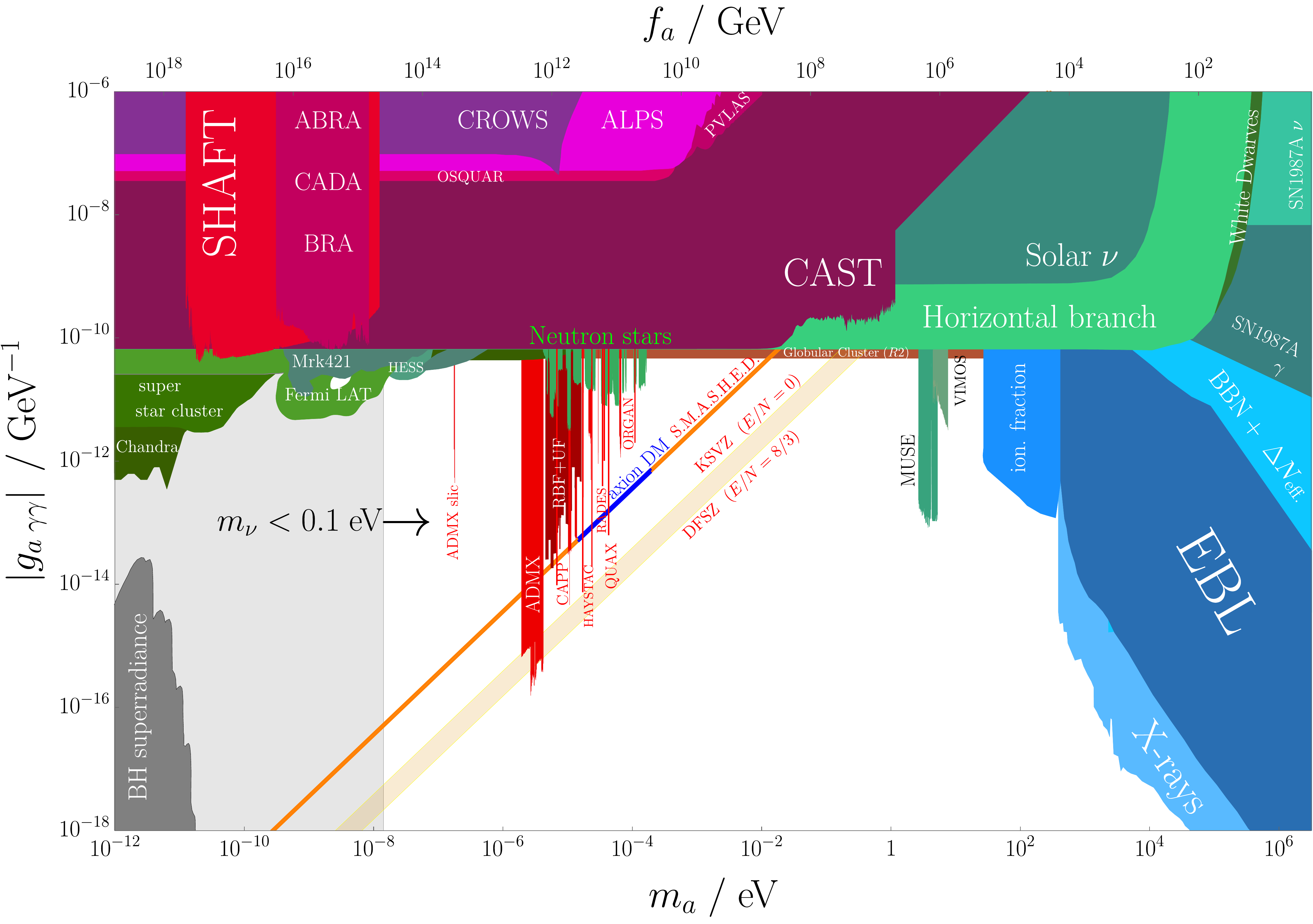}
    \caption{Depicted are the  \enquote{S.M.A.S.H.E.D.} band in orange and the  QCD-axion band for the KSVZ and DFSZ models in yellow  as well as a collection of cosmological, astrophysical and laboratory constraints. The white space is allowed and in the gray area the active neutrino masses would be too small.  The black arrow to the right indicates that the lower allowed limit on $m_a$ increases for larger values of $m_\nu$. The corresponding references can be found in appendix \ref{sec:lim}. }
    \label{fig:Axion-param-space-limits}
    \includegraphics[width=0.81\textwidth]{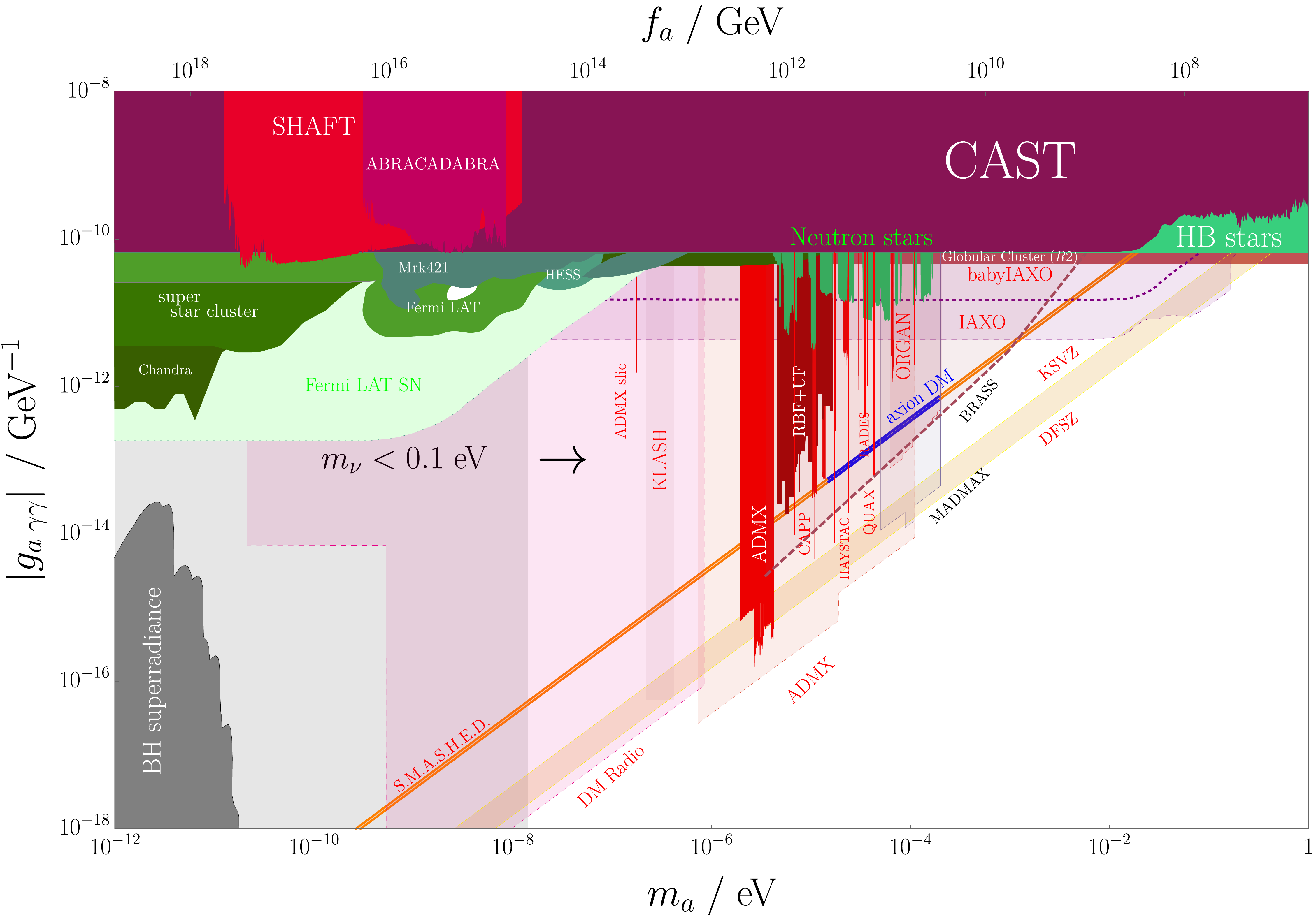}
    \caption{Depicted are the \enquote{S.M.A.S.H.E.D.} band in orange and the  QCD-axion band for the KSVZ and DFSZ models in yellow  as well as  cosmological, astrophysical and laboratory constraints together with  projected sensitivities. The black arrow to the right indicates that the lower allowed limit on $m_a$ increases for larger values of $m_\nu$. The corresponding references can be found in appendix \ref{sec:lim}.}
    \label{fig:Axion-param-space-projections}
\end{figure}

\noindent 
The most relevant coupling for the direct detection of axions in laboratory experiments is the axion-to photon coupling, which is given by \cite{KAPLAN1985215,Srednicki:1985xd}
\begin{equation}\label{eq:axphot}
    g_{a\gamma\gamma} = \frac{\alpha}{2\pi f_a}\left(\frac{E}{N}-1.92(4)\right),
\end{equation}
where the second term represents the model-independent irreducible contribution from the axion-pion mass mixing.
The color anomaly coefficient $N$ was specified in equation \eqref{eq:color-anomaly-coeff} and is found to be $N=1/2$  for this model. The electromagnetic anomaly coefficient $E$ is defined via \cite{DiLuzio:2020wdo}
\begin{equation}
     E = \sum_{\psi}\; \chi_\psi\; d_c(\psi) \; \text{Tr}\left(Q^2_{\text{EM}\;\psi}\right),
\end{equation}
where $ \chi_\psi$ is the PQ charge  of the fermion $\psi$, $d_c(\psi)$ is the dimension of its color representation and $Q_{\text{EM}\;\psi}$ its electric charge matrix.  For the original KSVZ model, where the exotic quarks have no hypercharge one finds $E/N=0$. Since we equip them with hypercharge for cosmological reasons (see table \ref{tab:charges-reps} for their charges) their contribution reads
\begin{equation}
    \left(\frac{E}{N}\right)_Q = 2 \cdot 3 \cdot \begin{cases} \frac{4}{9}\\ \frac{1}{9}\end{cases} = \begin{cases} \frac{8}{3}  \quad \text{for} \quad  Y_{Q_L} = \frac{2}{3},\\ \frac{2}{3} \quad \text{for} \quad  Y_{Q_L} = -\frac{1}{3}, \end{cases}
\end{equation}
where the factor of three comes from $d_c(Q_L)=3$. Only the first generation of exotic quarks contribute since we assume the PQ charges of the second and third generations cancel each other in order to fix the domain wall number. Similarly even though the SM leptons $L$ and $e_R$ have electric charge, they are vector-like under PQ and do not contribute to $E$. As a consequence of introducing non-trivial $\text{SU(2)}_\text{L}$ representations which are chiral under PQ for neutrino mass generation we obtain an additional contribution from the $T$ and $D$:
\begin{equation}
    \left(\frac{E}{N}\right)_{T,D} = 2 \cdot 3 \cdot (2+1) = 18  
\end{equation}
The factor of three takes the three generations of exotic leptons into account and since each triplet contains two charged fermions their contribution is twice as large as for the doublets.
Consequently the model dependent part of the axion to photon coupling
\begin{equation}\label{eq:ENtot}
    \left(\frac{E}{N}\right)_\text{tot} =   18  +\begin{cases} \frac{8}{3}  \quad \text{for} \quad  Y_{Q_L} = \frac{2}{3},\\ \frac{2}{3} \quad \text{for} \quad  Y_{Q_L} = -\frac{1}{3} \end{cases} 
\end{equation}
is significantly larger than in conventional models such as the DFSZ I (II) scenario, where the ratio reads $8/3\;(2/3)$. Thus our neutrino mass mechanism has the additional benefit of making the axion easier to detect in laboratory experiments.
Compared to other constructions in the literature this enhancement is rather small. For comparison clockwork based models such as \cite{Farina:2016tgd,Higaki:2015jag,Higaki:2016yqk} lead to an exponential enhancement of the coupling, a recent construction  with quantized magnetic charges \cite{Sokolov:2021ydn,Sokolov:2022fvs} can increase the axion to photon coupling by six orders of magnitude. Mirror sector models with $n$ copies of the SM and PQ sectors \cite{Hook:2018jle,DiLuzio:2021pxd} can increase the axion to photon coupling as a function of $n$ by lowering the axion mass compared to the usual QCD axion. Alternatively if one sticks to the particle content of the KSVZ model then the largest possible positive anomaly coefficient was found to be  $E/N = 170/3$ \cite{DiLuzio:2016sbl,DiLuzio:2017pfr}. A recent scan over possible representations found that  $E/N = - 166/3$  \cite{Plakkot:2021xyx} is the largest possible negative value within the range of the scan. Note that the two previously mentioned scenarios require three or eight quarks of different  SM representations.
We depict the experimentally allowed parameter space and a collection of limits in \ref{fig:Axion-param-space-limits} together with the projections form upcoming searches in \ref{fig:Axion-param-space-projections}. The  limits and projected limits were compiled in \cite{GITHUB} and they can be found in appendix \ref{sec:lim}. The orange band dubbed \enquote{S.M.A.S.H.E.D.} corresponds to the prediction of our model. Inside this band there is a  blue region called \enquote{axion DM} which reproduces the observed DM abundance for the cosmological history of the S.M.A.S.H. models and will be explained in section \ref{sec:axionDM}.
Experiments like QUAX   \cite{Alesini:2019ajt,Alesini:2020vny} or  HAYSTAC \cite{HAYSTAC:2018rwy,HAYSTAC:2020kwv} have already started to test the relevant parameter space depicted in blue. Other experiments like ORGAN \cite{McAllister:2017lkb,doi:10.1126/sciadv.abq3765} or RADES \cite{CAST:2020rlf} are  close to probing the axion DM parameter space as can be seen in \ref{fig:Axion-param-space-limits}. 
When it comes to next generation experiments we find that MADMAX  \cite{Beurthey:2020yuq}, the upgraded ADMX experiment    \cite{2010PhRvL.104d1301A,Stern:2016bbw,ADMX:2018gho,ADMX:2019uok,ADMX:2021nhd,ADMX:2018ogs,Bartram:2021ysp,Crisosto:2019fcj} as well as BRASS \cite{BRASS} have good chances of testing the aforementioned parameter region.  In appendix \ref{sec:landau} we demonstrate numerically that the new heavy fermions do not lead to phenomenologically relevant Landau poles in any of the SM gauge couplings following the treatments in \cite{MACHACEK198383,DiLuzio:2015oha,Plakkot:2021xyx}.

\subsection{Axion to fermion coupling}
\noindent The chiral rotations that remove the phase of the singlet field $\sigma$ from the mass terms induce the following derivative interactions for all PQ charged fermions $\Psi$ with chiral charges $\chi_{L,R}$
\begin{equation}\label{eq:axion-fermion-gen}
    \mathcal{L}_\text{int.} = i \frac{\partial_\mu a}{f_a}\sum_\Psi \overline{\Psi}\gamma^\mu \left(  \frac{\chi_L+\chi_R}{2}\cdot \mathbbm{1}_4- \frac{\chi_L-\chi_R}{2}\cdot \gamma_5 \right)\Psi.
\end{equation}
The only SM fermions that pick up an interaction at tree level are the three generations of charged and neutral leptons:
\begin{equation}\label{eq:axion-fermion}
      \mathcal{L}_\text{int.} = i \frac{\partial_\mu a}{f_a} \sum_j \left( \overline{e}_j \gamma^\mu e_j + \overline{\nu}_j \gamma^\mu \left(2\cdot \mathbbm{1}_4+\gamma_5\right) \nu_j\right)
\end{equation}
As expected the charged lepton coupling is vector-like. If we integrate the first term by parts we pick up a contribution of
\begin{equation}
    \sum_j \partial_\mu\left(\overline{e}_j \gamma^\mu e_j\right) 
\end{equation}
which vanishes for on shell leptons.  Of course as in the original KSVZ model a pseudoscalar coupling to  will be regenerated at loop level from the axion to photon coupling in \eqref{eq:axphot} \cite{Shin:1987xc}
\begin{equation}
    g_{ae} \simeq \alpha\;   g_{a\gamma\gamma}\; m_e\; \text{Log}\left(\frac{f_a}{m_e}\right),
\end{equation}
which is dimensionless as $g_{a\gamma\gamma}\sim 1/f_a$. Here we neglected the contribution from axion pion mixing in \eqref{eq:axphot} as it was subdominant to the inclusion of the heavy exotic fermions. One can see that this coupling $g_{ae}\simeq 3\times10^{-17}$ is very small due to its dependence on $ \alpha\; m_e/f_a$. There are more one loop contributions to $g_{ae}$ from one loop diagrams involving the other massive EW gauge bosons as well as the new exotic fermions. By recasting the result for a Majorana Type I Seesaw from  \cite{Shin:1987xc,Garcia-Cely:2017oco} we estimate
\begin{align}
    g_{ae}\simeq \frac{1}{16\pi^2} \left(\frac{m_\nu}{v_H}\right)^2 \frac{m_e}{f_a}.
\end{align}
This contributions is  suppressed by  both  $m_e /f_a \simeq 5\times 10^{-15}$ and $m_\nu^2 /v_H^2 \simeq 10^{-25}$ so we do not consider them further. Stellar cooling arguments for the sun and red giants exclude $g_{ae}\gtrsim \mathcal{O}\left(10^{-13}\right)$
\cite{Viaux:2013hca,Giannotti:2017hny,2018MNRAS.478.2569I,2018arXiv180210357S}, which is respected by our model.
Axions could also be emitted in laboratory experiments from the final state neutrino or charged lepton in pseudoscalar-meson decay. Since this will remove the chirality suppression of the two-body decay, these channels are sensitive probes for new physics. Existing analyses \cite{Barger:1981vd,Aditya:2012ay,Pasquini:2015fjv,Gallo:2021ame} (often) do not use the full derivative coupling in \eqref{eq:axion-fermion} but rely on Yukawa-interactions which are technically only  valid for on-shell fermions \cite{PhysRevLett.60.1793,Raffelt:1996wa}.
However since the full calculation can involve technical subtleties such as infrared-divergences which have to be cancelled via loop-corrections  \cite{Pasquini:2015fjv} we will limit ourselves to recasting the existing limits. For the emission from a neutrino line we replace the axion neutrino coupling with $g_\nu= m_\nu / f_a$.
Reference \cite{Pasquini:2015fjv} found that depending on the flavor-structure $g_\nu^2<\mathcal{O}\left(10^{-6-7}\right)$ which translates to
\begin{equation}
    f_a >  \mathcal{O}(10^3)\;\times m_\nu
\end{equation}
and is not restrictive at all.

\section{Unification}
The DFSZ \cite{Zhitnitsky:1980tq,Dine:1981rt} version of the original S.M.A.S.H. framework  avoids the exotic vector-like quarks by charging the SM quarks under PQ symmetry and was discussed in \cite{Ballesteros:2019tvf}. Since this variant of S.M.A.S.H. only requires an additional gauge singlet sterile neutrino $N$, one can embed the DFSZ-S.M.A.S.H. in  a basic Grand Unified Theory (GUT):
One can choose $\text{SU}(5)$ \cite{Georgi:1974sy,Georgi:1974my} by introducing $N$ as an additional singlet or pick the larger $\text{SO}(10)$ \cite{Fritzsch:1974nn,Georgi:1974my}, where $N$ fills the 16-dimensional spinorial representation together with the other 7 Weyl spinors for one generation of the SM. For a Type I Dirac Seesaw one can also find an $\text{SO}(10)$-embedding by introducing the SM gauge singlet vector-like neutrinos as $\text{SO}(10)$-singlets \cite{Peinado:2019mrn}. In our case of S.M.A.S.H.E.D. the situation is not as straight-forward, because we introduce vector-like fermions transforming non-trivially under the SM gauge group. This is why we would need to fill   additional multiplets of e.g. $\text{SO}(10)$, which comes at the price of introducing additional fermions for anomaly cancellation. One can see, that there is no obvious GUT-embedding of our setup and further work would be required to find one.

\section{Cosmology of S.M.A.S.H.}\label{sec:smaschcosmo}
We briefly recapitulate the most important aspects of the cosmological history in the S.M.A.S.H. framework \cite{Ballesteros:2016euj,Ballesteros:2016xej,Ballesteros:2019tvf}.
\subsection{Inflation and reheating}\label{sec:reh}
Scalar fields with a non-minimal coupling to scalar curvature \cite{STAROBINSKY198099,PhysRevD.40.1753,PhysRevD.41.1783,PhysRevD.52.4295,PhysRevD.59.064029,Bezrukov:2007ep,Bezrukov:2008ej,Bezrukov:2008ut} are chosen as the inflationary scenario. In a  two field model,  such as the present setup featuring the  neutral component of $H$ together with $\sigma$, the inflationary dynamics are more complicated, which is why the authors of \cite{Ballesteros:2016euj,Ballesteros:2016xej,Ballesteros:2019tvf} worked out limiting cases, in which effectively only one field is responsible for inflation. Because of the unitarity problem \cite{Lerner:2009na,Burgess:2010zq,Hertzberg:2010dc} for pure Higgs inflation (HI) \cite{Garcia-Bellido:2011kqb,Barvinsky:2008ia,Barvinsky:2009fy} reference \cite{Ballesteros:2016xej} considered the inflaton to be either arising from the field $\sigma$ (HSI scenario) or as a linear combination of the neutral component of $H$ and $\sigma$ (HHSI scenario).
One finds  the valleys of the potential, that are attractors for the inflationary trajectories, by inspecting the signs of the following quantities \cite{Ballesteros:2016xej}
\begin{equation}
    \kappa_H \equiv \lambda_{H\sigma} \xi_H - \lambda_H \xi_\sigma,\quad \kappa_\sigma\equiv \lambda_{H\sigma} \xi_\sigma - \lambda_\sigma \xi_H.
\end{equation}
The relevant ranges are   $\kappa_H >0\;\wedge \kappa_\sigma >0$ (here $\wedge$ is a logical \enquote{and}) for either HI or HSI, whereas HHSI needs $\kappa_H <0\;\wedge \kappa_\sigma <0$. Solving the unitarity problem requires $0<\xi_\sigma \lesssim 1$ \cite{Ballesteros:2016xej} for the coupling to gravity. In this context and because of vacuum stability (see \ref{sec.vac}) for HHSI one needs a trajectory that is parametrically close to the HSI one, which can be achieved in the limit $\xi_H\ll \xi_\sigma$ \cite{Ballesteros:2016xej}. Then one finds that the radial modes of the SM like Higgs and $\sigma$ lead to the following inflationary trajectory \cite{Ballesteros:2016xej}
\begin{equation}
    \frac{\rho_\sigma}{\rho_H} = \sqrt{-\frac{\lambda_H}{\lambda_{H\sigma}}}+\mathcal{O}\left(\frac{\xi_H}{\xi_\sigma}\right),
\end{equation}
where $\lambda_{H\sigma}<0$ is required for HHSI \cite{Ballesteros:2016xej}. On the other hand $\lambda_{H\sigma}>0$ selects HI and HSI. In the HSI scenario non-thermal axions get produced during reheating after the  non-thermal restoration of PQ symmetry (see the next subsection \ref{sec:axionDM}). This scenario has a reheating temperature of around $10^7 \;\text{GeV}$ that is so low that the axions never thermalize (see section \ref{sec:axNeff}), leading to an abundance of dark radiation $\Delta N_\text{eff.} \simeq 0.35-1.6$ \cite{Ballesteros:2016xej} which is excluded by observations \cite{Planck:2018vyg}.
Therefore only the HHSI scenario is viable. In this regime the effective quartic coupling for inflation reads
\begin{equation}\label{eq:tilde}
    \tilde{\lambda}_\sigma \equiv \lambda_\sigma - \frac{\lambda_{H\sigma}^2}{\lambda_H}
\end{equation}
and it is bounded by \cite{Ballesteros:2016xej}
\begin{equation}\label{eq:infbounds}
    5\times 10^{-10} < \tilde{\lambda}_\sigma < 5\times 10^{-13},
\end{equation}
where the upper limit comes from the amplitude of primordial scalar perturbations and the lower limit from the bound on the tensor to scalar ratio  \cite{BICEP2:2015xme,Planck:2018vyg}.
Reheating occurs via damped inflaton oscillations in a quartic potential. The dominant channel is the  production of EW gauge bosons from the inflaton's SM like Higgs component during zero crossings of the oscillating condensate. The EW gauge bosons have effective inflaton dependent masses and decay efficiently to SM fermions as the gauge boson masses increase away from the crossings. During the zero crossings  gauge boson production from the resulting fermion bath is also efficient. Due to their mass gain  and the fact that the gauge bosons decay to fermions before they can loose this energy to the condensate again, energy is efficiently drained from the inflaton to the SM fermion bath. Production of the heavy exotic fermions occurs as well, but since their decays to SM fermions are suppressed compared to the gauge bosons, their inclusion is negligible. The reheating  temperature can be estimated to be \cite{Ballesteros:2016xej}
\begin{equation}\label{eq:TRH}
    T_\text{RH} \simeq 10^9\; \text{GeV}\;\cdot  \left(\frac{\tilde{\lambda}_\sigma}{10^{-10}}\right)^\frac{5}{8}\cdot \left(\frac{g^2 \left|\lambda_{H\sigma}\right|/(4 \lambda_H)}{0.03}\right)^\frac{5}{8},
\end{equation}
where $g$ is a shorthand for the EW gauge couplings and we used typical parameters from the analysis of \cite{Ballesteros:2016xej}.
A recent reevaluation \cite{PhysRevD.106.063027} of the preheating dynamics revealed that one can no longer neglect the exponential growth of fluctuations of the SM like Higgs after non-thermal restoration of PQ symmetry, because there are large fluctuations in the singlet direction, whose imaginary part lowers the effective Higgs mass. The new results  \cite{PhysRevD.106.063027} indicate that the reheating temperature can be as large as
\begin{align}\label{eq:TRH2}
    T_\text{RH}= \left(10^{12}-10^{13}\right)\;\text{GeV}.
\end{align}
As it turns out the critical temperature for the PQ phase transition is $T_c \simeq 0.01\; f_a$ \cite{Ballesteros:2016xej}, which for $f_a<10^{11}\;\text{GeV}$  lies below the reheating temperature in \eqref{eq:TRH2} meaning that PQ symmetry is thermally restored in the HHSI case.

\subsection{Axion dark matter}\label{sec:axionDM}
It is well known that the cosmological evolution of the axion depends on the fate of PQ symmetry during inflation \cite{PhysRevLett.66.5}. In S.M.A.S.H. one finds that PQ symmetry is non-thermally restored during preheating \cite{Ballesteros:2016xej} via the growth of $\sigma$-excitations that destroy the coherence of the inflaton condensate as long as
\begin{equation}
    f_a \lesssim \SI{4e16}{\giga\electronvolt}.
\end{equation}
In case PQ symmetry was not restored only the misalignment mechanism \cite{PRESKILL1983127,ABBOTT1983133,DINE1983137} contributes to the axion DM abundance but this scenario leads to axion isocurvature fluctuations, which are tightly constrained \cite{Beltran:2006sq}. Requiring the absence of these fluctuations imposes the condition \cite{Fairbairn:2014zta}
\begin{equation}
    f_a < \SI{1.4e14}{\giga\electronvolt},
\end{equation}
implying that the non-restored scenario is ruled out. In the PQ restored scenario there are contributions from both axion misalignment and the decay of the network of topological defects  (made up of domain walls and axionic strings see subsection \ref{sec:KSVZ})
\cite{PhysRevD.24.2082,DAVIS1986225,PhysRevLett.48.1867,PhysRevD.35.1138,HARARI1987361,DAVIS1989167,LYTH1992279} which is unstable for $N_\text{DW} =1$. The contributions from the misalignment mechanism and the decay of the topological defects fits the observed DM relic density in the regime  \cite{Ballesteros:2016xej}
\begin{equation}\label{eq:faDM}
    \SI{3e10}{\giga\electronvolt}\lesssim f_a \lesssim \SI{1.2e11}{\giga\electronvolt},
\end{equation}
which corresponds to axion masses of \cite{Ballesteros:2016xej}
\begin{equation}
    \SI{50}{\micro\electronvolt}\lesssim m_a \lesssim \SI{200}{\micro\electronvolt}.
\end{equation}
A more recent study \cite{Buschmann:2021sdq} found a compatible range of masses in the window $\SI{40}{\micro\electronvolt}\leq m_a \leq\SI{180}{\micro\electronvolt}$. 
In the case that the entire axion relic abundance comes from misalignment only, the precise value of the required axion mass for post-inflationary PQ breaking varies from study to study and some examples are 
$m_a =\SI[separate-uncertainty = true]{14.6(1)}{\micro\electronvolt}$ \cite{Berkowitz:2015aua}, $m_a=\SI{18}{\micro\electronvolt}$ \cite{Fleury:2015aca} and $m_a=\SI[separate-uncertainty = true]{25.2(110)}{\micro\electronvolt}$ \cite{Buschmann:2019icd}. 
We show the corresponding parameter space $\SI{14}{\micro\electronvolt}\leq m_a \leq\SI{200}{\micro\electronvolt}$ for the axion to photon coupling as the blue interval labelled \enquote{axion DM} in the orange \enquote{S.M.A.S.H.E.D.}-band in figures \ref{fig:Axion-param-space-limits} and \ref{fig:Axion-param-space-projections}.
If one abandons the cosmological history of S.M.A.S.H. and assumes a sequestered sector for inflation, then pre-inflationary PQ breaking is possible again. In that regime only the misalignment mechanism is important and depending on the arbitrary initial misalignment angle
in principle all $f_a$ inside the \enquote{S.M.A.S.H.E.D.}-band could reproduce the DM relic abundance. For the remainder of this work we will assume the S.M.A.S.H. cosmology.
In our construction the axion DM could decay to neutrinos via the interaction \eqref{eq:axion-fermion} leading to the width
\begin{equation}
    \Gamma\left(a\rightarrow \nu_l \overline{\nu}_l\right) = \frac{m_a}{16 \pi}  \left(\frac{m_{\nu_l}}{f_a}\right)^2,
\end{equation}
where $\nu_l$ is the lightest neutrino with $m_a > 2 \;m_{\nu_l}$ and we neglected the phase space suppression from the final state neutrino masses. The condition $m_a > 2\; m_{\nu_l}$ can be recast as a bound on $f_a$
\begin{equation}
    f_a < \SI{3e7}{\giga\electronvolt} \;\cdot\left( \frac{\SI{0.1}{\electronvolt}}{m_{\nu_l}}\right).
\end{equation}
In order to be a good DM candidate the axion lifetime needs to be longer than about $249.6\;\text{Gyr}$ \cite{Simon:2022ftd} which leads to the bound
\begin{equation}
    f_a \gtrsim \SI{600}{\tera\electronvolt}\cdot \left(\frac{m_{\nu_l}}{\SI{0.1}{\electronvolt}}\right)^\frac{2}{3}.
\end{equation}
Since both conditions intimately rely on the unknown value of the lightest neutrino mass and not the overall neutrino mass scale, we do not depict them in figures \ref{fig:Axion-param-space-limits} and \ref{fig:Axion-param-space-projections}.
If we plug in our estimates for the lightest neutrino masses from the coming section \ref{sec:lightest} the bounds read
\begin{align}
    0.1 \;\text{GeV}&< f_a < 5\times 10^{17}\;\text{GeV} \quad \text{for} \quad m_{\nu_l} \simeq 6\times 10^{-12}\;\text{eV},\\
    175 \;\text{GeV}&< f_a < 6\times 10^{12}\;\text{GeV} \quad \text{for} \quad m_{\nu_l} \simeq 5\times 10^{-7}\;\text{eV}.
\end{align}
This agrees with the $f_a$ required for axion DM in \eqref{eq:faDM}.

\subsection{Vacuum stability and Leptogenesis}\label{sec.vac}
The S.M.A.S.H framework can also deal with the instability of the EW vacuum in the direction of the SM like Higgs by using the $\sigma$ field to implement the threshold stabilization mechanism \cite{Lebedev:2012zw,Elias-Miro:2012eoi}. Integrating out the radial mode of $\sigma$ shifts the quartic coupling of the SM like Higgs to 
\begin{equation}
    \tilde{\lambda}_H \equiv \lambda_H -\frac{\lambda_{H\sigma}^2}{\lambda_\sigma}
\end{equation}
at energies below $m_{h_\sigma}$. The mechanism works for typical values of $\lambda_{H\sigma}^2 / \lambda_\sigma \sim 10^{-2}$ \cite{Lebedev:2012zw}.
Absolute stability of the tree level vacuum requires that $\tilde{\lambda}_H>0$ as well as $\tilde{\lambda}_\sigma>0$ (see \eqref{eq:tilde}) \cite{Ballesteros:2016xej}.
Since  $\sigma$ couples to the exotic quarks and leptons, RGE effects from these interactions could destabilize the scalar potential in the $\sigma$-direction
if the following is not satisfied  (assuming a hierarchical fermion spectrum and using the triplets as a representative) \cite{Ballesteros:2016xej}
\begin{equation}\label{eq:boundM}
    M_T < 0.3 \cdot \lambda_\sigma^\frac{1}{4} \cdot f_a.
\end{equation}
Setting $\lambda_\sigma \simeq  5\times 10^{-10}$ (from \eqref{eq:infbounds}) and $f_a= 1.2\times 10^{11}\;\text{GeV}$ (from \eqref{eq:faDM}) as a first estimate  would then lead to $M_T<\SI{1.7e8}{\giga\electronvolt}$.
Since for gauge singlets such small masses would be problematic with successful  leptogenesis, which requires \cite{Buchmuller:2002rq}
\begin{equation}
    M_N \gtrsim \SI{5e8}{\giga\electronvolt},
\end{equation}
 the authors of \cite{Ballesteros:2016xej}  invoke a small amount of resonant enhancement to generate enough baryon asymmetry. As we will show in section \ref{eq:DiracLep} we do not need resonant enhancement for our model. 
Since we assume that the lightest triplet already has a mass of $\SI{e8}{\giga\electronvolt}$, the other triplets would not be allowed to remain heavier by a factor of 3-10 as is commonly assumed  because of \eqref{eq:boundM}. Therefore we would need to   include them in the cosmological analysis as well.
However since the inflationary bounds in \eqref{eq:infbounds} affect only $\tilde{\lambda}_\sigma$ defined in \eqref{eq:tilde} and not $\lambda_\sigma$ itself, we can relax the bound \eqref{eq:boundM} on $M_T$ by adjusting  $\lambda_{H\sigma} / \lambda_H$.
To be conservative we will assume that all exotic fermion masses are bounded from above not only by $f_a$ but by $\SI{e9}{\giga\electronvolt}$.

\section{Dirac-Leptogenesis in S.M.A.S.H.E.D.}\label{eq:DiracLep}

\subsection{Overview}
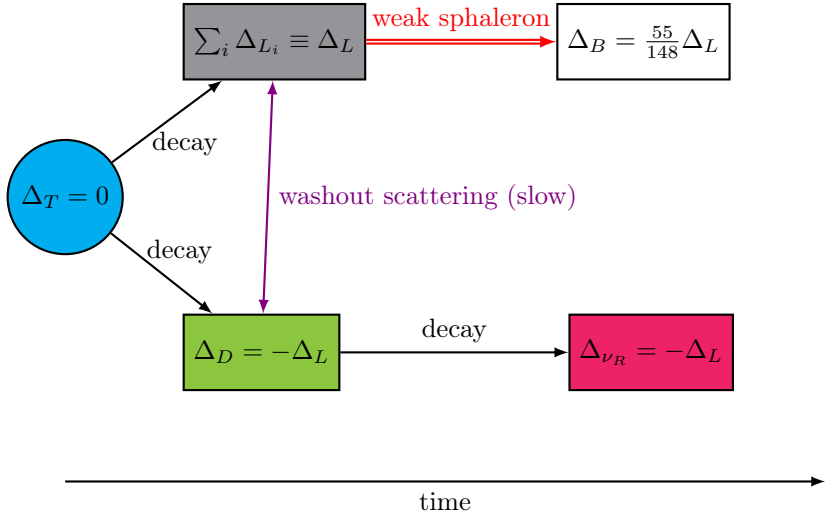
\begin{figure}[t!]
 \centering
  \tikzset{
  blackline/.style={thin, draw=black, postaction={decorate},
    decoration={markings, mark=at position 0.6 with {\arrow[black]{triangle 45}}}}
}

\begin{tikzpicture}[font=\small,thick]
 
% Start block
 \node[draw,
   circle,
   minimum size = 0.8 cm,
    fill=Cyan
] (block1) {$\Delta_T = 0$};

\node[draw,
    minimum width=1cm,
    minimum height=1cm,
    above right=of block1,
    fill=Gray
] (block2) { $\sum_i \Delta_{L_i} \equiv  \Delta_L$};

 \node[draw,
    minimum width=1cm,
    minimum height=1cm,
    below right=of block1,
    fill=LimeGreen
] (block3) { $\Delta_D = -\Delta_L$};

 \node[draw,
    minimum width=1cm,
    minimum height=1cm,
    right=3cm of block3,
    fill=WildStrawberry
] (block4) { $\Delta_{\nu_R} = -\Delta_L$};

 \node[draw,
    minimum width=1cm,
    minimum height=1cm,
    right=2.5cm of block2,
    fill=White
] (block5) { $\Delta_B = \frac{55}{148} \Delta_L$};

\coordinate[below = 3 cm of block1](start1);
\coordinate[right = 10 cm of start1](end1);

\draw[-latex ] (block1) -- (block2)  node[right,below,midway]{$\quad$ decay};
\draw[-latex ] (block1) -- (block3) node[right,above,midway]{$\quad$  decay};
\draw[-latex ] (block3) -- (block4) node[above,midway]{decay};
\draw[red,-latex,double] (block2) -- (block5) node[above,midway]{weak sphaleron};
\draw[-latex ] (start1) -- (end1)  node[below,midway]{time};
\draw[violet,latex-latex ] (block2) -- (block3)  node[right,midway]{washout scattering (slow)};

\end{tikzpicture}
  \caption{Schematic representation of our Dirac-Leptogenesis scenario.}
  \label{fig:asym-cartoon}
\end{figure}
\noindent The idea of Leptogenesis in theories without lepton number violation and Majorana masses was pionered by \cite{Dick:1999je} and applied to a plethora of proposed models in \cite{Murayama:2002je,Boz:2004ga,Cerdeno:2006ha,Thomas:2005rs,Thomas:2006gr,Chun:2008pg,Bechinger:2009qk,Chen:2011sb,Choi:2012ba,Heeck:2013vha,Borah:2016zbd,Gu:2016hxh,Gu:2017bdw,Gu:2019yvw,Mahanta:2021plx}.
Since the sphalerons only freeze out around the temperature of the EW crossover, which is far below the temperature range required for high scale leptogenesis, we work in the limit of unbroken $\text{SU(2)}_\text{L}$. We consider the lightest triplet with a mass $M_T\simeq  \SI{e8}{\giga\electronvolt} \ll M_T^{(2,3)}\simeq\mathcal{O}\left(\SI{e9}{\giga\electronvolt}\right)$ and assume that any preexisting asymmetry from the heavier triplets was washed out. Only the lightest exotic triplet and doublet fermions, hereafter denoted as $T$ and $D$, are present  in the plasma together with the SM. For each particle species we define
\begin{equation}
    \Sigma_\psi \equiv \frac{n_\psi +n_{\overline{\psi}} }{s}\quad  \text{and} \quad \Delta_\psi \equiv \frac{n_\psi -n_{\overline{\psi}} }{s}
\end{equation}
in terms of the number densities for particles and antiparticles $n_{\psi,\overline{\psi}}$ and the entropy density $s$. A schematic picture of our scenario can be found in figure \ref{fig:asym-cartoon}. The universe is initially symmetric with respect to $\text{B-L}$ enforcing a vanishing asymmetry $\Delta_T$ in the fermionic triplets. Their decays $T\rightarrow L  H$ and $T\rightarrow D H^\dagger$ generate equal and opposite asymmetries in the SM leptons $L$ and the heavy vector-like leptons $D$  so that $\Delta_L \equiv \sum_i \Delta_{L_i} = -\Delta_D$. Microscopically lepton number is conserved but since in the plasma the weak sphalerons couple only to the left chiral $L$ this will be the source of lepton number violation needed for the Sakharov criteria \cite{Sakharov_1991}. In appendix \ref{sec:sph-red-coeff} we  explain why the asymmetry in vector-like leptons can not be transferred into a baryon asymmetry by the weak sphalerons because they do not contribute to the violation of $\text{B+L}$  \cite{Cerdeno:2018dqk} via the weak interaction. Consequently only $\Delta_L$ can be converted into an asymmetry in baryons $\Delta_B$, where the corresponding conversion factor will also be determined in the aforementioned section. 
Since the vector-like leptons will be heavy with masses above the EW scale they will decay to $\nu_R$ transmitting their asymmetry to these gauge singlets, which do not couple to the weak sphalerons at all.
At late times we recover $\Delta_L = -\Delta_{\nu_R}$ demonstrating  lepton number conservation. In order to preserve the SM lepton asymmetry until the sphalerons freeze out at the electroweak phase transition we must enforce that the $2\rightarrow 2$ scattering processes that transform $L$ into $D$ are inefficient, which will be investigated in \ref{sec:wash}.

\subsection{CP violation}\label{sec:out-decays}
\begin{figure}[t]
 \centering
  \tikzset{
  blackline/.style={thin, draw=black, postaction={decorate},
    decoration={markings, mark=at position 0.6 with {\arrow[black]{triangle 45}}}},
    blueline/.style={thin, draw=blue, postaction={decorate},
    decoration={markings, mark=at position 0.6 with {\arrow[blue]{triangle 45}}}},
    redline/.style={thin, draw=red, postaction={decorate},
    decoration={markings, mark=at position 0.6 with {\arrow[red]{triangle 45}}}},
    greenline/.style={thin, draw=green, postaction={decorate},
    decoration={markings, mark=at position 0.6 with {\arrow[green]{triangle 45}}}},    
    graydashed/.style={dashed, draw=gray, postaction={decorate},
    decoration={markings}},
   yellowdashed/.style={dashed, draw=orange, postaction={decorate},
    decoration={markings}},
    photon/.style={decorate, draw=red,
    decoration={coil,amplitude=12pt, aspect=0}},
    reddashed/.style={thick, dashed, draw=red, postaction={decorate},
    decoration={markings}},
    photon/.style={decorate, draw=red,
    decoration={coil,amplitude=12pt, aspect=0}},
  gluon/.style={dashed, decorate, draw=black,
    decoration={coil, segment length=5pt, amplitude=8pt}}
  line/.style={thick, draw=black, postaction={decorate},
    decoration={markings}}
}

\NewDocumentCommand\semiloop{O{black}mmmO{}O{above}}
{%
\draw[#1] let \p1 = ($(#3)-(#2)$) in (#3) arc (#4:({#4+180}):({0.5*veclen(\x1,\y1)})node[midway, #6] {#5};)
}
% Syntax
%\semiloop[fermion][<draw options>]{<first node>}{<second node>}{<angle>}[<label>][<below, default: above>];

\begin{tikzpicture}[node distance=1cm and 1cm]

%tree level
\coordinate[label = left: $\color{blue} T_L^{(1)}$] (start1);
\coordinate[right=1cm of start1] (sigma1);
\coordinate[right=1cm of sigma1] (vertex1);
\coordinate[above right =1.2cm of vertex1,label=right: $L^{(i)}$] (L1);
\coordinate[below right =1.2cm of vertex1,label=right: $\color{gray} H$] (H1);

\coordinate[below=0.6cm of sigma1,label=below: $\color{orange} \braket{\sigma}$] (vevS1);

\draw[blueline] (start1)   -- (sigma1);
\draw[blueline] (sigma1)   -- (vertex1);
\draw[graydashed] (vertex1) -- (H1);
\draw[blackline] (vertex1) -- (L1);

\draw[yellowdashed] (sigma1)   -- (vevS1);

\coordinate[left=0.5cm of vertex1, label=above: $\color{blue} T_R^{(1)}$];

\fill (vertex1) circle (2pt);
\fill[blue] (sigma1) circle (2pt);

%vertex correction
\coordinate[right = 2 cm of vertex1, label=left: $\color{blue} T_L^{(1)}$](start2);
\coordinate[right = 1 cm of start2](vertex2);
\coordinate[below right =1.5cm of vertex2,label=below: $\color{green} D_R^{(1)}$] (Int1);
\coordinate[above right =1.5cm of vertex2,label=left: $\color{gray} H$] (Int2);

\fill (vertex2) circle (2pt);
\fill (Int1) circle (2pt);
\fill (Int2) circle (2pt);

\draw[blueline] (start2)   -- (vertex2);
\draw[graydashed] (vertex2) -- (Int2);
\draw[greenline] (vertex2) -- (Int1);

\coordinate[right=1.2cm of Int2,label=right: $L^{(i)}$] (L2);
\coordinate[right=1.2cm of Int1, label= right: $\color{gray} H$] (H2);
\draw[blackline] (Int2) -- (L2);
\draw[graydashed] (Int1) --(H2);

\coordinate[above= 1 cm of Int1] (sigma2);
\fill[blue] (sigma2) circle (2pt);
\draw[blueline] (Int1) -- (sigma2);
\draw[blueline] (sigma2) -- (Int2);
\coordinate[right = 0.8 cm of sigma2,label=right: $\color{orange} \braket{\sigma}$] (vevS2);
\draw[yellowdashed] (sigma2) -- (vevS2);

\coordinate[above right = 0.6cm of vevS2, label= left: $\color{blue} T_R^{(2,3)}$] (labelx);
\coordinate[below right = 0.6cm of vevS2, label= left: $\color{blue} T_L^{(2,3)}$] (labely);

\coordinate[left = 0.4cm of sigma2] (redhelp);
\coordinate[above = 1.5cm of redhelp] (red1);
\coordinate[below = 1.5cm of redhelp] (red2);
\draw[reddashed] (red1) --(red2);

% self energy 
\coordinate[right = 2 cm of vevS2, label=left: $\color{blue} T_L^{(1)}$] (start3);
\coordinate[right = 1cm of start3] (vertex3);
\draw[blueline] (start3) -- (vertex3);
\fill (vertex3) circle (2pt);

\coordinate[right= 0.75cm of vertex3] (redhelp2);
\coordinate[above = 1.5cm of redhelp2] (red3);
\coordinate[below = 1.5cm of redhelp2] (red4);
\draw[reddashed] (red3) --(red4);

\coordinate[right=1.5cm of vertex3] (vertex4);
\fill (vertex4) circle (2pt);
\semiloop[graydashed]{vertex3}{vertex4}{0}[$\color{gray}\quad  H$][above];
\semiloop[greenline]{vertex4}{vertex3}{180}[$\color{green}\qquad D_R^{(1)}$][below];

\fill (vertex4) circle (2pt);
\coordinate[right= 1cm of vertex4, label=above left: $\color{blue} T_L^{(2,3)}$] (sigma3);
\fill[blue] (sigma3) circle (2pt);
\coordinate[right= 1cm of sigma3, label = above left: $\color{blue} T_R^{(2,3)}$] (vertex5);
\draw[blueline] (vertex4) -- (sigma3);
\coordinate[below= 0.6 cm of sigma3, label=below: $\color{orange} \braket{\sigma}$] (vevS3);
\draw[yellowdashed] (sigma3) -- (vevS3);
\draw[blueline] (sigma3) -- (vertex5);
\fill (vertex5) circle (2pt);

\coordinate[above right =1.2cm of vertex5,label=right: $L^{(i)}$] (L3);
\coordinate[below right =1.2cm of vertex5,label=right: $\color{gray} H$] (H3);

\draw[graydashed] (vertex5) -- (H3);
\draw[blackline] (vertex5) -- (L3);

\end{tikzpicture}
  \caption{Feynman diagrams in the chiral basis for the CP violating decay of the lightest triplet fermion into SM leptons $L^{(i)}$. Internal lines intersecting the red dashed line are required to go on shell.}
  \label{fig:asym-L}
  \bigskip
    \tikzset{
  blackline/.style={thin, draw=black, postaction={decorate},
    decoration={markings, mark=at position 0.6 with {\arrow[black]{triangle 45}}}},
    blueline/.style={thin, draw=blue, postaction={decorate},
    decoration={markings, mark=at position 0.6 with {\arrow[blue]{triangle 45}}}},
    redline/.style={thin, draw=red, postaction={decorate},
    decoration={markings, mark=at position 0.6 with {\arrow[red]{triangle 45}}}},
    greenline/.style={thin, draw=green, postaction={decorate},
    decoration={markings, mark=at position 0.6 with {\arrow[green]{triangle 45}}}},    
    graydashed/.style={dashed, draw=gray, postaction={decorate},
    decoration={markings}},
   yellowdashed/.style={dashed, draw=orange, postaction={decorate},
    decoration={markings}},
    photon/.style={decorate, draw=red,
    decoration={coil,amplitude=12pt, aspect=0}},
    reddashed/.style={thick, dashed, draw=red, postaction={decorate},
    decoration={markings}},
    photon/.style={decorate, draw=red,
    decoration={coil,amplitude=12pt, aspect=0}},
  gluon/.style={dashed, decorate, draw=black,
    decoration={coil, segment length=5pt, amplitude=8pt}}
  line/.style={thick, draw=black, postaction={decorate},
    decoration={markings}}
}

\NewDocumentCommand\semiloop{O{black}mmmO{}O{above}}
{%
\draw[#1] let \p1 = ($(#3)-(#2)$) in (#3) arc (#4:({#4+180}):({0.5*veclen(\x1,\y1)})node[midway, #6] {#5};)
}
% Syntax
%\semiloop[fermion][<draw options>]{<first node>}{<second node>}{<angle>}[<label>][<below, default: above>];

\begin{tikzpicture}[node distance=1cm and 1cm]

%tree level
\coordinate[label = left: $\color{blue} T_R^{(1)}$] (start1);
\coordinate[right=1cm of start1] (sigma1);
\coordinate[right=1cm of sigma1] (vertex1);
\coordinate[above right =1.2cm of vertex1,label=right: $\color{green} D_R^{(1)}$] (DR1);
\coordinate[below right =1.2cm of vertex1,label=right: $\color{gray} H$] (H1);
\coordinate[below=0.6cm of sigma1,label=below: $\color{orange} \braket{\sigma}$] (vevS1);

\draw[blueline] (start1)   -- (sigma1);
\draw[blueline] (sigma1)   -- (vertex1);
\draw[graydashed] (vertex1) -- (H1);
\draw[greenline] (vertex1) -- (DR1);

\draw[yellowdashed] (sigma1)   -- (vevS1);

\coordinate[left=0.5cm of vertex1, label=above: $\color{blue} T_L^{(1)}$];

\fill (vertex1) circle (2pt);
\fill[blue] (sigma1) circle (2pt);

%vertex correction
\coordinate[right = 2 cm of vertex1, label=left: $\color{blue} T_R^{(1)}$](start2);
\coordinate[right = 1 cm of start2](vertex2);
\coordinate[below right =1.5cm of vertex2,label=below: $ L^{(i)}$] (Int1);
\coordinate[above right =1.5cm of vertex2,label=left: $\color{gray} H$] (Int2);

\fill (vertex2) circle (2pt);
\fill (Int1) circle (2pt);
\fill (Int2) circle (2pt);

\draw[blueline] (start2)   -- (vertex2);
\draw[graydashed] (vertex2) -- (Int2);
\draw[blackline] (vertex2) -- (Int1);

\coordinate[right=1.2cm of Int2,label=right: $\color{green} D_R^{(1)}$] (L2);
\coordinate[right=1.2cm of Int1, label= right: $\color{gray} H$] (H2);
\draw[greenline] (Int2) -- (L2);
\draw[graydashed] (Int1) --(H2);

\coordinate[above= 1 cm of Int1] (sigma2);
\fill[blue] (sigma2) circle (2pt);
\draw[blueline] (Int1) -- (sigma2);
\draw[blueline] (sigma2) -- (Int2);
\coordinate[right = 0.8 cm of sigma2,label=right: $\color{orange} \braket{\sigma}$] (vevS2);
\draw[yellowdashed] (sigma2) -- (vevS2);

\coordinate[above right = 0.6cm of vevS2, label= left: $\color{blue}  T_L^{(2,3)}$] (labelx);
\coordinate[below right = 0.6cm of vevS2, label= left: $\color{blue}  T_R^{(2,3)}$] (labely);

\coordinate[left = 0.4cm of sigma2] (redhelp);
\coordinate[above = 1.5cm of redhelp] (red1);
\coordinate[below = 1.5cm of redhelp] (red2);
\draw[reddashed] (red1) --(red2);

% self energy 
\coordinate[right = 2 cm of vevS2, label=left: $\color{blue} T_R^{(1)}$] (start3);
\coordinate[right = 1cm of start3] (vertex3);
\draw[blueline] (start3) -- (vertex3);
\fill (vertex3) circle (2pt);

\coordinate[right= 0.75cm of vertex3] (redhelp2);
\coordinate[above = 1.5cm of redhelp2] (red3);
\coordinate[below = 1.5cm of redhelp2] (red4);
\draw[reddashed] (red3) --(red4);

\coordinate[right=1.5cm of vertex3] (vertex4);
\fill (vertex4) circle (2pt);
\fill (vertex4) circle (2pt);
\semiloop[graydashed]{vertex3}{vertex4}{0}[$\color{gray}\quad  H$][above];
\semiloop[blackline]{vertex4}{vertex3}{180}[$\qquad  L^{(i)}$][below];

\coordinate[right= 1cm of vertex4, label=above left: $\color{blue} T_R^{(2,3)}$] (sigma3);
\fill[blue] (sigma3) circle (2pt);
\coordinate[right= 1cm of sigma3, label = above left: $\color{blue}T_L^{(2,3)}$] (vertex5);
\draw[blueline] (vertex4) -- (sigma3);
\coordinate[below= 0.6 cm of sigma3, label=below: $\color{orange} \braket{\sigma}$] (vevS3);
\draw[yellowdashed] (sigma3) -- (vevS3);
\draw[blueline] (sigma3) -- (vertex5);
\fill (vertex5) circle (2pt);

\coordinate[above right =1.2cm of vertex5,label=right: $\color{green} D_R^{(1)}$] (D3);
\coordinate[below right =1.2cm of vertex5,label=right: $\color{gray} H$] (H3);

\draw[graydashed] (vertex5) -- (H3);
\draw[greenline] (vertex5) -- (D3);

\end{tikzpicture}
  \caption{Feynman diagrams in the chiral basis for the CP violating decay of the lightest triplet fermion into the lightest exotic lepton $D^{(1)}$.  Internal lines intersecting the red dashed line are required to go on shell.}
  \label{fig:asym-DR}
\end{figure}
\noindent The relevant Feynman diagrams for generating the CP violating interference between the tree- and one-loop-level contributions to the decay $T\rightarrow L H$  are depicted in figure \ref{fig:asym-L}. In order to have a non-vanishing imaginary part for this interference term, the Cutkosky rules \cite{Cutkoskyrules} demand that the intermediate particles can go on shell.
Therefore the decay $T\rightarrow L H $ can only violate CP if the decay channel $T\rightarrow D H^\dagger$ is also open. Consequently we must demand that $M_D < M_T$ and take both channels into account. Another consequence of our construction is that the same vertices generating the asymmetry in $L$
also lead to an asymmetry in $D$ as can be seen from the analogous diagrams in figure \ref{fig:asym-DR}. We depicted the relevant kinematically allowed cut with a red dashed line in figures \ref{fig:asym-L} and \ref{fig:asym-DR}.
The tree-level decay widths are determined to be 
\begin{align}\label{eq:dec-tree}
  \Gamma\left(T\rightarrow L H \right) &\equiv   \sum_{\alpha,\beta}  \sum_i  \Gamma\left( T^a\rightarrow  L^\alpha_i H^\beta \right) = \frac{M_T}{32\pi}  \left(Y_{LT}^\dagger Y_{LT}\right)_{1 1},\\
    \Gamma\left(T\rightarrow D H^\dagger \right) &\equiv    \sum_{\alpha,\beta}   \Gamma\left( T^a\rightarrow  D^\alpha H^{\dagger\;\beta} \right)  = \frac{M_T}{32\pi}  \left|\left(Y_{TD}\right)_{11}\right|^2 \left(1-\delta^2\right)   \label{eq:dec-tree2},
\end{align}
where we summed over all $\text{SU(2)}_\text{L}$ indices of the final states  as well as SM lepton flavors $i$ and introduced $\delta\equiv M_D^2 / M_T^2$. Each component $T^a$ with $a=1,2,3$ has the same decay width. Note the phase space suppression for a massive final state fermion $\propto1-\delta^2$, which is different from the case for a massive final state scalar where one would have $\propto\left(1-\delta\right)^2$ instead. Decays to the doublets $D$ could be suppressed with respect to the leptonic mode by considering $\delta\lesssim 1$ \cite{Gu:2008yk}.  Both triplets and doublets will receive different thermal corrections to their tree level mass owing to their different gauge interactions so the mass ratio is not constant. However we will ignore thermal effects for this analysis. We define the CP-conserving branching ratios to be
\begin{equation}\label{eq:BRs}
   \text{B}_\text{L} \equiv \frac{\Gamma\left(T\rightarrow L H \right)}{\Gamma_\text{tot.}} = \frac{\Gamma\left(\overline{T}\rightarrow \overline{L} H^\dagger \right)}{\Gamma_\text{tot.}} = 1-\text{B}_\text{D}.
\end{equation}
Here  we also introduced the total decay width
\begin{align}\label{eq:gamma-tot}
    \Gamma_\text{tot} = \Gamma\left(T\rightarrow L H\right) +\Gamma\left(T\rightarrow D H^\dagger \right) =  \Gamma\left(\overline{T}\rightarrow \overline{L} H^\dagger\right)+\Gamma\left(\overline{T}\rightarrow \overline{D} H\right) 
\end{align}
in terms of the  tree-level decay widths from \eqref{eq:dec-tree}. 
Using CPT invariance together with unitarity one can show that for the matrix elements  
\begin{equation}\label{eq:CPT}
     \left|M(T\rightarrow L H)\right|^2 -  \left|M(\overline{T} \rightarrow \overline{L} H^\dagger)\right|^2
     = \left|M(\overline{T} \rightarrow \overline{D} H)\right|^2 - \left|M(T\rightarrow D H^\dagger)\right|^2,
\end{equation}
holds true, which implies that the asymmetries for both channels are equal and opposite 
\begin{align}
    \varepsilon  &\equiv \frac{\Gamma\left(T\rightarrow L H\right)-\Gamma\left(\overline{T} \rightarrow \overline{L} H^\dagger\right)}{2\Gamma_\text{tot}}= - \frac{\Gamma\left(T\rightarrow D H^\dagger\right)-\Gamma\left(\overline{T} \rightarrow \overline{D} H\right)}{2\Gamma_\text{tot}},
\end{align}
because the phase-space factors   divide out. Let us emphasize that the combined  asymmetry  of  all three SM lepton flavors equals the asymmetry stored in one generation of $D$.
We compute all imaginary parts using the Cutkosky prescription \cite{Cutkoskyrules}.
We split the asymmetry into a piece for the vertex correction and a piece for the self-energy correction (second and third diagrams in figures \ref{fig:asym-L} and \ref{fig:asym-DR})
\begin{equation}\label{eq:asym}
    \varepsilon_{L,D} =  \sum_{k\neq 1} \frac{\text{Im}\left[\left(Y_{LT}^\dagger Y_{LT} \right)_{1k} \left(Y_{TD}\right)_{k1} \left(Y_{TD}^\dagger\right)_{11} \right]}{\left(Y_{LT}^\dagger Y_{LT}\right)_{1 1}+\left|\left(Y_{TD}\right)_{11}\right|^2 \left(1-\delta^2\right) }\left(I^V_{L,D} + I^S_{L,D}\right)
\end{equation}
They read for the Lepton doublet with $x_k \equiv M_T^{(k)\;2} / M_T^2$
\begin{align}
  I_L^V = -\frac{\sqrt{x_k}}{8\pi}  \left(1-\delta - \left(1+x_k\right)\text{Log}\left(1+\frac{1-\delta}{x_k}\right) \right)  ,\quad 
    I_L^S =  \frac{\sqrt{x_k}}{8\pi} \frac{1-\delta^2}{1-x_k}.
\end{align}
For the $D$ we have $I_L^V=-I_D^V$ and $I_L^S=-I_D^S$. 
Note that because of the Cutkosky rule both expressions rely on a $D$ from the $T$ decay going on-shell so they vanish for $\delta\rightarrow 1$. We checked that the asymmetry parameter for $L$ reduces to the Type III Seesaw  result \cite{Hambye:2003rt} in the limits $\delta\rightarrow 0$ and $Y_{TD}\rightarrow Y_{LT}$:
\begin{equation}
    I_L^V + I_L^S \Big|_{\delta=0} =- \frac{\sqrt{x_k}}{8\pi}\left(1-\left(1+x_k\right)\text{Log}\left(1+\frac{1}{x_k}\right)-\frac{1}{1-x_k}\right).
\end{equation}
Unlike the case for a decaying gauge singlet, here there is a relative minus sign between the loop factors from the vertex- and self-energy-corrections \cite{Hambye:2003rt}.
The diagram involving an intermediate first generation triplet does not contribute in equation \eqref{eq:asym} because for $k=1$ we find the following combination of couplings
\begin{equation}
    \left(Y_{LT}^\dagger Y_{LT} \right)_{11} \left(Y_{TD}\right)_{11} \left(Y_{TD}^\dagger\right)_{11} = \sum_{i=1}^{3}    \left(Y_{LT}^\dagger\right)_{1i} \left( Y_{LT} \right)_{i1} \left|\left(Y_{TD}\right)_{11}\right|^2 = \sum_{i=1}^{3}  \left|\left( Y_{LT} \right)_{i1}\right|^2 \left|\left(Y_{TD}\right)_{11}\right|^2,
\end{equation}
which is purely real.

\subsection{Enhancement of the asymmetry parameter}\label{sec:boost}
 
We can find a simpler expression in the case of (infinitely) hierarchical triplets $x_k\gg1$:
\begin{equation}\label{eq:asymm-delta}
    \varepsilon_{L} \simeq -\frac{1}{16\pi } \sum_{k\neq 1} \frac{\text{Im}\left[\left(Y_{LT}^\dagger Y_{LT} \right)_{1k} \left(Y_{TD}\right)_{k1} \left(Y_{TD}^\dagger\right)_{11} \right]}{\left(Y_{LT}^\dagger Y_{LT}\right)_{1 1}+\left|\left(Y_{TD}\right)_{11}\right|^2 \left(1-\delta^2\right) } \sqrt{\frac{1}{x_k}} (1- \delta^2),
\end{equation}
which (for $\delta\ll1 $) is smaller by a factor of 3 compared to the singlet case due to relative sign between both contributions \cite{Hambye:2003rt}. If we assume that the couplings responsible for the decay of the lightest triplet in \eqref{eq:dec-tree} are real valued then we may trade them for the branching ratios to obtain
\begin{equation}\label{eq:BRS}
    \varepsilon_{L} \simeq -\frac{M_T}{\sqrt{3}\;16\pi}\sqrt{B_L B_D (1-\delta^2)}  \sum_{i=1}^{3}\sum_{k\neq1}\frac{ \text{Im}\left[\left(Y_{LT}\right)_{ik} \left(Y_{TD}\right)_{k1}\right]}{M_T^{(k)}},
\end{equation}
where the factor of $\sqrt{3}$ takes into account that we assume the same $\left(Y_{LT}\right)_{i1}$ for three generations $i$ of SM leptons. By maximising the branching ratios $B_L=B_D=1/2$ and neglecting the phase space suppression ($\delta\ll 1$) we find 
\begin{equation}\label{eq:eps-simpl}
    \varepsilon_{L} \simeq -\frac{M_T}{\sqrt{3}\;32\pi}  \sum_{i=1}^{3}\sum_{k\neq1}\frac{ \text{Im}\left[\left(Y_{LT}\right)_{ik} \left(Y_{TD}\right)_{k1}\right]}{M_T^{(k)}}
\end{equation}
which is completely independent of the couplings for the tree level decay in \eqref{eq:dec-tree}. The Yukawas determining the cosmological evolution of the triplets and their out of equilibrium conditions discussed in the next section in \eqref{eq:eff-Yuk} are therefore different from the  couplings appearing in the asymmetry. Unlike in  most Majorana Seesaw models we can not just use the Casas-Ibarra parametrization \cite{Casas:2001sr} to trade the Yukawa couplings for expressions from the available low energy neutrino data. This would allow one to set an upper bound on $\varepsilon$ a la  Davidson-Ibarra \cite{Davidson:2002qv,Hambye:2003rt}:
\begin{equation}\label{eq:vanilla-DI}
    \left|\varepsilon_L\right| \leq \varepsilon_L^\text{DI max} = \frac{1}{16 \pi} \frac{M_T \left(m_3-m_1\right)}{v_H^2} < \SI{3e-9}{} \cdot \left(\frac{M_T}{\SI{e8}{\giga\electronvolt}}\right)\cdot \left(\frac{m_3}{\SI{0.1}{\electronvolt}}\right),
\end{equation}
where $m_3$ and $m_1$ are the heaviest and lightest active neutrino respectively. This occurs since the triplet decay rate does not involve the coupling $Y_{DR}$  to $\nu_R$ needed to reconstruct the neutrino mass matrix. Instead we find the following \enquote{effective mass} parameter 
\begin{equation}
\tilde{m}_{i1} \equiv \sum_{k\neq1}  \left(Y_{LT}\right)_{ik} \left(Y_{TD}\right)_{k1} \frac{v_H^2}{M_T^{(k)}} \simeq \SI{60}{\kilo\electronvolt}\cdot  \sum_{k\neq1}   \left(\frac{\left|\left(Y_{LT}\right)_{ik}\right|}{\mathcal{O}(1)}\right) \cdot\left(\frac{\left|\left(Y_{TD}\right)_{k1}\right|}{\mathcal{O}(1)}\right) \cdot  \left(\frac{\SI{e9}{\giga\electronvolt}}{M_T^{(k)}}\right) 
\end{equation}
 playing the role of $m_3$ in our case. This parameter can be much larger than in the usual models, because it is missing a suppression factor of $Y_{DR} v_H / M_D$ compared to the active neutrino mass scale. Hence we will be able to generate a significantly larger asymmetry  than the $\mathcal{O}\left(10^{-9}\right)$ for $M_T\simeq\SI{e8}{\giga\electronvolt}$ without having to invoke the limit of resonant leptogenesis where $M_T^{(1)}\approx M_T^{(2,3)}$ \cite{Pilaftsis:2003gt}. To quantify this enhancement we utilize our optimized asymmetry from \eqref{eq:eps-simpl}  and decompose the Yukwawa couplings  into their real parts and phases:
\begin{equation}\label{eq:eps-approx}
    \left|\varepsilon_{L}\right| \simeq \frac{1}{32 \sqrt{3}\pi}  \sum_{i}\sum_{k\neq1}   \frac{M_T}{M_T^{(k)}} \left|\left(Y_{LT}\right)_{ik}\right| \left|\left(Y_{TD}\right)_{k1}\right| \left|\text{sin}\left(\left(\alpha_{LT}\right)_{ik}+ \left(\alpha_{TD}\right)_{k1} \right)\right|.
\end{equation}
Since the  absolute value of the sine is bounded from above we get an upper limit of
\begin{align}\label{eq:asym-uplimit}
    \left|\varepsilon_{L}\right| &\lesssim \left|\varepsilon_L^\text{max}\right|  =  \frac{1}{32 \sqrt{3} \pi}  \sum_{i}\sum_{k\neq1}   \frac{M_T}{M_T^{(k)}} \left|\left(Y_{LT}\right)_{ik}\right| \left|\left(Y_{TD}\right)_{k1}\right| \\
    &\simeq \SI{5.7e-4}{}\cdot \sum_{i}\sum_{k\neq1}  \left( \frac{ M_T /M_T^{(k)}}{1/10}\right) \cdot  \left(\frac{\left|\left(Y_{LT}\right)_{ik}\right|}{\mathcal{O}(1)}\right) \cdot\left(\frac{\left|\left(Y_{TD}\right)_{k1}\right|}{\mathcal{O}(1)}\right).
\end{align}
If we assume that the sum on the right hand side is flavor-independent and that there are no accidental cancellations we obtain an additional factor of six to that $\left|\varepsilon_{L}\right|\lesssim 3\times10^{-3}$.
Note that we did not assume any resonant enhancement from the self energy diagrams for mass degenerate generations of triplets \cite{Pilaftsis:2003gt}. Thus our scenario provides an alternative approach to resonant leptogenesis \cite{Pilaftsis:2003gt}  for enhancing the asymmetry parameter. Since \eqref{eq:asym-uplimit} depends only on the ratio of masses, we could even try to realize the neutrino masses by integrating out twice as many species of exotic fermions lowering their mass scale closer to the TeV range.
We conclude that due to our sequential Seesaw  needing two species of heavy mediators we were able to generate a lepton asymmetry that can potentially be up to six orders of magnitude larger than in conventional models for $M_T\simeq \SI{e8}{\giga\electronvolt}$. Note that this asymmetry still satisfies the perturbativity requirement  $ \left|\varepsilon_L\right|\ll1$ that is assumed when deriving the semi-classical Boltzmann-equations in \ref{sec:Boltz}. 

\subsection{Analytical estimates}\label{sec:analy}
 We can determine the asymmetry in SM leptons from $T$ decays to be \cite{Giudice:2003jh}
\begin{equation}
    \Delta_L =  3\cdot  \kappa \cdot \left|\varepsilon_L\right| \cdot \Sigma_T\left(T\gg M_T\right),
\end{equation}
where the factor of 3 comes from the three components of the decaying triplet \cite{Hambye:2003rt} and we have
\begin{equation}
    \Sigma_T\left(T\gg M_T\right) =  4\cdot \frac{135 \zeta(3)}{8\pi^4 g_*(T\gg M_T)}
\end{equation}
with a spin degeneracy factor of 4 because $T$ is a Dirac fermion and $g_*(T\gg M_T)=\mathcal{O}(100)$ is the effective number of degrees of freedom in entropy.
Using this we find that a baryon abundance of 
\begin{equation}\label{eq:sphal2}
    \frac{n_B}{s}  = c_\text{sph.} \Delta_L,
\end{equation}
 where the sphaleron redistribution coefficient in this model is determined in equation \eqref{eq:red-coeff} of appendix \ref{sec:sph-red-coeff}
\begin{equation}
    c_\text{sph.} =   \frac{55}{148}.
\end{equation}
Using that $s=7.04\;n_\gamma$ together with  conservation of $n_B/s$ one obtains a Baryon-to photon ratio $\eta_B$ today  of \cite{Giudice:2003jh}
\begin{equation}\label{eq:baryon-to-photon}
    \eta_B = \frac{n_B}{n_\gamma}\Big|_\text{today} \simeq  6.1\times 10^{-2} \cdot \kappa \cdot \left|\varepsilon_L\right|,
\end{equation}
In this context $\kappa$ is the so called efficiency factor, which penalizes the triplets staying close to thermal equilibrium, as this would violate Sakharov's first condition \cite{Sakharov_1991}. Its functional form can be approximated as
\begin{equation}
    \kappa \simeq \frac{\Sigma_T \left(T\ll M_T\right)}{\Sigma_T^\text{eq.}\left(T\gg M_T\right)} \Big|_{T=T_\text{FO}},
\end{equation}
where $T_\text{FO}$ is the temperature at which the last interactions that changes the leptonic asymmetry freezes out. Decaying particles far away from equilibrium are still as abundant as radiation at $T_\text{FO}\ll M_T$ so $\Sigma_T \left(T_\text{FO}\ll M_T\right)\simeq \Sigma_T^\text{eq.}\left(T_\text{FO}\gg M_T\right)$ which implies $\kappa\simeq1$ as long as they do not dominate the energy density of the universe. In case they do one can even have $\kappa\sim g_*\gg 1$ \cite{Strumia:2006qk}. As a consequence of their weak scale gauge interactions the triplets will have an initial thermal population after reheating so that $\kappa<1$. The penalizing effect can be seen if one considers a fully thermalized triplet with $\Sigma_T \left(T_\text{FO}\ll M_T\right)\simeq \Sigma_T^\text{eq.} \left(T_\text{FO}\ll M_T\right)$  which would imply $\kappa \sim e^{-\frac{M_T}{T_\text{FO}}}\ll 1$.
If we compare our estimate to the value extracted from BBN and Planck data \cite{Fields:2019pfx}
\begin{equation}\label{eq:etaB}
    \eta_B = \left(6.143\pm0.190\right)\times 10^{-10}
\end{equation}
we find that we need an efficiency factor of 
\begin{equation}\label{eq:eff-BM}
    \kappa\gtrsim  1.8\times 10^{-5} \cdot \left( \frac{5.7\times 10^{-4}}{\left|\varepsilon_L\right|}\right).
\end{equation}

\subsubsection{Efficiency parameter}\label{sec:eff}
\noindent The task is now to show that our model can realize a cosmological history that leads to the required value of $\kappa$. Before we do so numerically we will find the required parameters using analytical arguments. Around $T=M_T$ the distribution function of a thermalized fermion would change from its Fermi-Dirac shape to a Maxwell-Boltzmann distribution leading to the non-relativistic expressions for its number density etc. On the other hand a fermion that  decoupled at $T\gg M_T$ would keep its relativistic Fermi-Dirac distribution, provided that the fermion self interactions also have decoupled. Hence $T=M_T$ is the right epoch to quantify the deviation from thermal equilibrium needed to satisfy the Sakharov conditions. We define the customary decay parameter from the decay widths in \eqref{eq:dec-tree}
\begin{equation}\label{eq:decay-param}
    K \equiv \frac{\Gamma\left(T\rightarrow L H\right)+\Gamma\left(T\rightarrow D H^\dagger\right)}{ H(T)}\Big|_{T=M_T}
\end{equation}
and introduce the effective Yukawa coupling
\begin{equation}
    Y \equiv \sqrt{\left(Y_{LT}^\dagger Y_{LT}\right)_{1 1}+\left|\left(Y_{TD}\right)_{11}\right|^2 \left(1-\delta^2\right)}
\end{equation}
Then we re-express the Yukawa coupling in terms of the decay parameters
\begin{equation}\label{eq:eff-Yuk}
    Y \simeq 4\times10^{-4}\cdot  \sqrt{\frac{K}{100}} \cdot \left(\frac{g_*(M_T)}{100}\right)^\frac{1}{4}  \cdot \sqrt{\frac{M_T}{\SI{e8}{\giga\electronvolt}}}
\end{equation}
for later convenience. If the triplet had no gauge interactions  the Sakharov criteria would be satisfied for an out of equilibrium decay with $K\ll1$ implying $\kappa\simeq1$. The efficiency for a decaying fermion with EW gauge interactions can be estimated as \cite{Hambye:2003rt,Strumia:2006qk}
\begin{equation}
    \kappa \simeq \text{Min}\left(1,\frac{1}{K}, \frac{M_T}{10^{12-13}\;\text{GeV}}\text{Max}\left(1,K\right)\right).
\end{equation}
Since we consider $M_T\ll \SI{e12}{\giga\electronvolt}$ the efficiency will always be less than unity. In principle there are three  regimes and we can understand the parametrics in terms of the last process to decouple from equilibrium:
\begin{enumerate}
    \item \textbf{Weak washout from inverse decays:}\\  $K\ll 1 \; \left(Y\ll 3.5\times10^{-5}\right)$  for which $\kappa \sim M_T / 10^{12-13}\;\text{GeV}\sim  10^{-4-5}$
\end{enumerate}

\noindent The gauge interactions lead to annihilation processes $T\overline{T} \leftrightarrow W W,\; F F$, where $W$ are the $\text{SU(2)}_\text{L}$ gauge bosons and $F=H,L,Q$ are the SM Higgs and Fermion doublets. Naively one would expect that this violates Sakharov's first condition, however one should not forget that the gauge scatterings will eventually drop out of equilibrium.
The  thermally averaged scattering rate $ \Gamma\left(T \;\overline{T} \leftrightarrow W W, F \overline{F}\right) = \gamma\left(T \;\overline{T} \leftrightarrow W W, F \overline{F}\right) / n_T^\text{eq.}$ can obtained from the expression \eqref{eq:gauge-scat} in the appendix and we find
\begin{equation}\label{eq:gaugfo}
    \frac{M_T}{T_\text{FO}^\text{gaug.}}  \simeq 17\quad \text{for} \quad M_T = \SI{e8}{\giga\electronvolt}.
\end{equation}
The  regime $K\ll 1$, also known as the regime of weak washout from inverse decays $L_i H,\; D H^\dagger \rightarrow T$, corresponds to a situation where the (inverse) decays are slow when the $T$s freeze out from gauge interactions at typical values of $M_T / T_\text{FO}^\text{gaug.} \simeq \mathcal{O}(20)$. Consequently the  frozen out abundance of $T$s decays out of equilibrium. The factor $\kappa\sim M_T /10^{12-13}\;\text{GeV} \sim 10^{-4-5}$ measures the abundance of triplets surviving the gauge annihilations.

\begin{enumerate}
    \setcounter{enumi}{1}
    \item \textbf{Strong washout from inverse decays}\\ $K\gtrsim 10^5\; \left(Y\gtrsim 1.3\times10^{-2}\right)$   for which  $\kappa\sim 1/K \lesssim 10^{-5}$
\end{enumerate}
In the regime of strong washout from inverse decays $K\gg 1$, the gauge interactions can freeze out before the inverse decays drop out of equilibrium. One can show that the approximate freeze out temperature is \cite{Giudice:2003jh}
\begin{equation}\label{eq:dec-FO}
    \frac{M_T}{T_\text{FO}^\text{dec.}} \simeq 5 \times \sqrt{\text{Log}(K)}  
\end{equation}
We find that the inverse decays  freeze out after the gauge scatterings for $K\gtrsim10^5$ and the efficiency factor corresponds to the one in the strong washout regime $\kappa \sim 1/K\lesssim 10^{-5}$ for vanilla leptogenesis.
\begin{enumerate}
    \setcounter{enumi}{2}
    \item \textbf{Intermediate regime}\\ $1<K<10^5\; \left( 3.5\times10^{-5}<Y<1.3\times10^{-2}\right)$   for which\\  $\kappa\sim \text{Min}\left(1/K, K \cdot  M_T/ 10^{12-13}\;\text{GeV}  \right)$
\end{enumerate}
In the intermediate regime   there is a non-negligible amount of decays during the annihilation phase. These lead to less efficient  gauge-annihilations  thus creating an efficiency $\kappa\sim \text{Min}\left(1/K, K \cdot  M_T/ 10^{12-13}\;\text{GeV}  \right) $, which can be  larger by up to a  factor of $K$ than the weak washout regime.
This means that for $K \sim 100$ we get $\kappa \sim K\cdot M_T / 10^{12-13}\;\text{GeV}\sim 10^{-2-3}$, which is the largest enhancement possible as larger $K\gtrsim1000$ will lead to $\kappa \sim 1/ K \lesssim 10^{-3}$.\\
\\
Comparing with \eqref{eq:eff-BM} we conclude that the weak washout regime $\left(\kappa \sim  10^{-4-5}\right)$ might be efficient enough to realize leptogenesis, whereas the strong washout regime  $\left(\kappa\lesssim 10^{-5}\right)$
being less efficient by up to a factor of ten and might not work even with the largest possible asymmetries. The intermediate regime can have a larger efficacy $\left(\kappa\sim 10^{-2-3}\right)$ than for weak washout so leptogenesis will definitely work in this regime.
For more precise estimates we will determine the relevant Boltzmann equations in \ref{sec:Boltz} and solve them in section \ref{sec:num}.\\
The numerical results of \cite{Hambye:2003rt} for the case of Majorana triplets indicate that for the considered range of triplet masses $M_T=\SI{e8}{\giga\electronvolt}$ the maximal efficiency is $\kappa_\text{max.} \simeq \mathcal{O}\left(10^{-3}\right)$, which agrees with the previous conclusions. Note that this number was calculated for self-conjugate fermions, that also have a different washout scattering process as will be discussed in section \ref{sec:wash}. The low efficiency together with the Davidson-Ibarra bound on the asymmetry parameter in \eqref{eq:vanilla-DI} is the reason why Majorana triplet leptogenesis is only possible for $M_T>\SI{1.4e10}{\giga\electronvolt}$ \cite{Hambye:2003rt}. However in section \ref{sec:num} we will numerically demonstrate that our Dirac-scenario is successful  even for $M_T=\SI{e8}{\giga\electronvolt}$ due to larger asymmetry parameter and the fact that the decaying particle is not self conjugate.\\
\\
Additionally there could be annihilations of the triplet into the PQ scalars $T\overline{T}\leftrightarrow h_\sigma h_\sigma, a\; a, h_\sigma a$ \cite{Langacker:1986rj,Hannestad:2005ex}. We find that they are slow compared to the Hubble rate at $T=M_T$ as long
as 
\begin{equation} 
    f_a > \mathcal{O}\left(\SI{e10}{\giga\electronvolt}\right) \cdot \left(\frac{M_T}{\SI{e8}{\giga\electronvolt}}\right)^\frac{3}{4} \cdot \left(\frac{100}{g_*(M_T)}\right)^{\frac{1}{8}},
\end{equation}
which is in agreement with the parameters needed for axion DM in \eqref{eq:faDM}.

\subsubsection{Washout scattering and heavy vector-like leptons}\label{sec:wash}
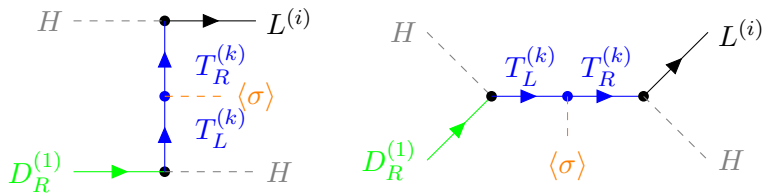
\begin{figure}[t]
 \centering
  \tikzset{
  blackline/.style={thin, draw=black, postaction={decorate},
    decoration={markings, mark=at position 0.6 with {\arrow[black]{triangle 45}}}},
    blueline/.style={thin, draw=blue, postaction={decorate},
    decoration={markings, mark=at position 0.6 with {\arrow[blue]{triangle 45}}}},
    redline/.style={thin, draw=red, postaction={decorate},
    decoration={markings, mark=at position 0.6 with {\arrow[red]{triangle 45}}}},
    greenline/.style={thin, draw=green, postaction={decorate},
    decoration={markings, mark=at position 0.6 with {\arrow[green]{triangle 45}}}},    
    graydashed/.style={dashed, draw=gray, postaction={decorate},
    decoration={markings}},
   yellowdashed/.style={dashed, draw=orange, postaction={decorate},
    decoration={markings}},
    photon/.style={decorate, draw=red,
    decoration={coil,amplitude=12pt, aspect=0}},
    reddashed/.style={thick, dashed, draw=red, postaction={decorate},
    decoration={markings}},
    photon/.style={decorate, draw=red,
    decoration={coil,amplitude=12pt, aspect=0}},
  gluon/.style={dashed, decorate, draw=black,
    decoration={coil, segment length=5pt, amplitude=8pt}}
  line/.style={thick, draw=black, postaction={decorate},
    decoration={markings}}
}

\NewDocumentCommand\semiloop{O{black}mmmO{}O{above}}
{%
\draw[#1] let \p1 = ($(#3)-(#2)$) in (#3) arc (#4:({#4+180}):({0.5*veclen(\x1,\y1)})node[midway, #6] {#5};)
}
% Syntax
%\semiloop[fermion][<draw options>]{<first node>}{<second node>}{<angle>}[<label>][<below, default: above>];

\begin{tikzpicture}[node distance=1cm and 1cm]

%vertex correction

\coordinate[](start2);
\coordinate[right = 1.2 cm of start2](start3);
\coordinate[below = 1 cm of start3] (Int1);
\coordinate[above = 1 cm of start3] (Int2);

\fill (Int1) circle (2pt);
\fill (Int2) circle (2pt);

\coordinate[above = 1 cm of start2,label=left: $\color{gray} H$ ](help1);
\coordinate[below= 1 cm of start2,label=left: $\color{green} D_R^{(1)}$](help2);

\draw[graydashed] (help1) -- (Int2);
\draw[greenline] (help2) -- (Int1);

\coordinate[right=1.2cm of Int2,label=right: $L^{(i)}$] (L2);
\coordinate[right=1.2cm of Int1, label= right: $\color{gray} H$] (H2);
\draw[blackline] (Int2) -- (L2);
\draw[graydashed] (Int1) --(H2);

\fill[blue] (start3) circle (2pt);
\draw[blueline] (Int1) -- (start3);
\draw[blueline] (start3) -- (Int2);
\coordinate[right = 0.8 cm of start3,label=right: $\color{orange} \braket{\sigma}$] (vevS2);
\draw[yellowdashed] (start3) -- (vevS2);

\coordinate[above right = 0.6cm of vevS2, label= left: $\color{blue} T_R^{(k)}$] (labelx);
\coordinate[below right = 0.6cm of vevS2, label= left: $\color{blue} T_L^{(k)}$] (labely);

% self energy 
\coordinate[right = 2 cm of vevS2] (start4);
\coordinate[right=1.5cm of start4] (vertex4);

\coordinate[above left = 1.2 cm of vertex4,label=left: $\color{gray} H$ ](help3);
\coordinate[below left = 1.2 cm of vertex4,label=left: $\color{green} D_R^{(1)}$](help4);

\draw[graydashed] (help3) -- (vertex4);
\draw[greenline] (help4) -- (vertex4);

\fill (vertex4) circle (2pt);
\coordinate[right= 1cm of vertex4, label=above left: $\color{blue} T_L^{(k)}$] (sigma3);
\fill[blue] (sigma3) circle (2pt);
\coordinate[right= 1cm of sigma3, label = above left: $\color{blue} T_R^{(k)}$] (vertex5);
\draw[blueline] (vertex4) -- (sigma3);
\coordinate[below= 0.6 cm of sigma3, label=below: $\color{orange} \braket{\sigma}$] (vevS3);
\draw[yellowdashed] (sigma3) -- (vevS3);
\draw[blueline] (sigma3) -- (vertex5);
\fill (vertex5) circle (2pt);

\coordinate[above right =1.2cm of vertex5,label=right: $L^{(i)}$] (L3);
\coordinate[below right =1.2cm of vertex5,label=right: $\color{gray} H$] (H3);

\draw[graydashed] (vertex5) -- (H3);
\draw[blackline] (vertex5) -- (L3);

\end{tikzpicture}
  \caption{Diagrammatic representation of the relevant washout processes  $D H \leftrightarrow L H $   involving leptons $L$ and exotic leptons $D$.  The diagrams for $L \overline{D} \leftrightarrow H H$ are obtained via crossing symmetry. }
  \label{fig:washout-diagrams}
\end{figure}
\noindent 
The defining feature of the original Dirac-Leptogenesis scenario is that the efficient conversion of the asymmetries in  $L$ into   $\nu_R$ would spoil baryogenesis, as sphalerons do not act on the $\nu_R$. However for a coupling $\overline{L} H^\dagger \nu_R$ this only occurs much after the sphaleron freeze-out due to the tiny Yukawa coupling connecting both to the SM Higgs $H$ \cite{Dick:1999je}. In our scenario before EWSB these fermions couple to three $H$ instead of one via the dimension six operator in \eqref{eq:dim-six} so the following three body scattering is possible
\begin{equation}
    \Gamma\left( \overline{L}\;\nu_R \leftrightarrow H H^\dagger H\right) \sim \frac{1}{32\pi^2} \left(\frac{m_\nu}{v_H^3}\right)^2 T^5.
\end{equation}
The above was estimated using dimensional analysis and the prefactor is an educated guess to take the three-body phase space into account. Two body annihilations of two Higgses into $\overline{\nu_L} \nu_R$ are only possible after the SM Higgs gets a vev so the sphalerons are already frozen out and this process is irrelevant for leptogenesis. Three body annihilations decouple from the bath at around
\begin{equation}
    T^{2\rightarrow 3}_\text{FO} \simeq \SI{2.3e6}{\giga\electronvolt}\cdot \left(\frac{\SI{0.1}{\electronvolt}}{m_\nu}\right)^\frac{2}{3},
\end{equation}
where we emphasize that this is just an order of magnitude estimate. As demonstrated earlier the triplets only decouple from gauge and Yukawa interactions around $T_\text{FO}\simeq M_T / \mathcal{O}(20) \simeq \SI{5e6}{\giga\electronvolt}$ (see equations \eqref{eq:gaugfo} and \eqref{eq:dec-FO}) for any asymmetry to develop. We expect the $2\rightarrow 3$ processes to be slow enough in this regime to disregard them compared to the next washout processes.\\
\\
The scattering processes  $D H  \leftrightarrow L H $ depicted in figure \ref{fig:washout-diagrams} and $L \overline{D} \leftrightarrow H H$   are  less suppressed compared to the previous process because they are two body reactions and lack the factors of $Y_{DR}/ M_D$ to form the neutrino mass.
The existence of these scatterings is also required by the Cutkosky rule from the diagrams in \ref{fig:asym-L}, \ref{fig:asym-DR} and our estimate for the rate is (see \eqref{eq:cross} in the appendix for the cross section)
\begin{equation}
    \Gamma\left(L_i \overline{D} \leftrightarrow HH\right)\sim \Gamma\left(L_i H \leftrightarrow D H\right)\sim \frac{1}{16\pi}\left(Y_{LT}Y^\dagger_{LT}\right)_{ii} \left(Y^\dagger_{TD}Y_{TD}\right)_{11} \frac{T^3}{M_T^{(k)\;2}}.
\end{equation}
We demand that this rate is slow compared to the Hubble rate at $T=M_T / 20 $, which as explained before is the relevant temperature scale for leptogenesis, provided that
\begin{equation}\label{eq:washout-bound}
    \left(Y_{LT}\right)_{i1} \left(Y_{TD}\right)_{11} <  5\cdot  10^{-4}\cdot  \sqrt{\frac{M_T}{\SI{e8}{\giga\electronvolt}}} \cdot \left(\frac{g_*\left(M_T / 25\right)}{100}\right)^\frac{1}{4}.
\end{equation}
Since we  assume that $M_T\gg M_D$  it may not be valid to neglect the $D$ mass at this temperature $T\ll M_T$. 
If we set $ Y \simeq\left(Y_{LT}\right)_{i1}\simeq \left(Y_{TD}\right)_{11}$ the bound implies $Y\lesssim 10^{-2}$, which is compatible with $K\lesssim 10^4$ see equation \eqref{eq:eff-Yuk}. The two benchmarks $K\ll 1$ and $K=100$ are therefore safe from this washout process and we expect it to matter only for the region $K\sim 10^5$.
For exchange of a heavier triplet $T^{(k)}\;(k\neq 1)$ the estimate changes to
\begin{equation}\label{eq:washout-bound2}
     \left(Y_{LT}\right)_{ik} \left(Y_{TD}\right)_{k1} <  1.5\cdot 10^{-3}\cdot \sqrt{\frac{M_T^{(k)}/M_T}{10}}\cdot  \sqrt{\frac{M_T}{\SI{e8}{\giga\electronvolt}}} \cdot \left(\frac{g_*\left(M_T / 25\right)}{100}\right)^\frac{1}{4}.
\end{equation}
We note that the bound \eqref{eq:washout-bound} can be satisfied for the range $\left(Y_{LT}\right)_{ik}\sim 0.5$ needed for the active neutrino masses in \eqref{eq:nu-mass} if one considers $\left(Y_{TD}\right)_{k1}\lesssim 10^{-2}$ for $k\neq 1$ . Now this is devastating for the asymmetry as \eqref{eq:asym-uplimit} gets reduced by three orders of magnitude and an efficiency of at least $\kappa\simeq 2\times10^{-2}$ would bee needed, which might be out of reach.
However so far we have assumed that the doublet $D$ is relativistic at the temperature $M_T / 20$, which does not have to be true. Since the processes $L \overline{D} \leftrightarrow H H$ and $L H \leftrightarrow D H $ require an on-shell $D$ they are Boltzmann-suppressed at $T<M_D$ and therefore decouple exponentially with $e^{-\frac{M_D}{T}}$. The $D$ are kept in equilibrium until $M_D / 20$ by  their EW gauge interactions. This leads to a  quantitatively different  behaviour when compared to the analogon for vanilla leptogenesis: Here the so called $\Delta L = 2$ processes $L L \leftrightarrow H H $ and  $L H \leftrightarrow \overline{L} H^\dagger $ (via an intermediate sterile neutrino $N$ ) can  destroy the asymmetry in $L$ down to temperatures of the electroweak scale. This occurs because their reaction rate densities only involve relativistic fermions as initial and final states so the rate is suppressed by a factor of $\left(T / M_N \right)^2$  \cite{Barbieri:1999ma}.
We rewrite the scattering rate using a non-relativistic number density
\begin{equation}
    \Gamma\left(L_i \overline{D} \leftrightarrow HH\right)\sim \Gamma\left(L_i H \leftrightarrow D H\right)\sim  \frac{\left(Y_{LT}Y^\dagger_{LT}\right)_{ii} \left(Y^\dagger_{TD}Y_{TD}\right)_{11}}{16\pi M_T^{(k)\;2}} \left(\frac{M_D T}{2\pi}\right)^{\frac{3}{2}} e^{-\frac{M_D}{T}}
\end{equation}
 and find that we can  relax  the constraint on the Yukawas  down to
\begin{equation}
    \left(Y_{LT}\right)_{i1} \left(Y_{TD}\right)_{11} \lesssim \mathcal{O}(0.1) \quad \text{for} \quad \delta = \frac{M_D^2}{M_T^2} \gtrsim 0.08,
\end{equation}
and similarly for $ \left(Y_{LT}\right)_{ik} \left(Y_{TD}\right)_{k1}$ both in agreement with our previous assumption $\delta\ll1$. 
Therefore we need a doublet that is less than one order of magnitude lighter than the lightest triplet and for concreteness we will set $\delta=0.1$.\\
\\
Since the doublets $D$ contain electrically neutral particles and eventually decouple from their gauge interactions, they are a good candidate for thermal WIMP dark matter. However we want to use the axion   to generate the observed DM relic abundance (see section \ref{sec:axionDM}) and additional (meta-)stable fermions could overclose the universe. This is why we have to demand that the decays of the $D$ to lighter particles occur rapidly enough. The most straightforward decay channel is to $\nu_R$ and  total decay rate  reads
\begin{equation}\label{eq:d-dec}
    \Gamma\left(D\rightarrow\nu_R H\right) \equiv   \sum_i  \Gamma\left( D^\alpha \rightarrow  \nu_{R\;i} H^\alpha\right) = \frac{M_D}{16 \pi}\left(Y_{DR} Y_{DR}^\dagger \right)_{1 1},
\end{equation}
where the rate is equal for both components of the doublets and we can also define a decay parameter
\begin{equation}\label{eq:dec-paramDR}
    K_{DR}\equiv \frac{  \Gamma\left(D\rightarrow\nu_R H\right)}{H(T)}\Big|_{T=M_D}
\end{equation}
to see that
\begin{equation}\label{eq:doub-dec}
    \left(Y_{DR}\right)_{1i} \simeq 7\times10^{-4}\cdot \sqrt{\frac{K_{DR}}{10}} \cdot    \left(\frac{g_*(M_D)}{100}\right)^\frac{1}{4} \cdot  \sqrt{\frac{M_D}{\SI{3e7}{\giga\electronvolt}}}.
\end{equation}
For the remainder of this work we set $K_{DR} =10$.

\subsection{Boltzmann equations}\label{sec:Boltz}

The definition of the  cross sections can be found in appendix \ref{sec:rates} and the parameterization for CP-violating rates densities can be found in appendix \ref{sec:ratesCP}.
Of central importance are the decays and inverse decays encoded in \cite{Kolb:1979qa}
\begin{align}
     \gamma_\text{tot.} &\equiv 3 \;\Sigma_T^{a\;\text{eq.}} \frac{\text{K}_1(z)}{\text{K}_2(z)} \Gamma_\text{tot.}= \frac{3 g_T M_T^3}{2 \pi^2 z} \text{K}_1(z) \Gamma_\text{tot.},
\end{align}
where $\Gamma_\text{tot.}$ was defined in \eqref{eq:gamma-tot} and $\text{K}_{1,2}(z)$ denotes the special Bessel functions of the first and second kind. Here $ \Sigma_T^{\text{eq.}}=3\; \Sigma_T^{a\;\text{eq.}}$ is the total number density of all three triplet components $a=1,2,3$.
We used the  rescaled temperature  $z\equiv M_T/T$. In this context we denote the thermally averaged density of the CP-conserving decay width $\Gamma\left(D\rightarrow H \nu_R\right)$ from \eqref{eq:d-dec} as $ \gamma_{DR}$ and the gauge scattering rate \eqref{eq:gauge-scat} as $\gamma_A$.
Special care needs to be taken with the inclusion of the washout scattering processes arising from the same Yukawas as the decays: 
The reactions $L\overline{D} \leftrightarrow H H$ occurring in the $t$-and $u$-channels can be computed by the usual methods  and will be denoted as $\gamma_{T_{t+u}}$.
As detailed in appendix \ref{sec:ratesCP} we need to remove the contribution of the on-shell triplet from the $s$-channel diagrams for the reactions $LH\leftrightarrow DH$ and their conjugates. This is necessary as the decays and inverse-decays of the intermediate on-shell triplets are already accounted for by the decay term $\propto \gamma_\text{tot.}$. Without this subtraction the Boltzmann equations would show unphysical behaviour such as generating asymmetries in equilibrium. One way to understand the appearance of this double counting problem is that the Boltzmann equations are essentially classical, whereas the cross sections and decay widths are computed using quantum physics. As outlined in \ref{sec:ratesCP} the CP-conserving rate density will be given by \cite{Hambye:2005tk}
\begin{equation}\label{eq:subdef}
    \gamma_{T_{s+t}}^\text{sub.} = \gamma_{T_{s+t}}- B_L B_D \gamma_\text{tot.}. 
\end{equation}
This structure can be understood by using a Breit-Wigner propagator for the unstable intermediate triplet and using the narrow width approximation for the propagator of $T^a$ on resonance \cite{Giudice:2003jh}
\begin{equation}\label{eq:submat}
     \left|D_s^{a}(s)\right|^{\text{sub.}\,\;2} \equiv  \left|\frac{1}{s-M_T^2+i M_T \Gamma_\text{tot.}}\right|^2 - \frac{\pi \delta(s-M_T^2)}{M_T \Gamma_\text{tot.}}.
\end{equation}
The coupling $Y_{LT}^2$ then reconstructs $B_L$. The coupling $Y_{TD}^2$ together with $1-\delta^2$ arising from expanding the numerator of the matrix element for $\Gamma_\text{tot.} / M_T \ll 1$ reconstructs $B_D$. For more details consult appendix \ref{sec:rates}. Furthermore we only include the lightest intermediate triplet in the rates $\gamma_{T_{s+t}}^\text{sub.}+\gamma_{T_{t+u}}$, as the effects of the heavier triplets are suppressed by $ M_T^2/M_T^{(2,3)\;2} \lesssim 0.11$
for $M_T^{(2,3)}=(3-10)\;M_T$.\\
\\
Our treatment for the asymmetry generation ignores all flavor and finite temperature effects. The impact of spectator processes responsible for washout, such as the ones from the SM lepton Yukawa couplings, is encoded in the sphaleron redistribution coefficient from appendix \ref{sec:sph-red-coeff}. We linearize the asymmetries in the small CP violating parameter $\varepsilon_L$ and find the following system of coupled non-linear Boltzmann equations:
\begin{eqnarray}\label{eq:boltz}
     z H s \frac{\text{d}\Sigma_T}{\text{d} z} &=& -2 \gamma_A \left(\frac{\Sigma_T^2}{\Sigma_T^{\text{eq.}\;2}}-1\right) - 2 \gamma_\text{tot.} \left(\frac{\Sigma_T}{\Sigma_T^\text{eq.}}-1\right),\\
    z H s \frac{\text{d}\Delta_T}{\text{d} z} &=& - 2 \gamma_\text{tot.} \left(\frac{\Delta_T}{\Sigma_T^\text{eq.}}- \text{B}_\text{L} \frac{\Delta_L}{\Sigma_L^\text{eq.}}-\text{B}_\text{D}\frac{\Delta_D}{\Sigma_D^\text{eq.}}     \right) \label{eq:deltaT}\nonumber,\\
    z H s \frac{\text{d}\Delta_L}{\text{d}z} &=& 2 \gamma_\text{tot.} \left[ \varepsilon_L \left(\frac{\Sigma_T}{\Sigma_T^\text{eq.}}-1\right) +\text{B}_\text{L} \left(\frac{\Delta_T}{\Sigma_T^\text{eq.}}-\frac{\Delta_L}{\Sigma_L^\text{eq.}}\right) \right] - 2 \left(\gamma_{T_{s+t}}^\text{sub.}+\gamma_{T_{t+u}}\right)\left(\frac{\Delta_L}{\Sigma_L^\text{eq.}}-\frac{\Delta_D}{\Sigma_D^\text{eq.}}\right),\label{eq:deltaL}\nonumber\\
    z H s \frac{\text{d}\Delta_D}{\text{d}z} &=& 2 \gamma_\text{tot.} \left[ -\varepsilon_L \left(\frac{\Sigma_T}{\Sigma_T^\text{eq.}}-1\right) +\text{B}_\text{D} \left(\frac{\Delta_T}{\Sigma_T^\text{eq.}}-\frac{\Delta_D}{\Sigma_D^\text{eq.}}\right) \right] + 2 \left(\gamma_{T_{s+t}}^\text{sub.}+\gamma_{T_{t+u}}\right)\left(\frac{\Delta_L}{\Sigma_L^\text{eq.}}-\frac{\Delta_D}{\Sigma_D^\text{eq.}}\right) \label{eq:deltaD}\nonumber\\
    &-& 2 \gamma_{DR}\left(\frac{\Delta_D}{\Sigma_D^\text{eq.}}-\frac{\Delta_{\nu_R}}{\Sigma_{\nu_R}^\text{eq.}}\right)\nonumber,\\
     z H s \frac{\text{d}\Delta_{\nu_R}}{\text{d}z} &=& 2 \gamma_{DR}\left(\frac{\Delta_D}{\Sigma_D^\text{eq.}}-\frac{\Delta_{\nu_R}}{\Sigma_{\nu_R}^\text{eq.}}\right)\label{eq:deltaNuR}\nonumber.
\end{eqnarray}

\noindent 
 A couple of more comments are in order:
First of all it is worth pointing out that in case of thermal equilibrium $\Sigma_T= \Sigma_T^\text{eq.}$ and vanishing initial asymmetries $\Delta_T=\Delta_L=\Delta_D=0$ no lepton asymmetry is produced in either channel. This is in complete accordance with Sakharov's criteria \cite{Sakharov_1991} and provides a useful consistency check.
Since the asymmetry in $D$ is already $\mathcal{O}\left(\varepsilon_L\right)$ it can be transmitted to the $\nu_R$ bath via CP conserving decay $\gamma_{DR}$. We do not include a Boltzmann equation for $\Sigma_{L},\Sigma_{D},\Sigma_{\nu_R}$ as they follow their equilibrium distributions for the relevant temperatures (see section \ref{sec.nuRtherm} for the interactions of $\nu_R$).  In principle one should also include all   scattering processes involving  one gauge vertex and one of the Yukawa interactions like e.g. $T\; W \rightarrow  L H,\; D^\dagger H$ or scattering of $T$ with the charged SM fermions. These will be important at temperatures $T\gtrsim M_T$ because they require on shell triplets and contribute to the thermalization of $T$. However since we already include gauge annihilations for the thermalization and further estimated that leptogenesis occurs far later at $T\lesssim M_T / 20$  their addition is negligible.\\
\\
For particles which are not self conjugate the Boltzmann equations depends on $\Sigma$ and $\Delta$ so new effects are possible when compared to the equations for e.g. decaying heavy  Majorana neutrinos \cite{Hambye:2005tk}. This is why there is an equation for the asymmetry in $T$ in \eqref{eq:boltz}, which only depends on the other asymmetries.
If we add the equations for all four asymmetries we find  
\begin{equation}
    z H s \frac{\text{d}}{\text{d} z} \left(\Delta_T + \Delta_L +\Delta_D+ \Delta_{\nu_R}\right)= 0.
\end{equation}
This sum rule implies
\begin{equation}
    \Delta_T(z) + \Delta_L(z) +\Delta_D(z)+ \Delta_{\nu_R}(z)  = \text{const.},
\end{equation}
where the constant is independent of $z$. For our case the appropriate initial conditions are vanishing initial asymmetries for each species so that we deduce that the constant term is actually zero. The sum rule can be represented in terms of chemical potentials 
\begin{equation}\label{eq:lept-cons}
    \mu_T + \mu_L +\mu_D +\mu_{\nu_R} = 0,
\end{equation}
which can be interpreted as the conservation of the total lepton number in both the SM and PQ sectors. We find that the structure of our Boltzmann equations agrees with the ones presented in \cite{Cerdeno:2006ha}, which also respect lepton number conservation.

\subsection{Numerical results}\label{sec:num}
\begin{figure}[t]
\centering
    \includegraphics[width=0.49\textwidth]{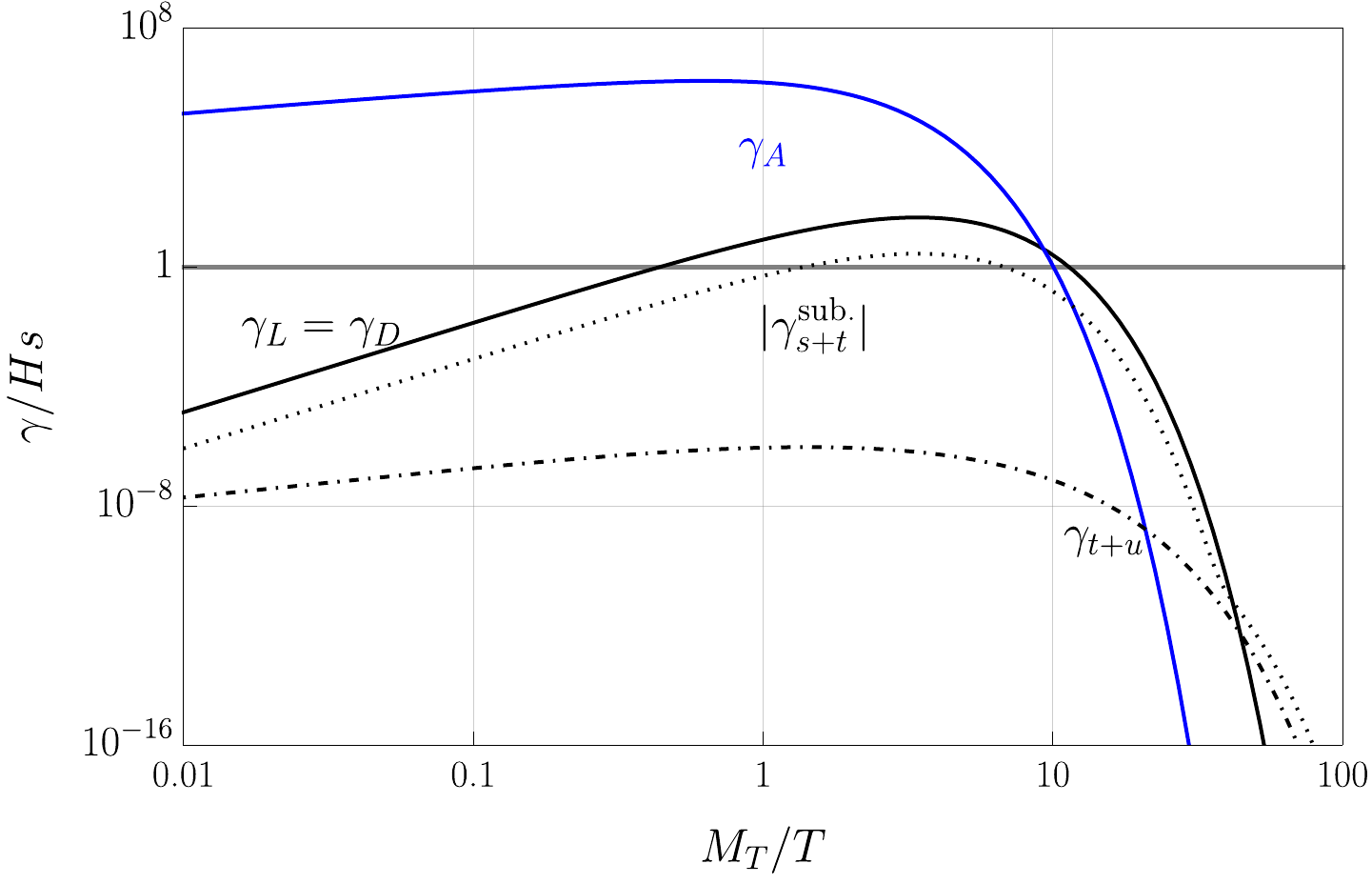}
    \includegraphics[width=0.49\textwidth]{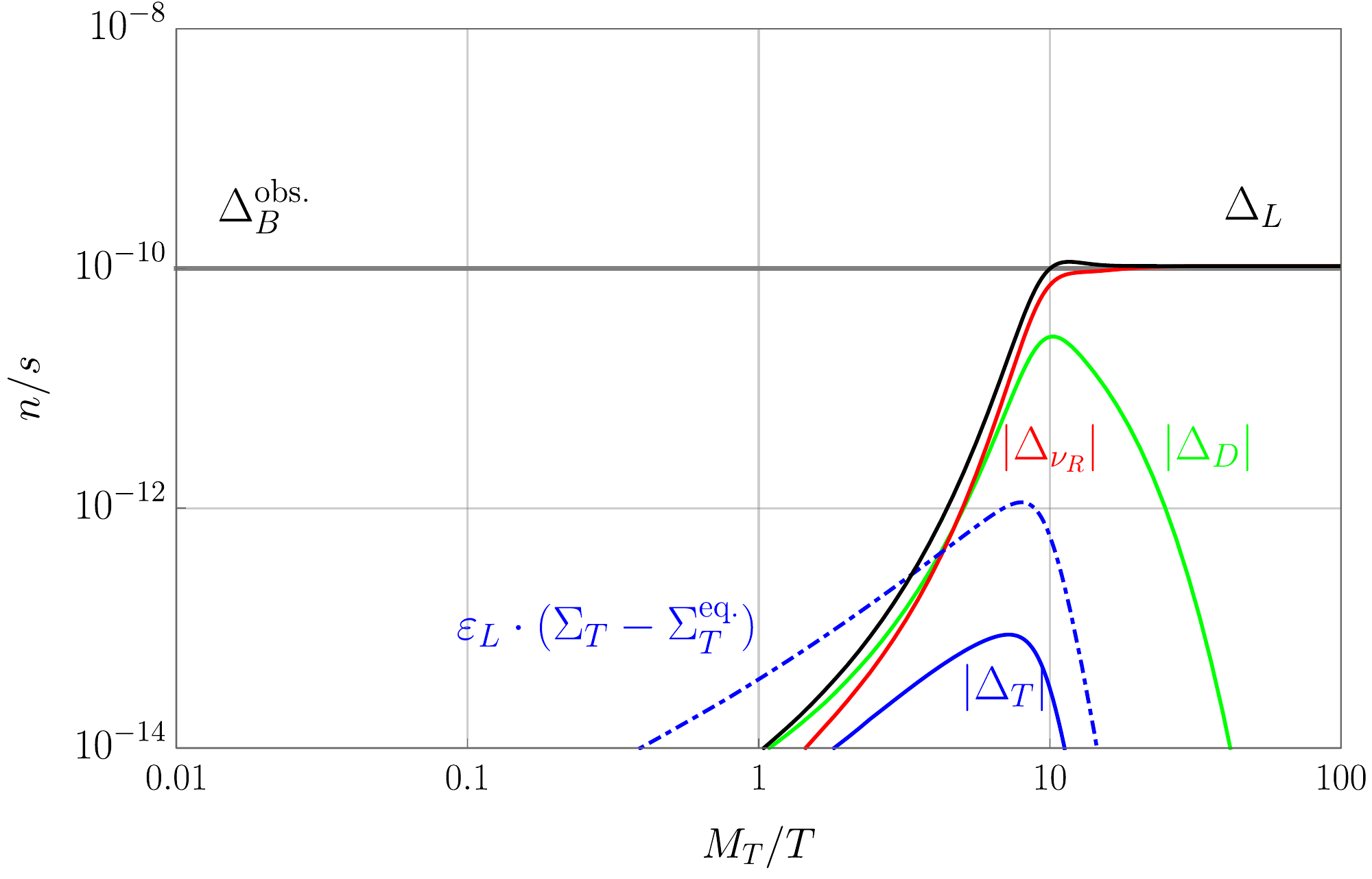}
    \caption{Rate densities \textit{(left)} for $K= 100,\; B_L = B_D = 1/2\; \left(Y_{LT} =  Y_{TD} = 9.1\times 10^{-4}\right),\;\varepsilon_L = 2.7\times 10^{-5}$ and leptonic yields \textit{(right)}. }
	\label{fig:Intermed}
\end{figure}

Due to the larger number of parameters compared to standard leptogenesis we refrain from scanning over the parameter space and rather present select benchmark points illustrating the phenomenology. To do so we integrate the Boltzmann equations between $z\equiv M_T / T \in [1/100,100]$ and use the initial conditions that all asymmetries vanish and that $\Sigma_T$ follows its equilibrium value due to its unavoidable gauge interactions for $T<10^{12}\;\text{GeV}$.
For concreteness we set
\begin{equation}
    M_T = 10^8\;\text{GeV} \quad \text{and} \quad M_D = 3\times 10^7 \;\text{GeV},
\end{equation}
which corresponds to $\delta=0.1\ll1$. We will investigate the three regimes outlined in \ref{sec:eff} and use $K=0.01, 100$ and $10^5$ as benchmark values for the decay parameter. For the most part we will consider $B_L = B_D = 1/2$ but in the strong washout regime $(K=10^5)$ we find interesting new effects for $B_L\neq B_D$. In case of $B_L=B_D$ we can vary the asymmetry parameter $\varepsilon_L$ independently of the couplings for the tree-level decay, which are encoded in $B_{L,D}$. For the scenario $B_L \neq B_D$ we will use the parameterization $\varepsilon_L = \varepsilon_L^\text{max.} \sqrt{4 B_L B_D (1-\delta^2)}$ from \eqref{eq:BRS}.
Additionally we take the decays of $D$ to $\nu_R$ to be fast and fix $K_{DR}=10$ which corresponds to $Y_{DR} = 7\times 10^{-4}$. We use the analytical estimate in \eqref{eq:gauge-scat} from the appendix for the gauge interactions of the triplet, which overestimates the correct rate for $z\lesssim 0.1$ by a factor of two but works excellently for the non-relativistic regime where the freeze-out occurs.
The left plots in figures \ref{fig:Gauge}-\ref{fig:DecayRev} show the evolution of the decay rate densities $\gamma_{L,D} \equiv B_{L,D} \gamma_\text{tot.}$ as well as the gauge scatterings $\gamma_A$ and the washout scattering terms  $\gamma_{T_{s+t}}^\text{sub.},\;\gamma_{T_{t+u}}$
divided by $H s$, because that is the combination of parameters appearing in the Boltzmann equations \eqref{eq:boltz}. A value of $\gamma / H s <1$ is slow on cosmological time scales and the corresponding process will be inefficient.\footnote{Note that $\gamma / H s$ has the same units as $\Gamma / H$, where $\Gamma$ is the decay width or scattering rate  $\Gamma \equiv \gamma / n_\psi^\text{eq.}$ for a particle $\psi$ with equilibrium number density $n_\psi^\text{eq.}$ in the initial state. Since for non-relativistic $\psi$ the density  $n_\psi^\text{eq.}$ will be Boltzmann suppressed instead of scaling like radiation  $n_\psi^\text{eq.}\sim n_\gamma\sim s$, the freeze-out temperature found from $\Gamma(T_\text{FO}) / H(T_\text{FO}) <1$  in section \ref{sec:eff} will in general be different than  the temperature when $\gamma / H s<1$.}
For all numerical benchmark points we find that the washout scatterings from $\gamma_{T_{t+u}}$ and $\gamma_{T_{s+t}}^\text{sub.}$ are always slow and orders of magnitude smaller than the smallest decay rate density (see e.g. figures \ref{fig:DecayOpt} and \ref{fig:DecayRev}) for $z\lesssim50$. Hence  we use $B_L B_R \gamma_\text{tot.}$ as a conservative estimate for the washout terms $\gamma_{T_{s+t}}^\text{sub.}+\gamma_{T_{t+u}}$, which is similar  to reference  \cite{Cerdeno:2006ha}, who employ $ \gamma_\text{tot.}$ instead. We chose $B_L B_R \gamma_\text{tot.}$ as the washout scatterings are bounded from above by $\text{Min}(B_L,B_D) \gamma_\text{tot.}$. Since the resulting asymmetries will all be $\sim \varepsilon_L$ we rescale $\Sigma_T(z)-\Sigma_T^\text{eq.}(z)$ by a factor of $\varepsilon_L$ to plot them in the same figure with the asymmetries. Additionally we rescale all leptonic asymmetries $\Delta_L,\Delta_{\nu_R}, \Delta_D, \Delta_T$ by the sphaleron redistribution coefficient from \eqref{eq:sphal2} to ease the visual comparison to the observed baryon asymmetry $\Delta_B^\text{obs} = \eta_B / 7.04\simeq 10^{-10}$ from \eqref{eq:etaB}.
 
\subsubsection{Weak washout regime}
\begin{figure}[t]
\centering
    \includegraphics[width=0.49\textwidth]{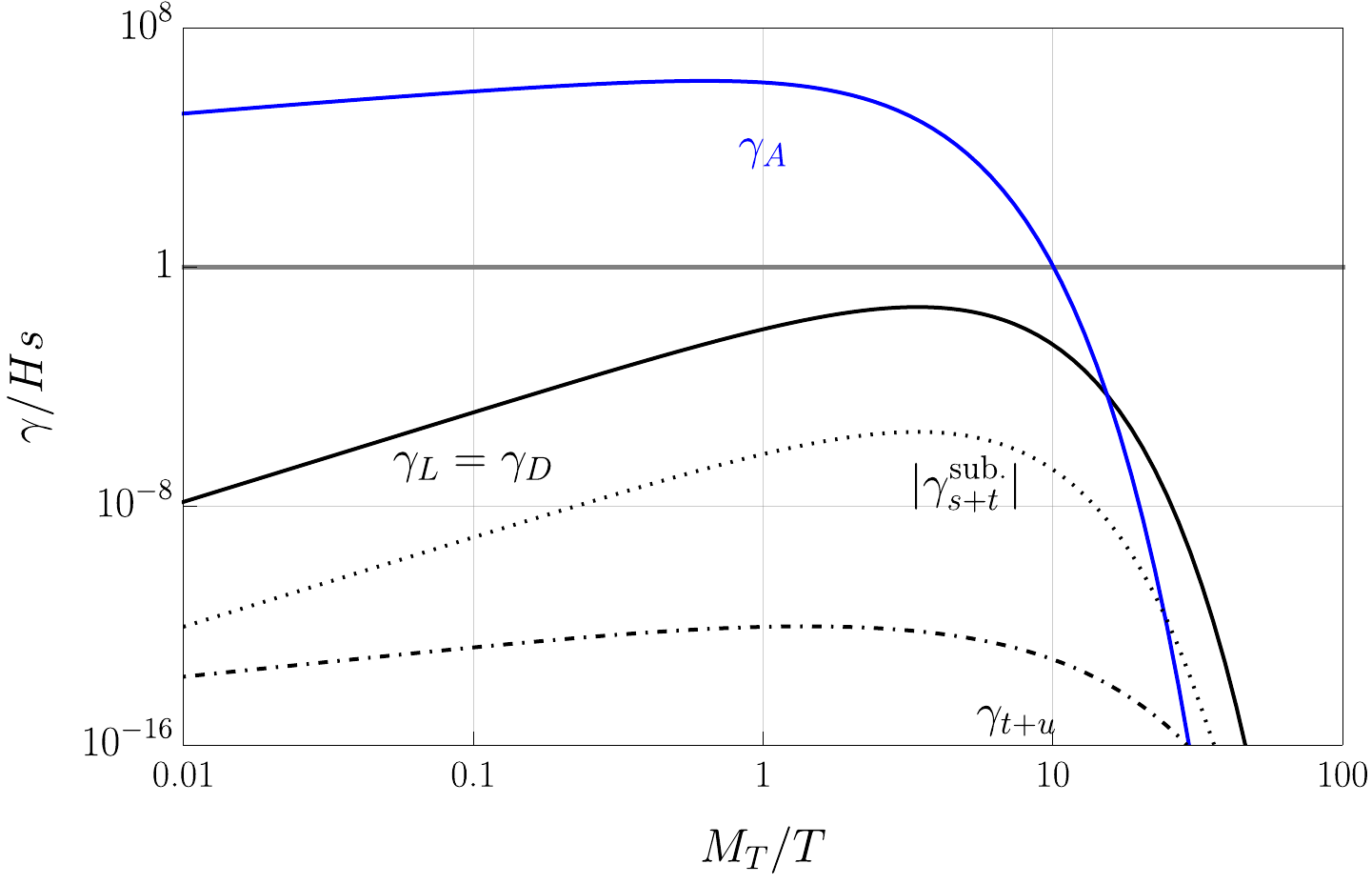}
    \includegraphics[width=0.49\textwidth]{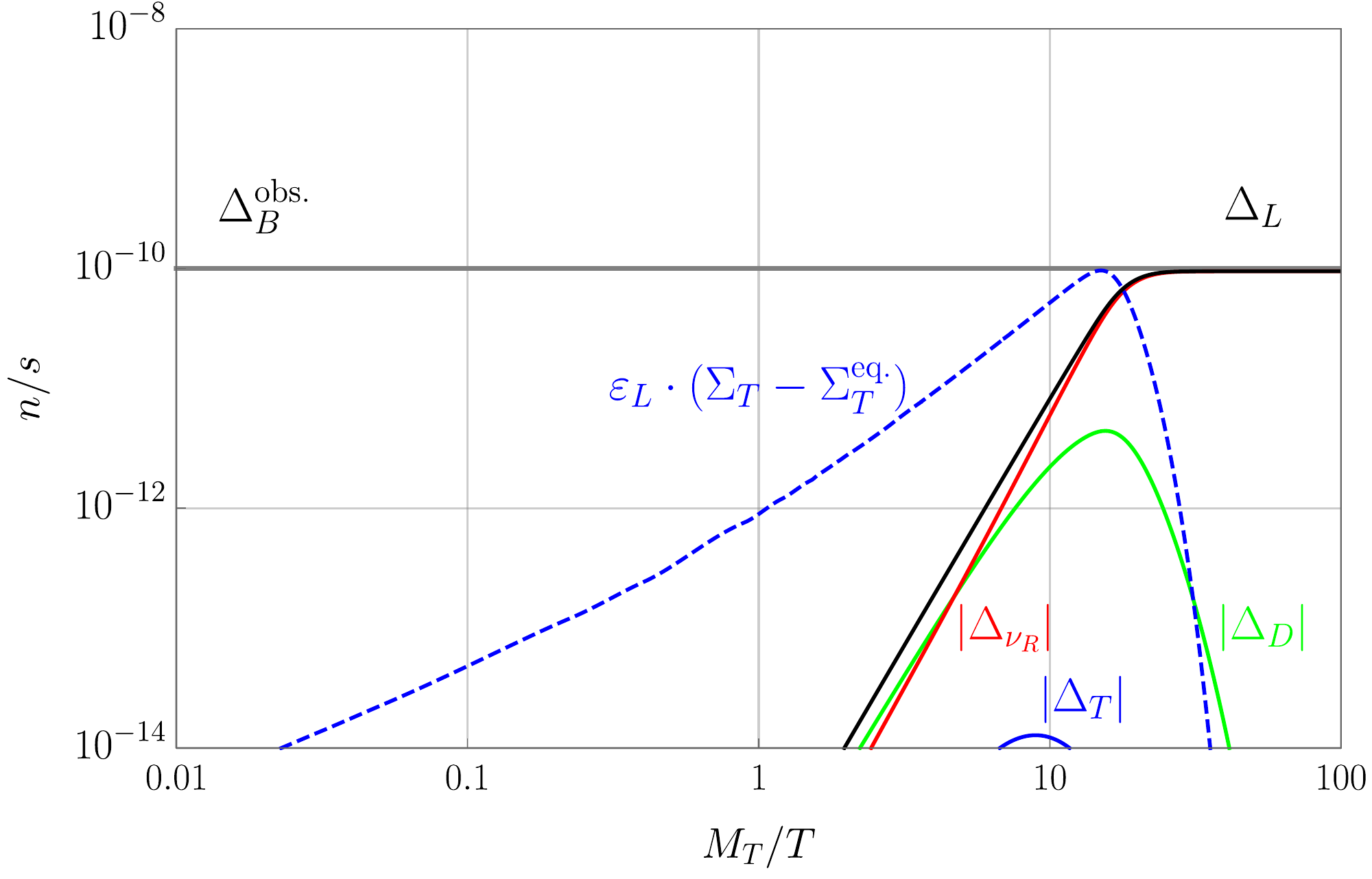}
	\caption{Rate densities \textit{(left)} for $K= 0.01,\; B_L = B_D = 1/2\; \left(Y_{LT} =  Y_{TD} = 9.1\times 10^{-6}\right),\;\varepsilon_L = 6.5\times 10^{-4}$ and leptonic yields \textit{(right)}. }
	\label{fig:Gauge}
\end{figure}
We fix $K= 0.01\;\text{and} \;B_L = B_D = 1/2$ corresponding to $Y_{LT} =  Y_{TD} = 9.1\times 10^{-6}$.
In the weak washout-regime the gauge annihilations are the last reaction to decouple form thermal equilibrium, as can be seen in the left plot of \ref{fig:Gauge}. Since decays and inverse decays are slow, washout will also be slow and the leptonic asymmetry can freeze-in \cite{Hall:2009bx} undisturbed. The right figure in \ref{fig:Gauge} illustrates the evolution of the asymmetries: First around $z<1$ equal and opposite amounts of $\Delta_L$ and $\Delta_D$ are generated. The  asymmetries track the deviation of the triplet abundance from equilibrium $\Sigma_T-\Sigma_T^\text{eq.}$. Then the  asymmetry in $D$ is   transmitted into $\Delta_{\nu_R}$ via   decays, which is why the corresponding red line starts later at $z\sim5$. Since the triplet is a Dirac fermion it can develop an asymmetry itself via inverse decays of  $\Delta_L,\Delta_D$. Owing to the fact that we take these decays to be slow the resulting $\Delta_T$ is small compared to the other asymmetries and the deviation from equilibrium.
As the gauge interactions decouple around $z\sim10$, the triplet abundance does not get restored from here on out and since the out-of-equilibrium decays become faster than the out-of-equilibrium gauge annihilations for $z \sim 20$, the frozen out abundance decays away, explaining the sharp decrease in the triplet abundance $\Sigma_T$ at $z\gtrsim 20$. The asymmetry production asymptotes to its final value around this time as there are no more triplets left to decay.  After all the triplets and doublets have decayed away only $L$ and $\nu_R$ remain. As expected from lepton number conservation we find that the asymptotic values satisfy $\Delta_L = -\Delta_{\nu_R}$.
We stop the evolution at $z=100$ corresponding to $T=M_T/100$ long before the sphaleron decoupling because the leptonic asymmetries are conserved after the $T,D$ have decayed.
We can reproduce the observed baryon asymmetry today $\Delta_B^\text{obs} = \eta_B / 7.04\simeq 10^{-10}$ from \eqref{eq:etaB} for an asymmetry parameter of $\varepsilon_L = 6.5\times10^{-4}$ corresponding to an efficiency of $\kappa\simeq 2\times 10^{-5}$ from \eqref{eq:eff-BM} in line with our estimate $\kappa \simeq 10^{-4-5}$.

\subsubsection{Intermediate Regime}
In this regime the decays of the triplet are not negligible during the gauge annihilation phase.
Here we fix $K= 100\; \text{and} \; B_L = B_D = 1/2$ corresponding to $Y_{LT} =  Y_{TD} = 9.1\times 10^{-4}$.
The evolution of $\Delta_L$ is more complicated than in the weak washout-regime owing to the washout from faster inverse decays, which decouple around the same time as the gauge interactions. This is also why a larger $\Delta_T$ more comparable to $\varepsilon_L\cdot\left(\Sigma_T(z)-\Sigma_T^\text{eq.}(z)\right)$ is generated when compared to the previous benchmark. We fit the observed baryon asymmetry for $\varepsilon_L = 2.7\times 10^{-5}$ analogous to $\kappa\simeq 4\times10^{-4}$. This efficiency is indeed larger by an order of magnitude than in the weak washout regime, but not quite as large as our analytical estimate from section \ref{sec:eff}.

\subsubsection{Strong washout regime}
\begin{figure}[t]
\centering
    \includegraphics[width=0.49\textwidth]{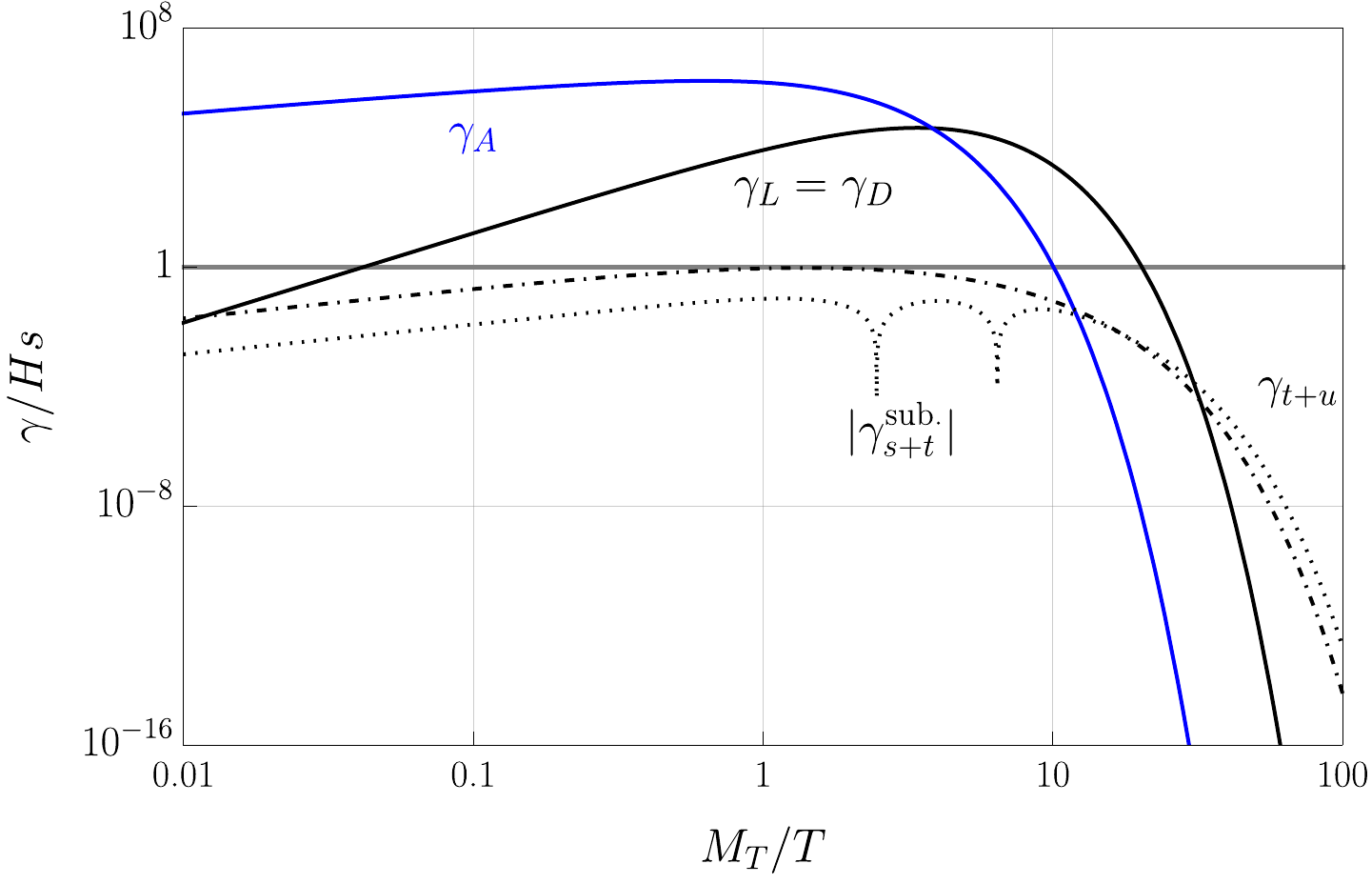}
    \includegraphics[width=0.49\textwidth]{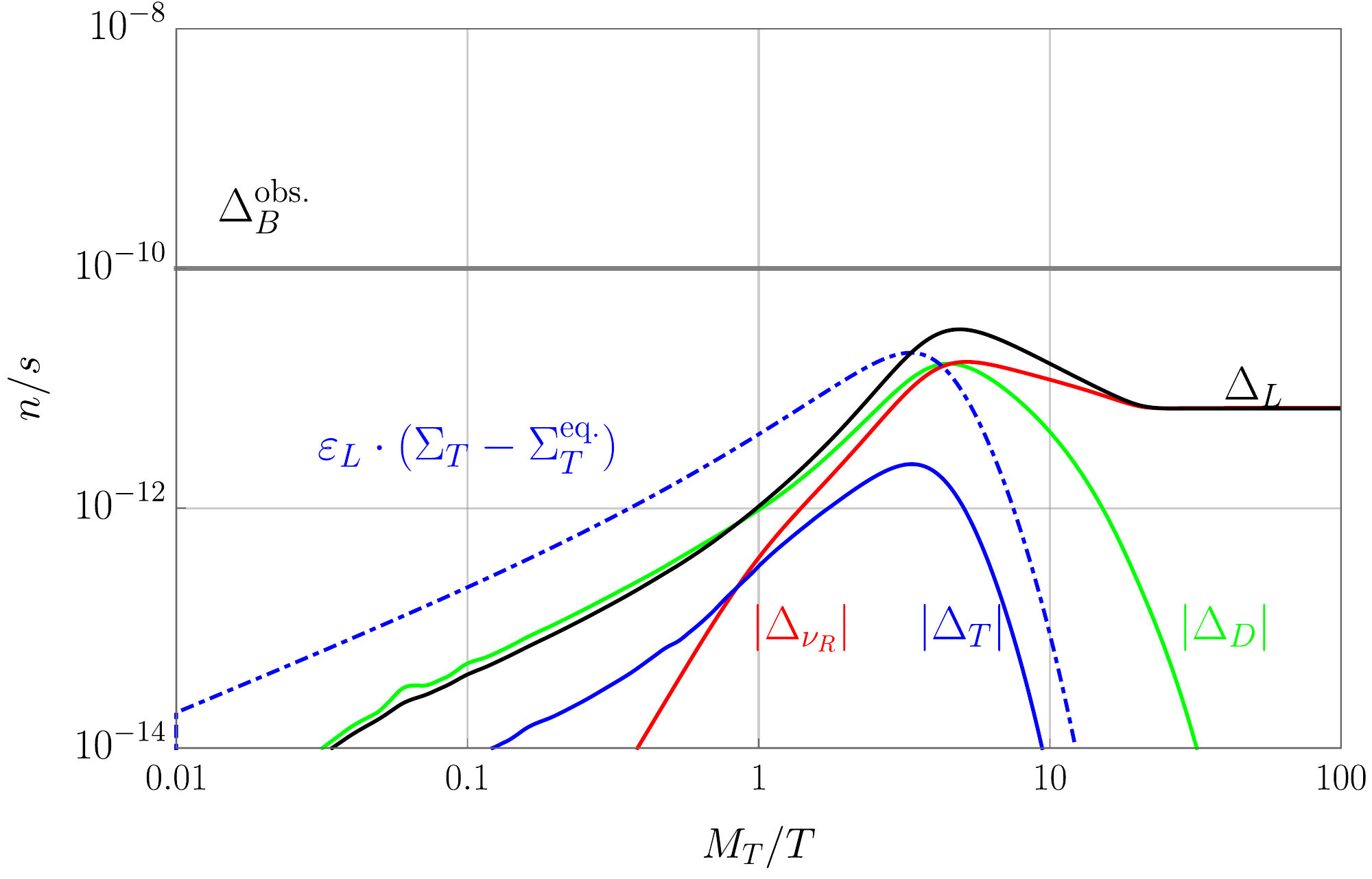}
	\caption{Rate densities \textit{(left)} for $K= 10^5,\; B_L = B_D = 1/2\; \left(Y_{LT} =  Y_{TD} = 2.9\times 10^{-2}\right),\;\varepsilon_L = 3\times 10^{-3}$ and leptonic yields \textit{(right)}. }
	\label{fig:Decay}
    \includegraphics[width=0.49\textwidth]{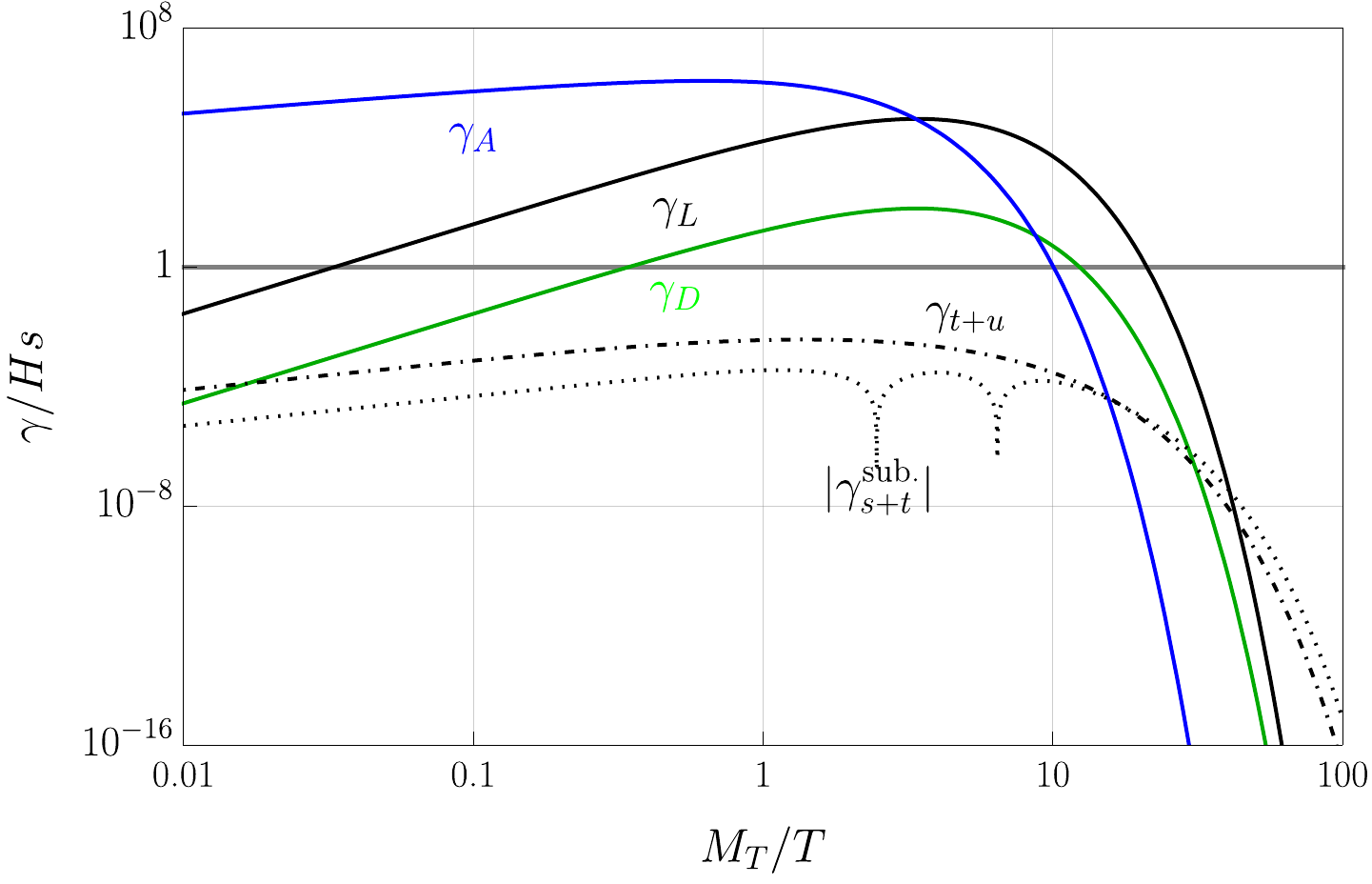}
    \includegraphics[width=0.49\textwidth]{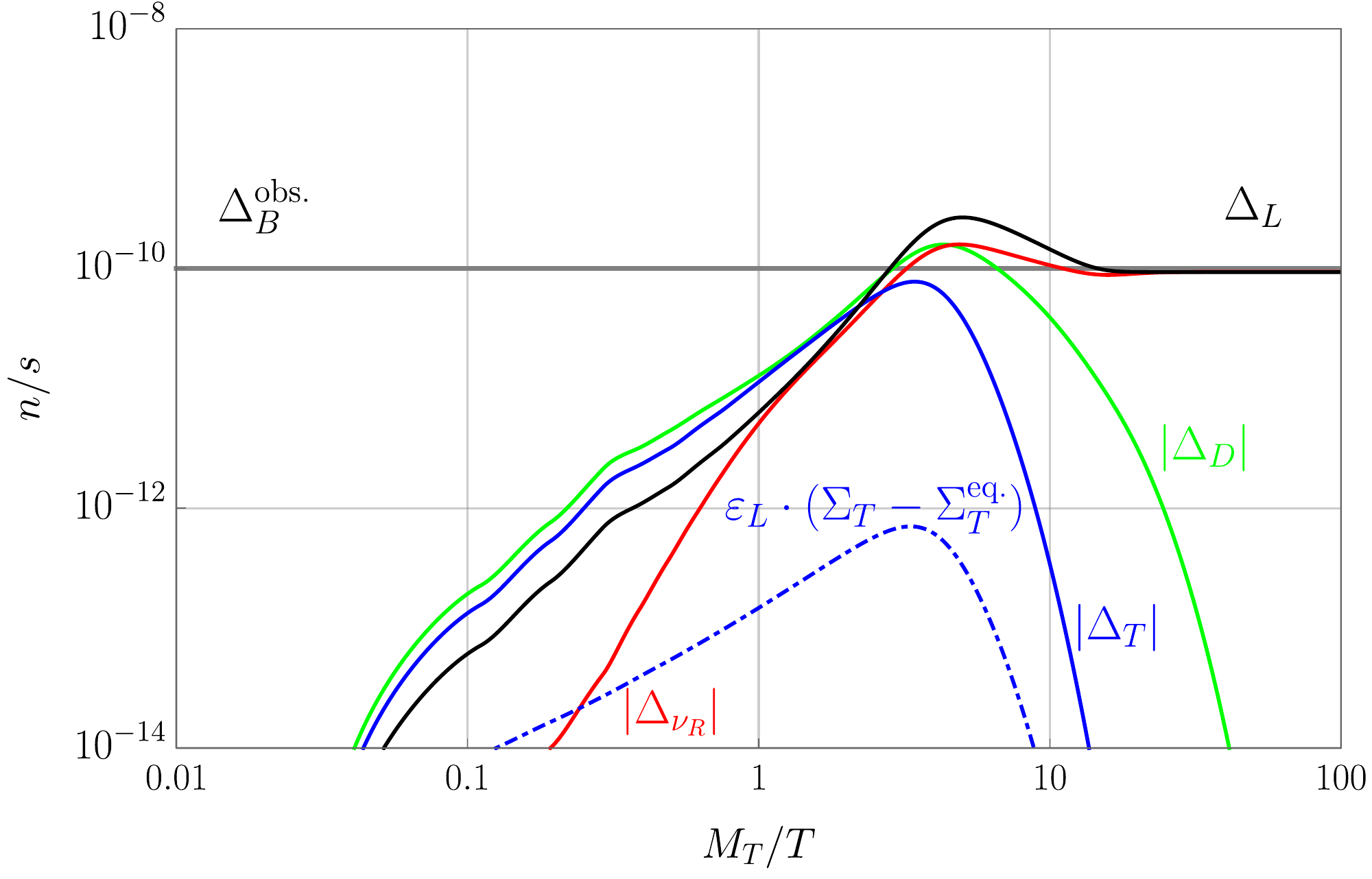}
	\caption{Rate densities \textit{(left)} for $K= 10^5,\; B_L =0.999,\; B_D = 10^{-3}\; \left(Y_{LT} =4.1\times 10^{-2},\; Y_{TD} = 1.2\times 10^{-3}\right),\;\varepsilon_L = 1.7\times 10^{-3}\sqrt{4 B_L B_D (1-\delta^2)}=4.7\times10^{-5}$ and the leptonic yields  \textit{(right)}. }
	\label{fig:DecayOpt}
    \includegraphics[width=0.49\textwidth]{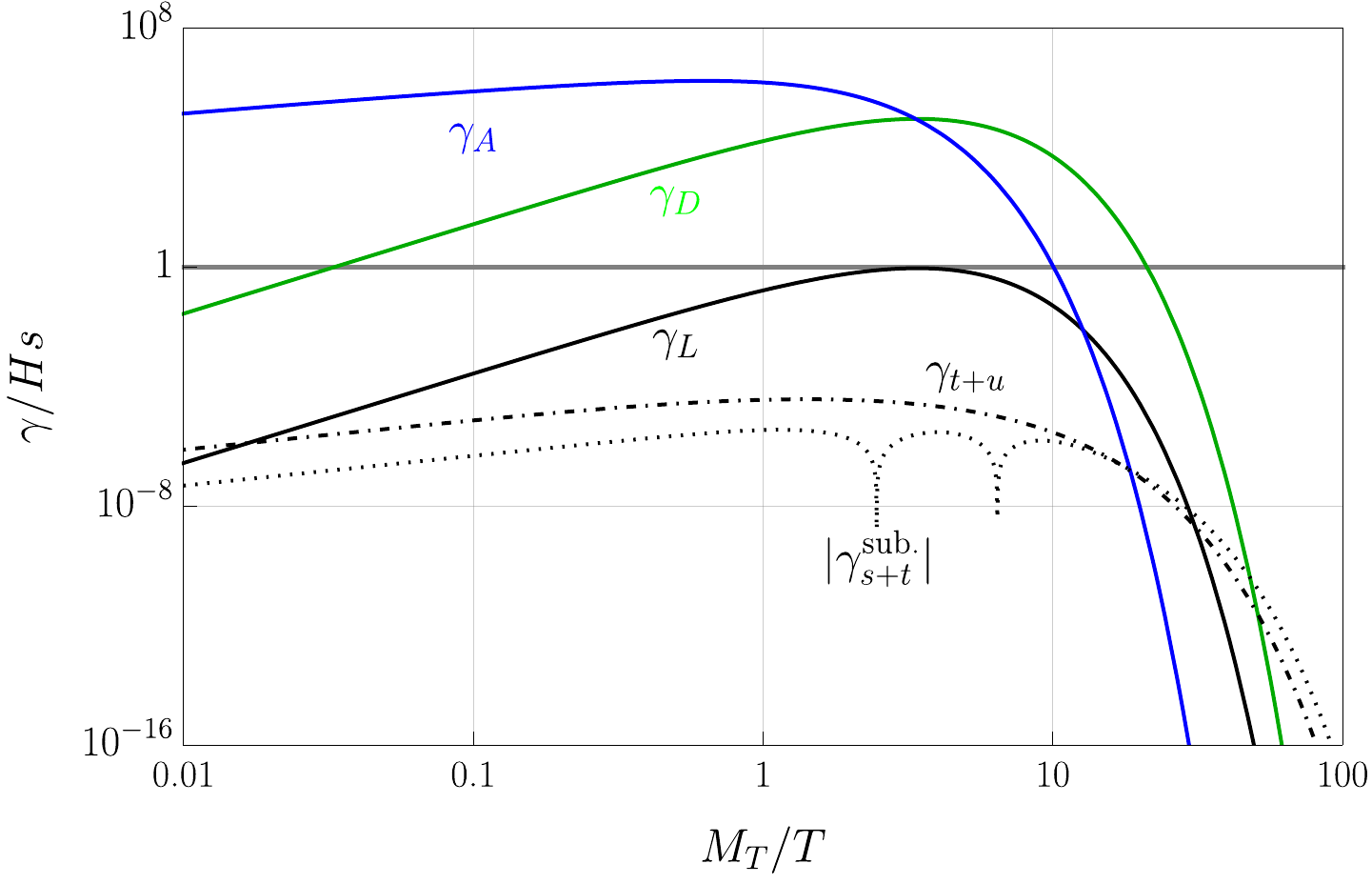}
    \includegraphics[width=0.49\textwidth]{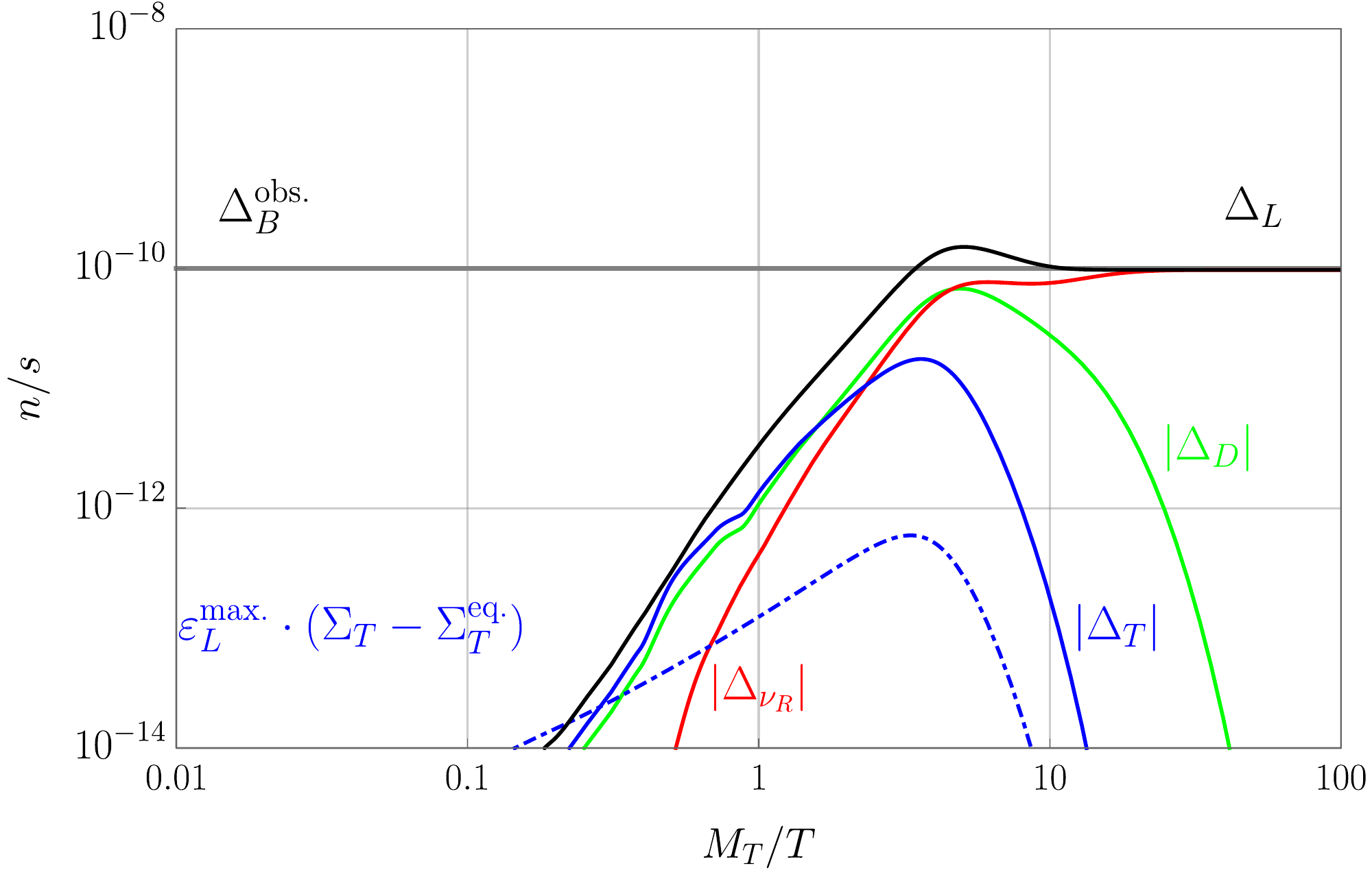}
	\caption{Rate densities \textit{(left)} for $K= 10^5,\; B_L =10^{-5},\; B_D = 0.99999\; \left(Y_{LT} =1.3\times 10^{-4},\; Y_{TD} = 4.1\times 10^{-2}\right),\;\varepsilon_L = 9\times 10^{-5}\sqrt{4 B_L B_D (1-\delta^2)}=2.2\times10^{-7}$ and the leptonic yields \textit{(right)}. }
	\label{fig:DecayRev}
\end{figure}
\noindent For the strong washout regime we fix $K=10^5$.
In the first scenario we retain $B_L=B_D= 1/2$ which can be realized for $Y_{LT} =  Y_{TD} = 2.9\times 10^{-2}$. The left plot in \ref{fig:Decay} shows that indeed the decays are the last interaction to decouple now. Decays and inverse decays are faster than both the annihilations and the Hubble rate only for  $z\sim 6-30$. The right plot in the same figure illustrates the evolution of the asymmetries for the maximum possible value of $\varepsilon_L = 3\times10^{-3}$ from \eqref{eq:asym-uplimit}.
All asymmetries and $\Sigma_T-\Sigma_T^\text{eq.}$ reach their maxima at $z\sim 6$, when the (inverse) decays overtake the gauge annihilations, and  decrease afterwards. Once $\gamma_{L}$ decouples at $z\sim 30$ the asymmetry in $\Delta_L$ approaches a constant instead of continuing to decrease. This occurs because inverse decays depleting $\Delta_L$ are slow now (and eventually there are no more triplets left to decay producing $\Delta_L$), analogous to the well known freeze-out scenario for thermal dark matter production. This is in agreement with Sakharov's conditions \cite{Sakharov_1991}, since the lepton asymmetry would continue to  decrease to zero if the inverse decays depleting them remained in equilibrium forever after $z  \sim 6$. Here even for the maximum of $\varepsilon_L$ we can not reproduce the observed baryon asymmetry.
This happens because of too much washout from inverse decays and  as a consequence we have a small efficacy $\kappa\sim 1/K \sim 10^{-5}$. However this is not the end for the strong washout regime. As explained in the previous subsections the larger amount of washout will produce a larger $\Delta_T$ and we will make use of this fact to obtain the required efficiency. This behaviour was first observed for decaying scalar triplets in the context of Type II Seesaw leptogenesis \cite{Hambye:2005tk}. The authors of \cite{Hambye:2005tk} found  that the lepton asymmetry produced by the decay of a non-self-conjugate particle with two decays modes is washed out only if both decay modes are faster than the Hubble rate. Our previous choice $B_L=B_D$ actually leads to the most amount of washout \cite{Hambye:2005tk}. We will demonstrate this for two concrete examples:\\
\\
The first benchmark for the same decay parameter has $B_L =0.999 \gg  B_D = 10^{-3}$ implying $Y_{LT} =4.1\times 10^{-2},\; Y_{TD} = 1.2\times 10^{-3}$. The behavior depicted in the right plot of figure \ref{fig:DecayOpt} can be understood as follows: First equal and opposite asymmetries in $L,D$ are produced. However since $B_L\gg B_D$ we find that $\Delta_L$ is washed out by the fast inverse decays $L H \rightarrow T$, whereas a large $\Delta_D$ develops undisturbed. Of course the asymmetry in $D$ is transmitted to $\nu_R$ via the fast decays. The sum rule for lepton number conservation in \eqref{eq:lept-cons} then enforces that 
\begin{equation}
    \Delta_D + \Delta_{\nu_R} = - \Delta_L - \Delta_T,
\end{equation}
which means that the asymmetry in the $D,\nu_R$ subsystem is compensated by an equally large asymmetry of opposite sign in the $L,T$ subsystem. When $\Delta_T$ eventually decays it gets predominantly transferred to $\Delta_L$ again because of $B_L\gg B_D$. The baryon asymmetry is successfully generated for $\varepsilon_L = 1.7\times 10^{-3}\sqrt{4 B_L B_D (1-\delta^2)}=4.7\times10^{-5}$, where we used a value close to the maximum possible asymmetry of $3\times10^{-3}$. The difference to the $B_L=B_D$ case can also be understood if one notices that the triplet asymmetry in \ref{fig:Decay} is smaller than their deviation from equilibrium $\varepsilon_L\cdot\left(\Sigma_T(z)-\Sigma_T^\text{eq.}(z)\right)$. For $B_L\gg B_D$ on the other hand we can see from \ref{fig:DecayOpt} that $\Delta_T$ actually becomes much larger so that $\Delta_L$ starts to track its behaviour.\\
\\
The second benchmark has $B_L =10^{-5}\ll B_D = 0.99999$ implying $Y_{LT} =1.3\times 10^{-4},\; Y_{TD} = 4.1\times 10^{-2}$ and was depicted in figure \ref{fig:DecayRev}: For $B_D\gg B_L$ a large initial asymmetry in $L$ can freeze-in and is not washed out. The triplet asymmetry is predominantly produced by inverse decays $D H \rightarrow T$ now. For equal $K$ we find that $\Delta_T$ is about an order of magnitude smaller for $B_D \gg B_L$ compared to $B_L \gg B_D$, because the inverse decay of $D$ to $T$ has to compete with its fast decay to $\nu_R$. Lepton number implies that 
\begin{equation}
    \Delta_L  = -  \Delta_D - \Delta_{\nu_R}  - \Delta_T.
\end{equation}
When the triplets decay away $\Delta_T$ decays primarily to $\Delta_D$ because of $B_D\gg B_L$, and $\Delta_D$ decays to $\Delta_{\nu_R}$ so again   $\Delta_{\nu_R}=-\Delta_L$ is produced. Note that here we had to rescale $\Sigma_T(z)-\Sigma_T^\text{eq.}(z)$ by the larger $\varepsilon_L^\text{max}$ and not $\varepsilon_L$ to fit it in the same plot with the other yields, since otherwise it would have been smaller than $10^{-14}$. This again illustrates that $\Delta_T$ becomes the driving force for this mode instead. This benchmark fits the baryon asymmetry for $\varepsilon_L = 9\times 10^{-5}\sqrt{4 B_L B_D (1-\delta^2)}=2.2\times10^{-7}$. From this we see that the required $\varepsilon_L /\sqrt{4 B_L B_D (1-\delta^2)} $ can be made smaller the more hierarchical the branching ratios  $B_L/B_D$ are.\\
\\
We conclude by noting that Dirac-Leptogenesis with a decaying fermion naturally realizes all the ingredients needed for the previously mentioned  \enquote{quasi optimal efficiency}-scenario  \cite{Hambye:2005tk}:
Since the decaying fermion is of Dirac nature it can have an asymmetry itself and because of CPT and unitarity (see \eqref{eq:CPT}) it needs to have two separate decay modes to generate the leptonic asymmetry parameter. The efficiency increases if their branching ratios are different.

\subsection{Lightest neutrino mass}\label{sec:lightest}
\noindent
For the previously mentioned benchmarks $M_T=10^8\;\text{GeV}\; \text{and}\; M_D = 3\times 10^{7}\;\text{GeV}$ we can use the Yukawa couplings $Y_{LT}, Y_{TD}$ required for the different washout scenarios in subsection \ref{sec:eff} to estimate the lightest neutrino mass. The third Yukawa was fixed to $Y_{DR}\simeq 7\times10^{-4}$ to allow for fast decays of the vector-like doublets see \eqref{eq:doub-dec}. Assuming no accidental flavor cancellations we find
\begin{align}
    m_{\nu_l} &\simeq 5\times 10^{-3}\;\text{eV}\;\cdot Y_{LT} Y_{TD}\\ 
    &= \begin{cases} 6\times 10^{-12}\;\text{eV}\quad \text{for weak washout} \quad Y_{LT}\simeq Y_{TD} \simeq 3.5\times 10^{-5}, \\ 5\times 10^{-7}\;\text{eV}\quad \text{for strong washout} \quad Y_{LT}\simeq Y_{TD} \simeq 1.3\times 10^{-2},\end{cases}
\end{align}
where the couplings refer only to the lightest doublet and triplet and we assumed equal branching ratios. The lightest neutrino mass eigenstate is substantially lighter than the cosmological limit on the total neutrino mass of $\sum_\nu m_\nu \lesssim 0.12\;\text{eV}$ \cite{Planck:2018vyg} and can be treated as massless for all intents and purposes. This outcome is generic in leptogenesis scenarios \cite{Hambye:2006zn} due to the small couplings required for out of equilibrium decay.

\section{Dark radiation}\label{sec:darkrad}
In the SM  the number of relativistic neutrinos is found to be \cite{Gnedin:1997vn,Mangano:2005cc,deSalas:2016ztq,Froustey:2020mcq,Akita:2020szl,Bennett:2020zkv}
\begin{equation}
    N_\text{eff.} = 3.0440 \pm 0.0002,
\end{equation}
where a small deviation from the value expected for three generations arises as the neutrino decoupling from the SM bath around MeV temperatures is not instantaneous.
The abundance of dark radiation is typically parameterized in terms of the effective  number of additional neutrinos $\Delta N_\text{eff.}$.
The value inferred from the observed abundance of light elements produced during Big Bang Nucleosynthesis  \cite{Planck:2018vyg} is
\begin{equation}
    N_\text{eff}^\text{BBN} = 2.95^{+0.56}_{-0.52}.
\end{equation}
Combined analyses of the current Planck CMB data together with  Baryon Accoustic oscillations found \cite{Planck:2018vyg}
\begin{equation}
    N_\text{eff}^\text{Planck+BAO} = 2.99^{+0.34}_{-0.33},
\end{equation}
which can be translated into
\begin{equation}\label{eq:deltneff}
    \Delta N_\text{eff.}^\text{Planck+BAO} \simeq 0.28.
\end{equation}
The upcoming CMB Stage IV experiment \cite{Abazajian:2019eic,annurev-nucl-102014-021908} and NASA's PICO proposal \cite{NASAPICO:2019thw}  have a sensitivity forecast of
\begin{equation}\label{eq:stageIV}
    \Delta N_\text{eff}^\text{proj.}=0.06.
\end{equation}
There is also the planned CORE experiment by the ESA \cite{CORE:2017oje} as well as the South Pole Telescope (SPT) \cite{SPT-3G:2014dbx} and the Simons observatory \cite{SimonsObservatory:2018koc}, which both aim to reach  $\Delta N_\text{eff.} \lesssim 0.12$.

\subsection{Contribution of the axion}\label{sec:axNeff}
The QCD axion does not reach thermal equilibrium via its couplings to the SM leptons: Reactions like $\nu\; \overline{\nu} \leftrightarrow Z\; a$ and $\nu \; e^+ \leftrightarrow W^+ \; a$ would only ever thermalize at temperatures far below the $Z$- and $W^\pm$-boson masses, because the rates are suppressed with $m_\nu^2 / f_a^2$. Three body processes like  $\nu\; \overline{\nu} \leftrightarrow\nu\; \overline{\nu}\; a$ avoid the production of heavy on-shell EW gauge bosons but are too slow to ever matter due to the previously mentioned tiny couplings. Production of two axions via $\nu\; \overline{\nu} \leftrightarrow a\; a$ is even more suppressed. 
References \cite{Masso:2002np,Graf:2010tv,Salvio:2013iaa} showed that the axion decouples from its unavoidable strong interactions with the quark-gluon-plasma  at
\begin{equation}
    T_a^\text{dec.} \simeq \SI{1.7e9}{\giga\electronvolt}\cdot \left(\frac{f_a}{10^{11}\;\text{GeV}}\right)^{2.246}.
\end{equation} 
In the HHSI scenario PQ symmetry is broken  at $T_c\simeq 0.01\; f_a$ \cite{Ballesteros:2016xej}, which occurs  after reheating for the typical range of $f_a<10^{11}\;\text{GeV}$  and the reheating temperature given by \eqref{eq:TRH2}. 
Since both the  critical temperature and the reheating temperature are larger than   the decoupling temperature $ T_a^\text{dec.}$, the axions will have had a thermal abundance in the early universe. Their contribution to the amount of dark radiation can then be estimated to be  \cite{Ballesteros:2016xej}
\begin{equation}
    \Delta N_\text{eff.}\simeq 0.027 \cdot \left(\frac{100}{g_*\left( T_a^\text{dec.}\right)}\right)^\frac{4}{3},
\end{equation}
which is an order of magnitude below the current bound of \eqref{eq:deltneff}.

\subsection{Contribution of the right handed neutrinos}\label{sec.nuRtherm}
\noindent  The production of right handed neutrinos is driven by the $Y_{DR}$ Yukawa coupling to the heavy doublet fields $D$ in equation \eqref{eq:main-model}. Since the doublets have $\text{SU(2)}_\text{L}\otimes \text{U(1)}_\text{Y}$ gauge interactions they will develop a thermal abundance at high temperatures and can produce $\nu_R$ from decays and scattering. Even after the temperature drops below the lightest $D$ mass there are still processes like $H^\dagger H \leftrightarrow \overline{\nu_R} \nu_R$ producing $\nu_R$ and keeping them in thermal equilibrium via off-shell $D$-exchange. On dimensional grounds we can estimate the interaction rate for the aforementioned process for $T\gg M_D$ as
\begin{equation}
    \Gamma(T\gg M_D) \sim \frac{ Y_{DR}^4 T}{16\pi}
\end{equation}
and find that it comes into thermal equilibrium at a temperature of 
\begin{equation}
    T_\text{coupl.} \simeq \SI{1.2e12}{\giga\electronvolt} \cdot \left(\frac{Y_{DR}}{0.1}\right)^4 \cdot  \sqrt{\frac{100}{g_*( T_\text{coupl.})}}.
\end{equation}
We therefore expect a thermalized population of $\nu_R$ after reheating at around $10^9\;\text{GeV}$ (see \eqref{eq:TRH}). Since this process is not Boltzmann-suppressed with the heavy $D$-mass it can still be effective until temperatures $T\ll M_D$. In this regime we estimate the interaction rate to be
\begin{equation}
    \Gamma(T\ll M_D) \sim \frac{Y_{DR}^4}{16 \pi M_D^2} T^3,
\end{equation}
where we neglected the Higgs mass and find that it drops out of thermal equilibrium at
\begin{equation}\label{eq:nuR-FO}
    T_\text{FO} \simeq \SI{10}{\tera\electronvolt} \cdot  \left(\frac{0.1}{Y_{DR}}\right)^4 \cdot \left(\frac{M_D}{\SI{e8}{\giga\electronvolt}}\right)^2 \cdot \sqrt{\frac{g_*(T_\text{FO})}{100}}.
\end{equation}
As this temperature is above the electroweak crossover it was self-consistent to neglect the mass of $H$. For a freeze-out before EWSB we can estimate the contribution of $N_\nu$ decoupled $\nu_R$ generations to the present day energy density in radiation as \cite{Abazajian:2019oqj}
\begin{equation}\label{eq:N-eff-main}
    \Delta N_\text{eff} \simeq 3\;\cdot 0.027\cdot2 \cdot \frac{7}{8}\cdot  \left(\frac{100}{g_*(T_\text{FO})}\right)^\frac{4}{3} =  0.142 .
\end{equation}
We did not include the predicted asymmetry in $\nu_R$  because it is of the order of the small baryon asymmetry and negligible compared to the equilibrium abundance.
Together with the contribution from the axion we have $ \Delta N_\text{eff} \simeq 0.17$ which is allowed by current data see \eqref{eq:deltneff} and will be probed by next generation experiments. An intriguing way to make our model of Dirac neutrinos compatible with the projected sensitivity  \eqref{eq:stageIV} would be to invoke additional entropy dilution after the decoupling of the right handed neutrinos, which would suppress $ \Delta N_\text{eff} $ by a factor $\Delta>1$. This mechanism has been used in the past to dilute the overabundance of  thermalized keV-scale sterile neutrino dark matter in gauge theories \cite{Asaka:2006ek,Bezrukov:2009th} needing $\Delta\sim 100$. Another application of entropy dilution  is to bring the gravitino abundance in accordance with the reheating temperature required for vanilla leptogenesis for $\Delta\sim 10^3 - 10^4$ \cite{Hasenkamp:2010if}.
The main ingredient would be a long-lived particle decaying far from equilibrium leading to an intermediate matter dominated epoch \cite{Scherrer:1984fd}. 
Since this particle needs to decouple while relativistic to produce sufficient entropy it should not have gauge interactions. This leaves only the radial mode of the PQ breaking field $h_\sigma$, which could be long-lived via its decays to the SM like Higgs. However the required scalar potential couplings would spoil \eqref{eq:infbounds} so that inflation and reheating in the HHSI scenario (see section \ref{sec:reh}) would cease to work. Hence we do not consider a long-lived $h_\sigma$ further and treat the rather large value of $ \Delta N_\text{eff} \simeq 0.17$ from the axion and three $\nu_R$ as an observational signature to distinguish our scenario from models involving either only a light scalar or only right handed neutrinos.\\
\\
The smoking gun signature for this model would be observation of such a large $ \Delta N_\text{eff}$ together with a signal in experiments probing the axion to photon coupling that is enhanced by an order of magnitude compared to the regular QCD axion band (see \eqref{eq:ENtot}).

\section{Summary}\label{sec:sum}
\begin{itemize}
    \item \textbf{Dirac neutrino masses:}\\
    We constructed Dirac neutrino mass model in the Seesaw  spirit, where all heavy particles have masses from the PQ breaking scale and no new scalar besides   the PQ breaking singlet is needed. To do so we introduce heavy leptons in the form of electroweak doublets $D$ and triplets $T$. Unlike most models we generate the neutrino masses not via a dimension five operator but rather at dimension six.
    The only fundamental scales of this model are the EWSB scale $v_H$ and the PQ scale $v_\sigma$ which can be identified with the axion decay constant $f_a$. The lightest Dirac neutrino is approximately massless due to the Yukawa couplings required for leptogenesis.
    \item \textbf{Axion to photon coupling:}\\
    Our model involves vector-like fermion that are anomalous with respect to PQ symmetry. They boost the coupling of the QCD axions to a pair of photons by around an order of magnitude (see equation \eqref{eq:ENtot}) relative to conventional models and can be probed by current experiments such as HAYSTAC, ORGAN and QUAX or future searches by MADMAX, BRASS or ADMX. The new fermions do not lead to phenomenologically relevant Landau Poles for the SM gauge couplings.
    \item \textbf{preserving the attractive features of S.M.A.S.H.:}
    The model is compatible with the cosmological history outlined of the original S.M.A.S.H. scenario \cite{Ballesteros:2016euj,Ballesteros:2016xej,Ballesteros:2019tvf} such as successful  inflation, reheating and  axion DM from both misalignment and topological defect decay. Our new heavy fermions do not spoil the stability of the scalar potential.
    \item \textbf{Dirac-Leptogenesis:}
    We found an alternative for resonant leptogenesis \cite{Pilaftsis:2003gt} when it comes to enhancing the leptonic asymmetry parameter $\varepsilon_L$ in \eqref{eq:asym-uplimit} by up to six orders of magnitude. Whereas a Majorana triplet fermion needs to have a mass of at least $10^{10}\;\text{GeV}$ our enhanced asymmetry allows for successful leptogenesis even with $10^8\;\text{GeV}$ masses. The phenomenology is qualitatively and quantitatively different from the case for Majorana fermions since the triplets can develop asymmetries themselves via washout processes. Choosing different branching ratios for the triplet decays to $L$ and $D$ allows for the \enquote{quasi optimal efficiency}-scenario first discussed for decaying scalar triplets in \cite{Hambye:2005tk}.
    We identified four parameter regions that reproduce the observed baryon to photon ratio.    
    \item \textbf{Dark radiation:}\\
    Our setup involves an axion and three right handed neutrinos that were thermalized in the early universe producing a large value of $\Delta N_\text{eff.}\simeq 0.17$ which will be probed and potentially excluded by next generation CMB experiments such as CMB-S4, PICO, SPT or the Simons observatory.  $\Delta N_\text{eff.}$ can be used to distinguish our construction from models involving only a light scalar $(\Delta N_\text{eff.}\simeq 0.028)$ like the original S.M.A.S.H. or only right handed neutrinos ($\Delta N_\text{eff.}\simeq 0.142$ for three generations).
\end{itemize}

\acknowledgments
This work benefited from the use of \verb|PackageX| \cite{Patel:2015tea,Patel:2016fam}.
We would like to think Ciaran O'Hare for compiling the available axion limits \cite{GITHUB} and for useful correspondence. Furthermore we are grateful to Andreas Trautner for providing valuable feedback on this manuscript.

\begin{appendices}
\section{Collection of limits on the axion to photon coupling}\label{sec:lim}
Constraints on the axion to photon coupling were compiled in  \cite{GITHUB} and can be grouped into the following categories
\begin{itemize}
    \item \textbf{Haloscopes} looking for DM axions from the galactic DM halo\\
    such as ABRACADABRA  \cite{Ouellet:2018beu,Salemi:2021gck}, ADMX   \cite{2010PhRvL.104d1301A,Stern:2016bbw,ADMX:2018gho,ADMX:2019uok,ADMX:2021nhd,ADMX:2018ogs,Bartram:2021ysp,Crisosto:2019fcj}, BRASS \cite{BRASS}, CAPP \cite{Lee:2020cfj,Jeong:2020cwz,CAPP:2020utb}, DM-Radio \cite{DMRadio}, HAYSTAC \cite{HAYSTAC:2018rwy,HAYSTAC:2020kwv}, KLASH \cite{Alesini:2017ifp}, MADMAX \cite{Beurthey:2020yuq}, ORGAN \cite{McAllister:2017lkb,doi:10.1126/sciadv.abq3765}, QUAX \cite{Alesini:2019ajt,Alesini:2020vny}, RADES \cite{CAST:2020rlf},  RBF \cite{PhysRevLett.59.839}, SHAFT \cite{Gramolin:2020ict} and UF \cite{PhysRevD.42.1297} 
 
    \item \textbf{Helioscopes} looking for axions produced inside the sun\\
    such as CAST \cite{CAST:2007jps,CAST:2017uph}, babyIAXO or IAXO \cite{2013ITAS...23T0604S,Ge:2020zww}.
 
    \item \textbf{Light shining through walls (LSW)} and similar experiments\\
    such as ALPS \cite{Ehret:2010mh,Ortiz:2020tgs}, CROWS \cite{Betz:2013dza}, OSQAR \cite{OSQAR:2015qdv} and PVLAS \cite{DellaValle:2015xxa}.
 
    \item \textbf{Cosmological probes}\\
    such as  extragalactic background light (EBL), ionisation fraction, X-rays \cite{Cadamuro:2011fd} or BBN and $\Delta N_\text{eff.}$ \cite{Depta:2020wmr}.
 
    \item (indirect) \textbf{Astrophysical bounds}\\
    such as Black hole superradiance \cite{Mehta:2020kwu}, the Chandra X-ray telescope \cite{Wouters:2013hua,Reynolds:2019uqt,Reynes:2021bpe,Dessert:2021bkv}, the Fermi Large Area Telescope (LAT) \cite{Fermi-LAT:2016nkz,Meyer:2016wrm,Meyer:2020vzy,Calore:2021hhn}, super star clusters \cite{Dessert:2020lil}, the cosmic distance ladder \cite{Buen-Abad:2020zbd}, the HESS cherenkov telescope \cite{HESS:2013udx}, horizontal branch stars \cite{Ayala:2014pea}, White dwarfs \cite{Dolan:2021rya}, Globular Cluster ($R_2$) \cite{Dolan:2022kul},
    the blazar Markarian 421 (Mark 421) \cite{Li:2020pcn}, neutron stars \cite{Foster:2020pgt,Darling:2020uyo,Battye:2021yue}, observations of the solar neutrino flux \cite{2015JCAP...10..015V}, supernova SN 1987A \cite{Jaeckel:2017tud,Caputo:2021rux}, radio telescopes \cite{Blout_2001} and optical telescopes like  MUSE \cite{Regis:2020fhw} and VIMOS \cite{Grin:2006aw}
\end{itemize}

\section{Landau poles for the SM gauge couplings}\label{sec:landau}
\noindent As shown in \cite{DiLuzio:2016sbl,DiLuzio:2017pfr}  the hypercharged exotic quarks with masses $\SI{5e11}{\giga\electronvolt}$ do not lead to a Landau pole for the $\text{U}_\text{Y}$ gauge coupling (or any other SM gauge coupling) below the Planck mass. Following the methods outlined in \cite{MACHACEK198383,DiLuzio:2015oha,Plakkot:2021xyx} we compute the coefficients of the two loop renormalization group equations (RGE)
\begin{equation}\label{eq:RGE}
    \frac{\text{d}}{\text{d}t} \alpha_i^{-1} = a_i + \frac{b_{ij}}{4\pi} \alpha_j, \quad \text{where} \quad \alpha_j \equiv \frac{g_j^2}{4\pi} \quad \text{and} \quad t \equiv \frac{1}{2\pi}\text{Log}\left(\frac{\mu}{m_Z}\right)
\end{equation}
where $i=1,2,3$ labels the gauge groups $\text{U}_\text{Y}$, $\text{SU(2)}_\text{L}$ and $\text{SU(3)}_\text{c}$ respectively. Here $\mu$ denotes the renormalization scale. We do not include the contribution from the Yukawa couplings for simplicity. The definitions of the constants can be found in \cite{DiLuzio:2015oha,Plakkot:2021xyx} and for the SM they read \cite{MACHACEK198383}
\begin{equation}
    \vec{a}^\text{SM} = 
    \begin{pmatrix}
    a_1\\
    a_2\\
    a_3
    \end{pmatrix}
    =
    \begin{pmatrix}
    4+\frac{1}{10}\\
    -22+4+\frac{1}{6}\\
    -11+4
    \end{pmatrix}
\end{equation}
as well as
\begin{equation}
    b^\text{SM} =
    \begin{pmatrix}
    0 & 0& 0\\
    0& -\frac{132}{3}& 0\\
    0& 0& -102
    \end{pmatrix}
    + N_\text{gen.} 
    \begin{pmatrix}
    \frac{19}{15} & \frac{3}{5} & \frac{44}{15}\\
    \frac{1}{5} & \frac{49}{3} & 4\\
    \frac{11}{30} & \frac{3}{2} & \frac{76}{3} 
    \end{pmatrix}
    +
    \begin{pmatrix}
    \frac{9}{50} & \frac{9}{10} & 0\\
    \frac{3}{10} & \frac{13}{6} & 0\\
    0&0&0
    \end{pmatrix}.
\end{equation}
The first matrix comes from the gauge bosons, the second matrix measures the contribution of $ N_\text{gen.}=3$ generations of fermions and the last matrix arises due to the SM Higgs boson.
Note that the matrix $ b^\text{SM}$ is transposed compared to reference \cite{MACHACEK198383} and we used the GUT normalization of $\text{SU(5)}$ or $\text{SO(10)}$ for the hypercharge \cite{DiLuzio:2015oha}
\begin{equation}
    Y_\text{norm.} = \sqrt{\frac{3}{5}} Y_\text{SM}.
\end{equation}
Motivated by our cosmological findings  we only include the lightest electroweak triplet of mass $M_T = \SI{e8}{\giga\electronvolt}$ with $Y_\text{SM}=0$ for which
\begin{equation}
    \vec{a}^\text{BSM}_T = \begin{pmatrix}
      0\\
      \frac{8}{3}\\
      0 
    \end{pmatrix},
    \quad 
    b^\text{BSM}_T = 
    \begin{pmatrix}
    0 & 0 & 0\\
    0 & \frac{128}{3} & 0\\
    0&0&0\\
    \end{pmatrix}
\end{equation}
and the doublet leptons with $Y_\text{SM}=-\frac{1}{2}$ at $M_D=\SI{3e7}{\giga\electronvolt}$ for which
\begin{equation}
    \vec{a}^\text{BSM}_D = \begin{pmatrix}
      \frac{2}{5}\\
      \frac{2}{3}\\
      0 
    \end{pmatrix},
    \quad 
    b^\text{BSM}_D = 
    \begin{pmatrix}
    \frac{9}{50} & \frac{9}{10} & 0\\
    \frac{3}{10} & \frac{49}{6} & 0\\
    0&0&0\\
    \end{pmatrix}.
\end{equation}
We include the threshold effects of the heavy fermions $\Psi$ by solving \eqref{eq:RGE} with
\begin{equation}
    \vec{a}^\text{BSM} = \vec{a}^\text{SM} + \Theta\left(\mu-M_\Psi\right) \vec{a}_{\Psi}^\text{BSM}, \quad b^\text{BSM} = b^\text{SM} + \Theta\left(\mu-M_\Psi\right)   b_{\Psi}^\text{BSM},
\end{equation}
where $\Theta$ denotes the Heaviside function and we use the following boundary conditions \cite{Mihaila:2012pz,Zyla:2020zbs}
\begin{equation}
    \alpha_1(m_Z) = 0.016923, \; \alpha_2(m_Z) = 0.03374, \; \alpha_3(m_Z) = 0.1173 \; \text{and} \; m_Z = \SI{91.188}{\giga\electronvolt}.
\end{equation}

\begin{figure}[t]
\centering
    \includegraphics[width=0.49\textwidth]{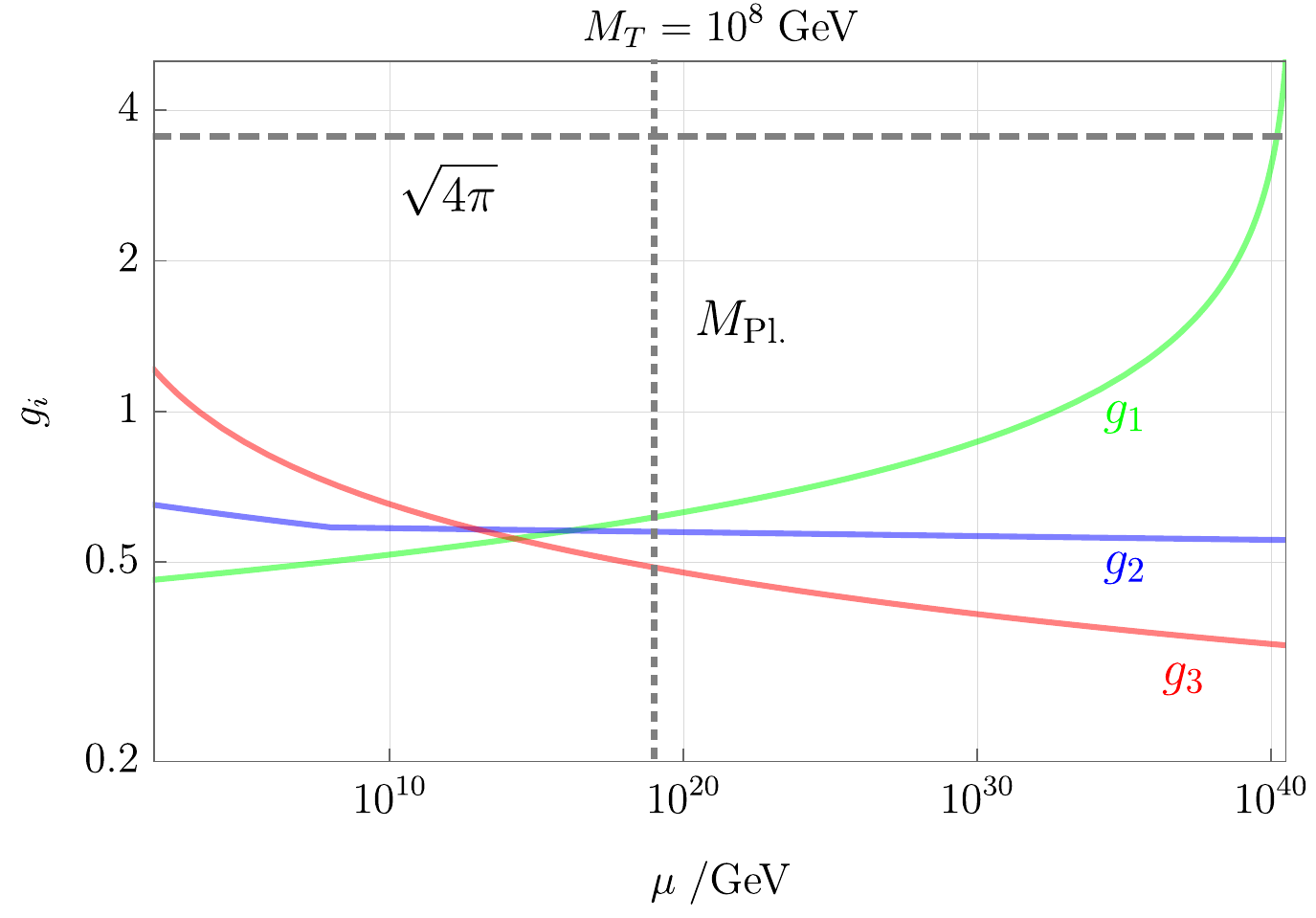}
    \includegraphics[width=0.49\textwidth]{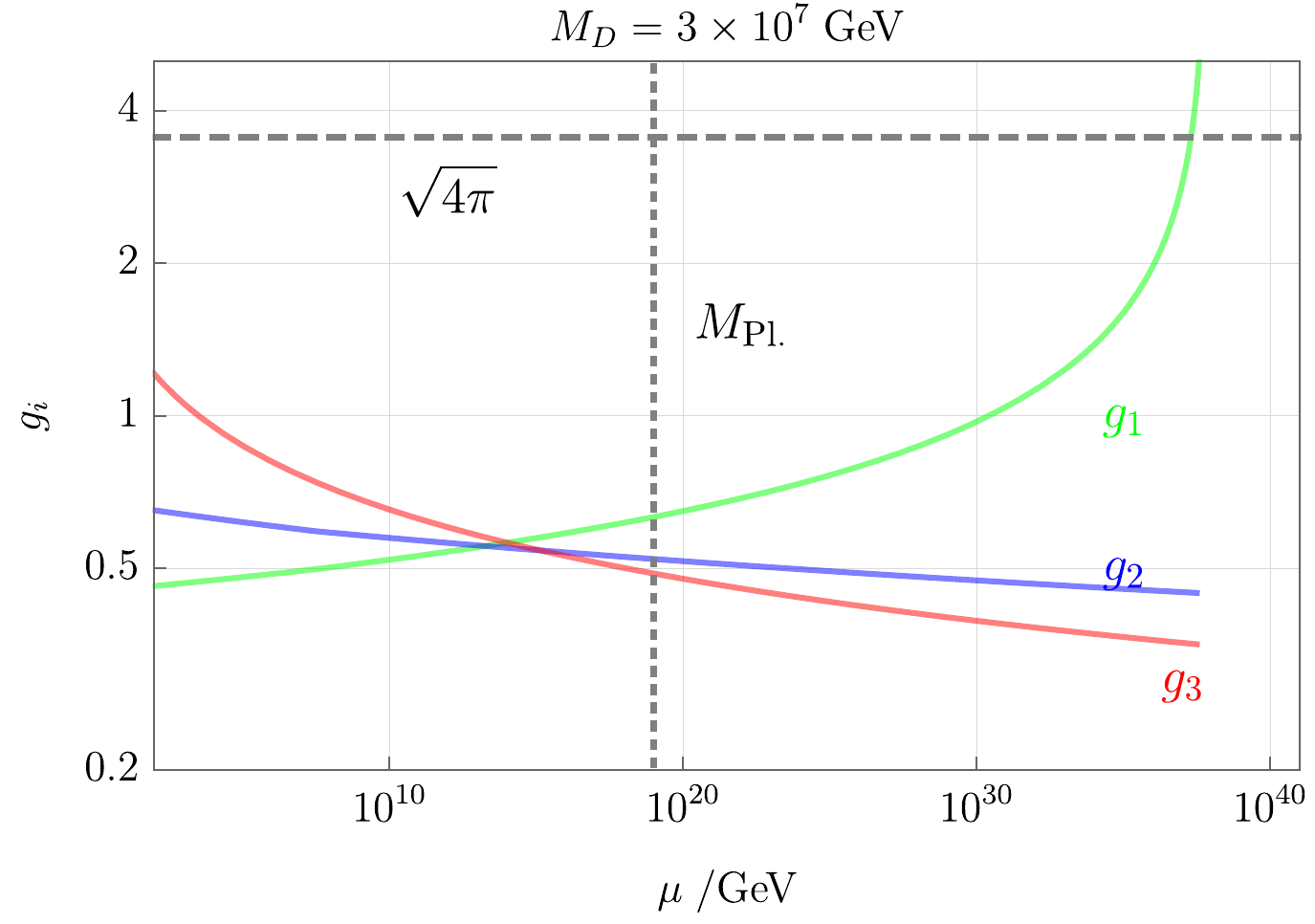}
	\label{fig:RGELandau}
	\caption{Two Loop RGE evolution of the SM gauge couplings as a function of the renormalization scale for the inclusion of a hypercharge zero $\text{SU}(2)_\text{L}$ triplet with $M_T= \SI{e8}{\giga\electronvolt}$ \textit{(left)} and a 
	hypercharge $1/2$ $\text{SU}(2)_\text{L}$ doublet fermion with a mass of $M_D=\SI{3e7}{\giga\electronvolt}$ \textit{(right)}.}
\end{figure}
\noindent
We reproduce  the Landau pole in $g_1$ around $10^{38}\;\text{GeV}$ found for  a vector-like quark  with $Y= -1/3$ with a representative mass  of $5\times 10^{11}\;\text{GeV}$ in \cite{DiLuzio:2016sbl,DiLuzio:2017pfr}. We depict the evolution of the three gauge couplings for adding the lightest triplet or doublet in  \ref{fig:RGELandau} with the masses $M_T=10^8\;\text{GeV}$ and $M_D=3\times 10^7\;\text{GeV}$ from section \ref{sec:num}. One can observe in  \ref{fig:RGELandau} that the Landau pole appears earlier for the lighter doublet \cite{DiLuzio:2015oha}. We stopped the numerical evaluation at the Landau poles because the system of differential equations starts to exhibit singular behaviour and we can not trust our calculation for larger energies anymore. This is why the lines for the doublet in \ref{fig:RGELandau} terminate earlier than for the triplet. In both scenarios the potential Landau pole in $g_1$ manifests far above the Planck scale, so it does not affect the phenomenology of our model. Let us emphasize that this analysis only serves as a first estimate and in principle all three generations of all new species and their Yukawa interactions should be included in the running. For a more realistic estimate we include  three generations of quarks with the same masses of $10^9 \;\text{GeV}$ for simplicity as well as three generations of triplets, where the two heavier ones have $10^9\;\text{GeV}$ masses, together with three doublets, with the two heavier ones at  $10^8\;\text{GeV}$ (this range of masses was motivated in section \ref{sec.vac}). 
Adding the entire fermionic particle content of the model induces a  Landau pole   at $10^{21}\;\text{GeV}$  again in $g_1$ and again above the Planck scale.  The landau pole appears at lower energies the more fermions we introduce because the positive coefficients in the RGEs \eqref{eq:RGE} increase with each additional fermion \cite{DiLuzio:2016sbl,DiLuzio:2017pfr}. We conclude that our three generations of exotic fermions do not lead to phenomenologically relevant Landau Poles for the SM gauge interactions.

\section{Sphaleron redistribution coefficient}\label{sec:sph-red-coeff}
\noindent Once an asymmetry in e.g. the SM leptons is created one has to take into account how this asymmetry is redistributed to the rest of the fermions and the SM Higgs via gauge and Yukawa interactions that are in equilibrium. These fast spectator processes lead to conservation laws for the individual number densities $n_\psi$, which for ultra-relativistic fermions (bosons) $\psi$ (and their anti-particles $\overline{\psi}$) can be expressed in terms of their chemical potentials $\mu_\psi$ via the relation
\begin{equation}
      n_\psi-n_{\overline{\psi}} = \frac{\mu_\psi g_\psi}{3}T^2 \begin{cases} \frac{1}{2}\quad \text{for fermions}\\1\quad \text{for bosons}\end{cases},
\end{equation}
whereas for a massive particles the appropriate relation would be  \cite{DREINER1993188}
\begin{equation}\label{eq:massive}
    n_\psi-n_{\overline{\psi}} = \frac{\mu_\psi g_\psi}{\pi^2} T^2 F_\pm\left(\frac{m_\psi}{T}\right), \quad \text{with} \quad F_\pm(x) \equiv \int_x^\infty \text{d}y\; \frac{y\sqrt{y^2-z^2}e^y}{(1\pm e^y)^2}
\end{equation}
where the + (-) applies for bosons (fermions). We will work in the regime $T_\text{EQ}\simeq \mathcal{O}\left(\SI{100}{\giga\electronvolt}\right)\ll T < 10^8\;\text{GeV}$
and since we need to be above the EW phase transition for the sphaleron transition to occur we can work in the regime of unbroken electroweak symmetry where the chemical potential of the gauge bosons is $\mu_W= 0$ and components of the same multiplet have the same chemical potential \cite{PhysRevD.42.3344}. For the SM we include 3 generations of $q,u,d,l,e$ and one Higgs $H$.
At temperatures below $10^8\;\text{GeV}$ all SM Yukawa interactions are in equilibrium \cite{Nardi:2005hs} and for simplicity we neglect all flavor effects and assign generation-independent chemical potentials. For the SM the appropriate conditions read \cite{PhysRevD.42.3344,Khlebnikov:1996vj}
\begin{itemize}
    \item hypercharge neutrality of the plasma
    \begin{equation}\label{eq:weaksph}
        3\left(\mu_q+2 \mu_u-\mu_d -\mu_l-\mu_e\right) + 2\mu_H = 0
    \end{equation}
    \item $\text{SU}(2)_\text{L}$ sphalerons
    \begin{equation}\label{eq:sph}
        3\mu_q +\mu_l = 0\quad \text{from} \quad \mathcal{O}_\text{sph.} = \Pi_{i=1}^3 l_i q_i q_i q_i
    \end{equation}
    \item $\text{SU}(3)_\text{c}$ sphalerons
    \begin{equation}\label{eq:strongsph}
        2 \mu_q -\mu-\mu_d = 0
    \end{equation}
    \item SM Yukawa interactions
    \begin{align}
        \mu_l-\mu_H-\mu_e &=0\quad \text{from} \quad \overline{L}He,\\
        \mu_q-\mu_H-\mu_d &=0\quad \text{from} \quad \overline{Q}Hd,\label{eq:q1}\\
        \mu_q +\mu_H -\mu_u&=0 \quad \text{from} \quad \overline{Q}\tilde{H}u.\label{eq:q2}
    \end{align}
\end{itemize}
Note that since all Yukawa interactions equilibrate the $\text{SU}(3)_\text{c}$ sphaleron condition \eqref{eq:strongsph} becomes redundant as it is jus the sum of \eqref{eq:q1} and \eqref{eq:q2}. As we have six potentials and only five conditions we can express five chemcial potentials in terms of a sixth. We choose the potential for the total baryon minus lepton number for three generations in the SM
\begin{equation}
    \mu_{\text{B}-\text{L}_\text{SM}} =   2\mu_q +\mu_u+\mu_d-2\mu_l-\mu_e,
\end{equation}
because it is not washed put by the weak sphalerons and recover the famous relations
\begin{equation}
    \mu_\text{B} = \frac{28}{79} \mu_{\text{B}-\text{L}_\text{SM}},\quad  \mu_{\text{L}_\text{SM}} = -\frac{51}{79} \mu_{\text{B}-\text{L}_\text{SM}}.
\end{equation}
Next we add the additional BSM particles. We start with three gauge singlets $\nu_R$. The previous conditions are unchanged, but since our theory conserves $\text{B-L}$ we have to impose this on the potentials
\begin{equation}
    \mu_{\text{B}-\text{L}_\text{tot}}= \mu_{\text{B}-\text{L}_\text{SM}} - \mu_{\nu_R} = 0,
\end{equation}
which by itself would lead to \cite{Dick:1999je}
\begin{equation}
    \mu_\text{B} =  \mu_{\text{L}_\text{tot}}= \frac{28}{79 }\mu_{\text{B}-\text{L}_\text{SM}}.
\end{equation}
However we also add the lightest two vector-like doublets $(D_L,D_R)$ with $Y= 1/2$ and lepton number 1. The heavier doublets will be Boltzmann suppressed. We assume that the interaction $\sigma \overline{D_L}D_R$ was in equilibrium at high temperatures and since after Peccei-Quinn breaking we have $\mu_\sigma = 0$ we conclude that $\mu_{D_L}=\mu_{D_R}\equiv\mu_D$. The hypercharge neutrality condition is modified to
\begin{equation}\label{eq:hyp-mod}
     3\left(\mu_q+2 \mu_u-\mu_d -\mu_l-\mu_e\right) + 2\mu_H + 2\mu_D= 0,
\end{equation}
where the factor of two for $\mu_H$ appears because the scalar Higgs has different quantum statistics compared to the fermions and the factor two for $\mu_D$ appears because of the two chiralities we add.
We furthermore demand that their coupling to $\nu_R$ is in equilibrium so that there is no population of stable heavy leptons (see section \ref{sec:wash})
\begin{equation}\label{eq:yuk-nuR}
    \mu_D-\mu_H-\mu_{\nu_R} =0\quad \text{from} \quad \overline{D_L}H\nu_R.
\end{equation}
Since the $D$s are leptonic doublets we expect them to couple to the weak sphaleron transition.
Naively one would expect to replace $\mu_l$ the condition \eqref{eq:sph} with $-\mu_D$, because of the opposite hypercharge. However the pair of vector-like fermions $(D_L,D_R)$ does not contribute to the $\text{U}(1)_\text{B+L}\otimes \text{SU}(2)_\text{L}^2$ anomaly. Reference \cite{Cerdeno:2018dqk} computed the effective sphaleron mediated operators and since the vector-like leptons do not lead to B+L violation the modified sphaleron vertex will still correspond to a $\Delta \left(\text{B+L}\right)=6$ transition like the original in  \eqref{eq:sph} and only involves a pair of vector-like leptons
\begin{equation}
    \mathcal{O}_\text{sph. I}^{\text{SM}+\text{VL}}=  \left(\Pi_{i=1}^3 l_i q_i q_i q_i\right) D_L \overline{D_R}.
\end{equation}
One can see that this vertex does not lead to new constraints on the chemical potential and that the asymmetry in $D$ can not be converted into a baryonic asymmetry.
Let us  continue with the conservation of the  total B-L 
\begin{equation}\label{eq:B-L-tot}
    \mu_{\text{B}-\text{L}_\text{tot}}= \mu_{\text{B}-\text{L}_\text{SM}} - \mu_{\nu_R} -\frac{4}{3}\mu_D = 0,
\end{equation}
where the factor of $4/3$ arises because of the two doublets $D_{L,R}$  and there is only one generation of heavy doublets in the plasma compared to the three generations of the SM quarks, leptons and $\nu_R$.  Moreover the Boltzmann equations in section \ref{sec:Boltz} respect the conservation of total lepton number see \eqref{eq:lept-cons} as well. Solving the system of equations of \eqref{eq:hyp-mod},\eqref{eq:yuk-nuR} and \eqref{eq:B-L-tot} together with the previously mentioned conditions on the SM chemical potentials we arrive at 
\begin{equation}\label{eq:red-coeff}
    \mu_\text{B} = \mu_{\text{L}_\text{tot}} = \frac{55}{148} \mu_{\text{B}-\text{L}_\text{SM}}, \quad  
    \mu_D = \frac{201}{592} \mu_{\text{B}-\text{L}_\text{SM}} \quad \text{and} \quad \mu_{\nu_R}= \frac{81}{148} \mu_{\text{B}-\text{L}_\text{SM}}.
\end{equation}
The conversion factor for generating B from $\text{B}-\text{L}_\text{SM}$ of $55/148\simeq 0.37$ is only slightly larger than the SM result  $28/79\simeq 0.35$.

\section{Cross sections and rate densities}\label{sec:rates}
The relevant cross sections are given in terms of the following dimensionless variables
\begin{equation}
   \delta \equiv \left(\frac{M_D}{M_T}\right)^2,\quad  x\equiv \frac{s}{M_T^2},\quad  r\equiv \sqrt{1-\frac{4}{x}},\quad \omega\equiv \frac{\Gamma_\text{tot.}}{M_T}
\end{equation}
and read
\begin{itemize}
    \item process $ L\overline{D} \rightarrow H  H$ possible in the $t-$ and $u$-channel:
    \begin{align}
        \sigma_{t+u} &= \frac{3}{8\pi M_T^2}\left(\frac{x-\delta}{x^2(1+x-\delta)} -\frac{\text{Log}(1+x-\delta)}{x^2(2+x-\delta)} \right)\label{eq:tu}\\
        &= \frac{3}{8\pi}\begin{cases}\frac{1}{2 M_T^2}\quad \text{for}\; s\ll M_T^2\\
        \frac{1}{s}\quad \text{for}\;s\gg M_T^2\end{cases}
    \end{align}
    \item  subtracted cross section (see section \ref{sec:Boltz}) for the  process $L H\rightarrow D H$ possible in the $s-$ and $t-$channel:
    \begin{align}
         \sigma_{s+t}^\text{sub.} &= \frac{ M_T^2}{32\pi x^2} (x^2-\delta^2) \sum^3_{a=1} \left|D_s^{a}(s)\right|^{\text{sub.}\,\;2}\\
         &+ \frac{3}{16\pi M_T^2 x^2} \frac{(x-\delta)((x-1)(-2 +(x-1)x+\delta)+(1+x)\omega^2 )}{(1+x-\delta)((x-1)^2+\omega^2)}\nonumber\\
         &+  \frac{3}{16\pi M_T^2 x^2} \frac{(-2+2x +\omega^2) \text{Log}(1+x-\delta) }{(1-x)^2+\omega^2}\nonumber\\
          &= \frac{3}{16\pi}\begin{cases}\frac{1}{2 M_T^2}\quad \text{for}\; s\ll M_T^2\\
        \frac{1}{s}\quad \text{for}\;s\gg M_T^2\end{cases}\label{eq:cross}
    \end{align}
    where for all $a=1,2,3$
    \begin{equation} 
        \left|D_s^{a}(s)\right|^{\text{sub.}\,\;2} \equiv   \left|\frac{1}{s-M_T^2+i M_T          \Gamma_\text{tot.}}\right|^2 - \frac{\pi \delta(s-M_T^2)}{M_T \Gamma_\text{tot.}}.
    \end{equation}
    For the numerical evaluation it is convenient to carry out the thermal average of the subtracted matrix element  \eqref{eq:submat} rather than to subtract the densities appearing in \eqref{eq:subdef}. We follow the methods of \cite{Giudice:2003jh} and use 
    the following representation of the $\delta$-distribution which decreases faster than the propagator away from the resonance
    \begin{equation}
       \delta(y) = \frac{2 \rho^3}{\pi\left(y^2 + \rho^2\right)} \quad \text{with} \quad y =\frac{s}{M_T^2}-1,  
    \end{equation}
    where $\rho\ll1$. Since we find  $\Gamma_\text{tot.} / M_T < 10^{-5}$ we are always in the narrow width regime and may set  $\rho =  \Gamma_\text{tot.} / M_T$ \cite{Giudice:2003jh}.
    \item process $T\overline{T}\rightarrow WW, F \overline{F}$, where $W$ are the $\text{SU}(2)_\text{L}$ gauge bosons and $F$ represents the SM fermion doublets \cite{Hambye:2003rt}:
    \begin{align}
        \sigma_W &= \frac{g_2^4}{\pi M_T^2 x\; r^2}  \left(3 r\left(1+\frac{2}{x}\right)  - r\left(4+ \frac{17}{x}\right) \right)\label{eq:gaug}\\
        &+\frac{3 g_2^4}{\pi M_T^2 x\; r^2} \left(1+\frac{4}{x}-\frac{4}{x^2}\right) \text{Log}\left(\frac{1+r}{1-r}\right)\nonumber
    \end{align}
 For non-relativistic triplets the gauge scattering rate density can be approximately written as \cite{Cirelli:2007xd,Strumia:2008cf} 
\begin{equation}\label{eq:gauge-scat}
    \gamma\left(T \;\overline{T} \leftrightarrow W W, F \overline{F}\right) = 4\times \frac{M_T T^3}{32\pi^3}e^{-2\frac{M_T}{T}}\left(c_s + 3 \frac{T}{M_T}\left(c_p+c_s\right) + \mathcal{O}\left( \frac{T^2}{M_T^2}\right)\right),
\end{equation}
where we inserted a factor of four by hand to take into account that Dirac triplets have twice as many internal degrees of freedom as Majorana ones (see \eqref{eq:rates}). Furthermore
\begin{equation}\label{eq:gauge-scat2}
    c_s = \frac{111}{8\pi}g_2^4 \quad \text{and} \quad c_p = \frac{51}{8\pi}g_2^4
\end{equation}
are the $s$- and $p$-wave coefficients from the non-relativistic velocity expansion.
\end{itemize}

\section{CP-violating rate densities}\label{sec:ratesCP}
\noindent $\gamma_\text{tot}$ is the thermal average of $\Gamma_\text{tot}$ in equation \eqref{eq:gamma-tot} computed via  \cite{Giudice:2003jh}
\begin{equation}\label{eq:decav}
    \gamma\left(\psi\rightarrow \dots\right) = \left(n_\psi^\text{eq.}+ n_{\overline{\psi}}^\text{eq.}\right) \frac{K_1(z)}{K_2(z)} \Gamma\left(\psi\rightarrow \dots\right)
\end{equation}
where we introduced $z\equiv m_\psi/T$ and $\text{K}_{1,2}(z)$ denotes the special Bessel functions of the first and second kind. The equilibrium number density of a particle $\psi$ reads \cite{Davidson:2008bu}
\begin{equation}
    n_\psi^\text{eq.} (z) = g_\psi \frac{T^3}{\pi^2}\begin{cases} \zeta(3) \quad \text{for bosons with}\quad T\gg m_\psi,  \\  \frac{3}{4} \zeta(3) \quad \text{for fermions with}\quad T\gg m_\psi,  \\
    \frac{z^2 \text{K}_{2}(z) }{2} \quad \text{for}\quad T\ll m_\psi,\end{cases}
\end{equation}
with $g_\psi$ being the spin degeneracy of $\psi$.
For scattering processes with a cross section $\sigma$ the appropriate thermally averaged density in the Maxwell-Boltzmann approximation is found to be \cite{Gondolo:1990dk,Davidson:2008bu}
\begin{equation}\label{eq:rates}
    \gamma\left(a+b\leftrightarrow i+j+\dots \right) = g_a g_b \frac{T}{32\pi^4}\int_{s_\text{min}}^\infty \text{d}s\; s^\frac{3}{2} \lambda\left(1,\frac{m_a^2}{s},\frac{m_b^2}{s}\right) K_1\left(\frac{\sqrt{s}}{T}\right)\; \sigma, \quad 
\end{equation}
with $g_{a,b}$ denoting the spin degeneracies of particles $a,b$ and 
\begin{equation}
\lambda(a,b,c) \equiv  (a-b-c)^2-4bc
\end{equation}
together with $s_\text{min} = \text{max}\left[\left(m_a+m_b\right)^2,\left(m_i+m_j+\dots\right)^2\right]$.
We parameterize the CP-violating thermally averaged decay widths in the following way \cite{Hambye:2005tk}
\begin{eqnarray}
    \gamma_\text{eq.}  \left(T\rightarrow L H\right) &=& \gamma_\text{eq.}  \left(\overline{L}H^\dagger \rightarrow \overline{T}\right) = \left(\text{B}_\text{L}+\varepsilon_L\right)\gamma_\text{tot}\\ 
    \gamma_\text{eq.}  \left(\overline{T}\rightarrow \overline{L}H^\dagger\right) &=& \gamma_\text{eq.}  \left(L H \rightarrow T\right) = \left(\text{B}_\text{L}-\varepsilon_L\right) \gamma_\text{tot}\\
     \gamma_\text{eq.}  \left(T\rightarrow DH\right) &=& \gamma_\text{eq.}  \left(\overline{D}H^\dagger \rightarrow \overline{T}\right) = \left(\text{B}_\text{D}-\varepsilon_L\right)\gamma_\text{tot}\\ 
    \gamma_\text{eq.}  \left(\overline{T}\rightarrow \overline{D}H^\dagger\right) &=& \gamma_\text{eq.}  \left(D H \rightarrow T\right) =\left(\text{B}_\text{D}+\varepsilon_L\right)\gamma_\text{tot}
\end{eqnarray}
which follows from CPT invariance and the definition of $\varepsilon$. 
The CP conserving branching ratios of the  decay widths are defined in \eqref{eq:BRs}.
Furthermore the same branching ratios apply to the CP conserving part of  $\gamma_\text{eq.} / \gamma_\text{tot.}$ because the factors from the thermal averages divide out.
Washout scattering mediated by $T$ occurs at the same order in the perturbative expansion as the generation of $\varepsilon$. The washout rate can be decomposed into two contributions $\gamma_{T_{s+t}}$ and $\gamma_{T_{t+u}}$. Because of CPT invariance the rate densities for the following reactions only possible in the $t$- and $u$-channel have to satisfy
\begin{equation}
    \gamma\left(L\overline{D} \rightarrow H H\right)    = \gamma\left(H^\dagger H^\dagger\rightarrow\overline{L}D\right) \equiv \gamma_{T_{t+u}}
\end{equation}
and the reactions possible in both the $s$- and $t$-channel  satisfy
\begin{eqnarray}
\gamma\left(\overline{D}H^\dagger\rightarrow \overline{L} H^\dagger\right)    &=& \gamma\left(LH\rightarrow DH\right),\\
\gamma\left(D H \rightarrow L H \right) &=& \gamma\left(\overline{L}H^\dagger\rightarrow \overline{D}H^\dagger\right).
\end{eqnarray}
Here the $s$-channel contribution to the washout scattering $\gamma_{T_{s+t}}$ involves intermediate on shell $T$s whose decays and inverse decays are already accounted for in the Boltzmann equation.
Therefore we have to perform real intermediate state (RIS) subtraction to remove the on shell contribution \cite{Kolb:1979qa,Giudice:2003jh,Buchmuller:2004nz} which can be expressed as \cite{Giudice:2003jh}
\begin{eqnarray}
    \gamma\left(LH\rightarrow DH\right)_\text{eq.} &=& \gamma_{T_{s+t}}- \gamma_\text{eq.}\left(L H \rightarrow T\right)\text{BR}\left(T\rightarrow DH\right),\\
    \gamma\left(\overline{L}H^\dagger\rightarrow \overline{D}H^\dagger\right)_\text{eq.} &=& \gamma_{T_{s+t}}- \gamma_\text{eq.}\left(\overline{L}H^\dagger\rightarrow \overline{T}\right)\text{BR}\left(\overline{T}\rightarrow \overline{D}H^\dagger\right),
\end{eqnarray}
and we expand the CP violating branching ratios  in  the subtracted rates to leading order in $\varepsilon_L$:
\begin{eqnarray}
    \gamma\left(LH\rightarrow DH\right)_\text{eq.} &=& \gamma_{T_{s+t}}- B_L B_D \gamma_\text{tot.} +\varepsilon_L\; \gamma_\text{tot.}+ \mathcal{O}\left(\varepsilon_L^2\right)\\
    \gamma\left(\overline{L}H^\dagger\rightarrow  \overline{D}H^\dagger\right)_\text{eq.} &=& \gamma_{T_{s+t}}- B_L B_D \gamma_\text{tot.}-\varepsilon_L\; \gamma_\text{tot.} + \mathcal{O}\left(\varepsilon_L^2\right)
\end{eqnarray}
Due to these relations we define the following object \cite{Hambye:2005tk}
\begin{equation}\label{eq:defsub}
    \gamma_{T_{s+t}}^\text{sub.} = \gamma_{T_{s+t}}- B_L B_D \gamma_\text{tot.} 
\end{equation}

\end{appendices}

\bibliographystyle{unsrt}
\bibliography{references}

\end{document}